\documentclass[11pt, a4paper]{article}
\pdfoutput=1

\usepackage{amsmath}
\usepackage{amsfonts}
\usepackage{amssymb}
\usepackage{bm}
\usepackage{bbm}
\usepackage{verbatim}
\usepackage{dsfont}
\usepackage{booktabs}
\usepackage{slashed}
\usepackage{nicefrac}
\usepackage{mathtools}
\usepackage{physics}
\usepackage{array}
\usepackage{bbding}

\usepackage{epsfig}
\usepackage{graphicx}
\usepackage[font=small]{caption}
\usepackage[listofformat=empty,subrefformat=empty]{subfig}

\usepackage{float}
\usepackage{overpic}
\usepackage{tikz}
\usepackage{tabularx}

\usepackage{enumerate}
\usepackage{hhline}
\usepackage{multirow}

\usepackage[nosort]{cite}
\usepackage{xspace}
\usepackage{setspace}

\usepackage[pdftitle={},
  pdfauthor={},
  pdfsubject={},
  bookmarksopen, bookmarksnumbered, bookmarksopenlevel=2, colorlinks=false, linkcolor=blue, citecolor=blue, urlcolor=blue]{hyperref}

\usepackage[nameinlink,capitalize]{cleveref}

\makeatletter{}
\g@addto@macro\bfseries{\boldmath}
\makeatother

\newcommand*{\bbZ}{\ensuremath{\mathbb{Z}}}

\newcommand*{\bbC}{\ensuremath{\mathbb{C}}}

\usepackage[normalem]{ulem}

\newcommand{\be}{\begin{equation}}
\newcommand{\ee}{\end{equation}}
\newcommand{\ba}{\begin{aligned}}
\newcommand{\ea}{\end{aligned}}

\newcommand{\wZ}{\widehat{X}} 

\newcommand{\frozen}{\text{\SnowflakeChevron}}

\def\half{{\frac{1}{2}}}

\newcommand{\Tor}{\text{Tor}\,}
\newcommand{\dA}{\partial_AB}
\newcommand{\dL}{\partial_LB}
\def\Z{{\mathbb{Z}}}

\newcommand{\affA}{\textsuperscript{\Snowflake}}
\newcommand{\affB}{\textsuperscript{\SnowflakeChevron}}
\newcommand{\affC}{\textsuperscript{\SnowflakeChevronBold}}

\usetikzlibrary{arrows}
\usetikzlibrary{arrows.meta}
\usetikzlibrary{shapes.geometric,calc,arrows, positioning,shapes.misc,decorations.markings}
\tikzset{
  big arrow/.style={
    decoration={markings,mark=at position 1 with {\arrow[scale=2,#1]{>}}},
    postaction={decorate},
    shorten >=0.4pt},
  big arrow/.default=black}

\pgfdeclarelayer{edgelayer}
\pgfdeclarelayer{nodelayer}
\pgfsetlayers{edgelayer,nodelayer,main}
\tikzstyle{none}=[inner sep=0pt]

\tikzstyle{NodeCross}=[draw, shape=circle, cross out, inner sep=0pt, minimum size=6pt,line width=0.25mm]
\tikzstyle{Circle}=[draw, shape=circle, black, fill=black, inner sep=0pt, minimum size=6pt]
\tikzstyle{circle}=[draw, shape=circle, black, fill=black, inner sep=0pt, minimum size=10pt]
\tikzstyle{Star}=[draw, shape=star, fill=red, star points=10, inner sep=0pt, minimum size=8pt]
\tikzstyle{CircleRed}=[draw, shape=circle, black, fill=red, inner sep=0pt, minimum size=4pt]
\tikzstyle{MidCircleRed}=[draw, shape=circle, black, fill=red, inner sep=0pt, minimum size=8pt]
\tikzstyle{StarP}=[draw={rgb,255: red,128; green,0; blue,128}, shape=star, fill={rgb,256: red,128; green,0; blue,128}, star points=8, inner sep=0pt, minimum size=12pt]
\tikzstyle{ShadedCircRed}=[draw=red, shape=circle, fill={rgb, 255: red,255; green,114; blue, 118}, inner sep=0pt, minimum size=80pt, line width=0.5mm, fill opacity=0.2]
\tikzstyle{ShadedCircRed2}=[draw=red, shape=circle, fill={rgb, 255: red,255; green,114; blue, 118}, inner sep=0pt, minimum size=10pt]
\tikzstyle{ShadedCircRed3}=[draw=black, shape=rectangle, fill={rgb, 255: red,255; green,114; blue, 118}, inner sep=0pt, minimum size=113pt, line width=0.25mm]
\tikzstyle{ShadedCirc}=[draw=red, shape=circle, fill=white, inner sep=0pt, minimum size=45pt,  fill opacity=1.0,  line width=0.5mm]
\tikzstyle{CircleBlue}=[draw, shape=circle, fill=blue, inner sep=0pt, minimum size=6pt]
\tikzstyle{BigCirclePurple}=[draw, shape=circle, fill={rgb,255: red,191; green,0; blue,191}, inner sep=0pt, minimum size=12pt]
\tikzstyle{CirclePurple}=[draw, shape=circle, fill={rgb,255: red,191; green,0; blue,191}, inner sep=0pt, minimum size=8pt]
\tikzstyle{EmptyCircle}=[draw, shape=circle, inner sep=0pt, minimum size=4pt]
\tikzstyle{GreenCircle}=[draw, shape=circle,  fill={rgb,255: red,80; green,200; blue,120}, inner sep=0pt, minimum size=8pt]
\tikzstyle{BrownCircle}=[draw, shape=circle,  fill={rgb,255: red,115; green,115; blue,115}, inner sep=0pt, minimum size=8pt]
\tikzstyle{CirclePurpleSmall}=[draw, shape=circle, fill={rgb,255: red,191; green,0; blue,191}, inner sep=0pt, minimum size=4pt]
\tikzstyle{BigCircleGreen}=[draw, shape=circle, fill={rgb,255: red,0; green,191; blue,0}, inner sep=0pt, minimum size=12pt]
\tikzstyle{BigCircleBlue}=[draw, shape=circle, fill={rgb,255: red,0; green,0; blue,191}, inner sep=0pt, minimum size=12pt]
\tikzstyle{BigCircleRed}=[draw, shape=circle, fill={rgb,255: red,191; green,0; blue,0}, inner sep=0pt, minimum size=12pt]
\tikzstyle{CircleBrown}=[draw, shape=circle, fill={rgb,255: red,210; green,105; blue,30}, inner sep=0pt, minimum size=8pt]
\tikzstyle{BigCircleGrey}=[shape=circle,  fill={rgb,255: red,120; green,120; blue,120}, inner sep=0pt, minimum size=10pt]
\tikzstyle{SmallCircleGrey}=[shape=circle,  fill={rgb,255: red,120; green,120; blue,120}, inner sep=0pt, minimum size=6pt]

\tikzstyle{DashedLine}=[-, densely dashed, line width=0.25mm]
\tikzstyle{DottedLine}=[-, dotted, line width=0.25mm]
\tikzstyle{ThickLine}=[-, line width=0.25mm]
\tikzstyle{ThickerLine}=[-, line width=0.4mm]
\tikzstyle{ArrowLineRight}=[-, -{Stealth[scale=1.25]}, line width=0.25mm, scale=5]
\tikzstyle{ArrowLineRed}=[-, draw={rgb,255: red,191; green,0; blue,0}, -{Stealth[scale=1.75]}, line width=0.1mm, scale=5]
\tikzstyle{RedLine}=[-, draw={rgb,255: red,191; green,0; blue,0}, fill=none, line width=0.5mm]
\tikzstyle{DashedLineThin}=[-, densely dashed, line width=0.125mm, fill=none, draw=black]
\tikzstyle{DottedRed}=[-, dotted, draw={rgb,255: red,191; green,0; blue,0}, fill=none, line width=0.25mm]
\tikzstyle{DashedRed}=[-, densely dashed, draw={rgb,255: red,191; green,0; blue,0}, fill=none, line width=0.25mm]
\tikzstyle{BlueLine}=[-, draw={rgb,255: red,0; green,0; blue,191}, fill=none, line width=0.5mm]
\tikzstyle{ArrowLineBlue}=[-, draw={rgb,255: red,0; green,0; blue,191}, -{Stealth[scale=1.75]}, line width=0.1mm, scale=5]
\tikzstyle{GreenDoubleArrow}=[<->, draw={rgb,155: red,0; green,255; blue,0},  line width= 0.5mm, scale=5]
\tikzstyle{RedDoubleArrow}=[<->, draw={rgb,255: red,255; green,0; blue,0},  line width= 0.5mm, scale=5]
\tikzstyle{BlueDottedLight}=[-, dotted, draw={rgb,255: red,0; green,0; blue,191}, fill=none, line width=0.3mm]
\tikzstyle{BrownLine}=[-, draw={rgb,255: red,210; green,105; blue,30}, fill=none, line width=0.5mm]
\tikzstyle{DottedBrownLine}=[-, dotted, draw={rgb,255: red,210; green,105; blue,30}, fill=none, line width=0.5mm]
\tikzstyle{DottedRed}=[-, dotted, draw={rgb,255: red,191; green,0; blue,0}, fill=none, dotted, line width=0.5mm]
\tikzstyle{DottedPurple}=[-, dotted, draw={rgb,255: red,191; green,0; blue,191}, fill=none, dotted, line width=0.5mm]
\tikzstyle{BlueDottedLight}=[-, dotted, draw={rgb,255: red,0; green,0; blue,191}, fill=none, line width=0.5mm]
\tikzstyle{ArrowLinePurple}=[-, draw={rgb,255: red,191; green,0; blue,191}, -{Stealth[scale=1.75]}, line width=0.5mm, scale=5]
\tikzstyle{DashedLineGreen}=[-, densely dashed, draw={rgb,255: red,74; green,103; blue,65}, line width=0.25mm]
\tikzstyle{LineGreen}=[-, draw={rgb,255: red, 74; green,200; blue,65}, line width=0.5mm]
\tikzstyle{ArrowLineGreen}=[-, draw={rgb,255: red,0; green,191; blue,0}, -{Stealth[scale=1.75]}, line width=0.5mm, scale=5]
\tikzstyle{GreenLine}=[-, draw={rgb,255: red,0; green,191; blue,0}, fill=none, line width=0.5mm]
\tikzstyle{PurpleLine}=[-, draw={rgb,255: red,191; green,0; blue,191}, fill=none, line width=0.5mm]
\tikzstyle{DPurpleLine}=[-, dotted, draw={rgb,255: red,191; green,0; blue,191}, fill=none, line width=0.5mm]
\tikzstyle{LightBlue}=[-, draw={rgb,255: red,0; green,150; blue,255}, fill=none, line width=0.5mm]
\tikzstyle{DLightBlue}=[-, dotted, draw={rgb,255: red,0; green,150; blue,255}, fill=none, line width=0.5mm]
\tikzstyle{ThickGreenLine}=[-, draw={rgb,255: red,0; green,255; blue,0}, fill=none, line width=0.75mm]
\tikzstyle{ThickDarkGreenLine}=[-, draw={rgb,255: red,80; green,200; blue,120}, fill=none, line width=0.75mm]

\tikzstyle{NodeCross}=[draw, shape=circle, cross out, inner sep=0pt, minimum size=10pt,line width=0.25mm]
\tikzstyle{SmallCircle}=[draw, shape=circle, black, fill=black, inner sep=0pt, minimum size=6pt]
\tikzstyle{BigCircle}=[draw, shape=circle, black, fill=black, inner sep=0pt, minimum size=10pt]
\tikzstyle{MidCircle}=[draw, shape=circle, black, fill=black, inner sep=0pt, minimum size=8pt]
\tikzstyle{SmallCircleRed}=[shape=circle, fill={rgb,255: red,191; green,0; blue,0}, inner sep=0pt, minimum size=6pt]
\tikzstyle{BigCircleRed}=[shape=circle, fill={rgb,255: red,191; green,0; blue,0}, inner sep=0pt, minimum size=10pt]
\tikzstyle{SmallCircleBlue}=[shape=circle, fill=blue, inner sep=0pt, minimum size=6pt]
\tikzstyle{BigCircleBlue}=[shape=circle, fill=blue, inner sep=0pt, minimum size=10pt]
\tikzstyle{SmallCirclePurple}=[shape=circle, fill={rgb,255: red,191; green,0; blue,191}, inner sep=0pt, minimum size=6pt]
\tikzstyle{BigCirclePurple}=[shape=circle, fill={rgb,255: red,191; green,0; blue,191}, inner sep=0pt, minimum size=10pt]
\tikzstyle{SmallCircleGreen}=[shape=circle, fill={rgb,255: red,80; green,200; blue,120}, inner sep=0pt, minimum size=6pt]
\tikzstyle{SmallCircleBrightGreen}=[shape=circle, fill={rgb,255: red,80; green,255; blue,120}, inner sep=0pt, minimum size=6pt]
\tikzstyle{BigCircleGreen}=[shape=circle,  fill={rgb,255: red,80; green,200; blue,120}, inner sep=0pt, minimum size=10pt]
\tikzstyle{SmallCircleBrown}=[shape=circle,  fill={rgb,255: red,210; green,105; blue,30}, inner sep=0pt, minimum size=6pt]
\tikzstyle{BigCircleBrown}=[shape=circle,  fill={rgb,255: red,210; green,105; blue,30}, inner sep=0pt, minimum size=10pt]
\tikzstyle{Star}=[draw, shape=star, fill=black, star points=8, inner sep=0pt, minimum size=10pt]
\tikzstyle{MidCircleGreen}=[shape=circle,  fill={rgb,255: red,80; green,200; blue,120}, inner sep=0pt, minimum size=8pt]

\tikzstyle{DashedLine}=[-, densely dashed, line width=0.25mm]
\tikzstyle{DottedLine}=[-, dotted, line width=0.25mm]
\tikzstyle{ThickLine}=[-, line width=0.25mm]
\tikzstyle{ArrowLineRight}=[-, -{Stealth[scale=1.75]}, line width=0.15mm, scale=5]
\tikzstyle{RedLine}=[-, draw={rgb,255: red,191; green,0; blue,0}, fill=none, line width=0.5mm]
\tikzstyle{DashedRedLine}=[-, densely dashed, draw={rgb,255: red,191; green,0; blue,0}, fill=none, line width=0.5mm]
\tikzstyle{DottedRed}=[-, dotted, draw={rgb,255: red,191; green,0; blue,0}, fill=none, dotted, line width=0.5mm]
\tikzstyle{BlueLine}=[-, draw=blue, fill=none, line width=0.5mm]
\tikzstyle{ThickBlueLine}=[-, draw=blue, fill=none, line width=0.75mm]
\tikzstyle{DashedBlueLine}=[-, densely dashed, draw=blue, fill=none, line width=0.5mm]
\tikzstyle{DottedBlue}=[-, dotted, draw=blue, fill=none, dotted, line width=0.5mm]
\tikzstyle{PurpleLine}=[-, draw={rgb,255: red,191; green,0; blue,191}, fill=none, line width=0.5mm]
\tikzstyle{DashedPurpleLine}=[-, densely dashed, draw={rgb,255: red,191; green,0; blue,191}, fill=none, line width=0.5mm]
\tikzstyle{DottedPurple}=[-, dotted, draw={rgb,255: red,191; green,0; blue,191}, fill=none, dotted, line width=0.5mm]\tikzstyle{GreenLine}=[-, draw={rgb,255: red,80; green,200; blue,120}, fill=none, line width=0.5mm]
\tikzstyle{DashedGreenLine}=[-, densely dashed, draw={rgb,255: red,80; green,200; blue,120}, fill=none, line width=0.5mm]
\tikzstyle{DottedGreen}=[-, dotted, draw={rgb,255: red,80; green,200; blue,120}, fill=none, dotted, line width=0.5mm]
\tikzstyle{BrownLine}=[-, draw={rgb,255: red,210; green,105; blue,30}, fill=none, line width=0.5mm]
\tikzstyle{DashedBrownLine}=[-, densely dashed, draw={rgb,255: red,210; green,105; blue,30}, fill=none, line width=0.5mm]
\tikzstyle{DottedBrown}=[-, dotted, draw={rgb,255: red,210; green,105; blue,30}, fill=none, dotted, line width=0.5mm]

\tikzset{snake it/.style={decorate, decoration=snake}}

\begin{document}

\baselineskip=18pt  
\numberwithin{equation}{section}  
\allowdisplaybreaks  

\thispagestyle{empty}

\vspace*{-2cm}

\vspace*{2.6cm}
\begin{center}
{\Large{\textbf{Coupling of TQFTs to QFTs and SymTFT interfaces \\[0.5em] from bordisms
}}} 
\\
\vspace*{1.5cm}
Max H\"ubner\affA, Ling Lin\affB, 
Daniel Robbins\affC and Subham Roy\affC \\

\vspace*{1.0cm}
{\it \affA Dipartimento di Fisica e Astronomia ``Galileo Galilei'', Universit\`{a} degli Studi di Padova \\
\& \\ INFN Sezione di Padova, Via F. Marzolo 8, 35131 Padova, Italy}\\
\vspace*{.3cm}
{\it \affB Dipartimento di Fisica e Astronomia, Universit\`{a} di Bologna, via Irnerio 46, Bologna, Italy\\ \& \\INFN, Sezione di Bologna, viale Berti Pichat 6/2, Bologna, Italy}\\
\vspace*{.3cm}
{\it  \affC Department of Physics,
University at Albany, SUNY}\\
\vspace*{.1cm}
{\it 1400 Washington Avenue, Albany, NY - 12222, USA}

\vspace*{0.8cm}
\end{center}
\vspace*{.5cm}

\noindent
M-theory local models may exhibit torsional cycles and other non-trivial topology which interpolate between quantum field theory sectors localized at singularities and the asymptotic regions of the local model. 
We show that such structures give rise, in standard decoupling limits, to additional topological sectors which can couple non-trivially to the localized QFT.  From a symmetry topological field theory (SymTFT) perspective, the physical boundary conditions associated to such local models are modified. 
We give a construction of SymTFT interfaces from geometric bordisms interpolating between the local singularity and the asymptotic boundary. We apply this construction to the circle compactification of a frozen $D_5$ singularity, identifying a topological interface between the $\mathbb{Z}_4$ SymTFT of the full theory and the $\mathbb{Z}_2$ SymTFT of its local $\mathfrak{su}(2)$ gauge sector. We show that the resulting coupling between the gauge theory and an additional topological quantum field theory accounts for the apparent mismatch between electric and magnetic symmetries of the frozen theory.

\newpage
\setcounter{tocdepth}{2}
\tableofcontents

\newpage


\section{Introduction}

A symmetry topological field theory (SymTFT) is an auxiliary $(d+1)$-dimensional topological quantum field theory that organizes the generalized global symmetries \cite{Gaiotto:2014kfa}  of a $d$-dimensional QFT \cite{Freed:2012bs, Gaiotto:2020iye, Apruzzi:2021nmk, Freed:2022qnc, Kaidi:2022cpf, vanBeest:2022fss, Kaidi:2023maf, Bhardwaj:2023ayw, Baume:2023kkf, Bhardwaj:2023bbf, Brennan:2024fgj, Antinucci:2024zjp, Bonetti:2024cjk, Bhardwaj:2024qrf, Cvetic:2024dzu, Gagliano:2024off, Jia:2025jmn, Lin:2025oml, Robbins:2025puq, Apruzzi:2025hvs, Bonetti:2025dvm, Jia:2025vrj, Bergman:2026lnz}.\footnote{For discussion on generalized global symmetries in general, see \cite{Sharpe:2015mja, Bah:2022xfv, McGreevy:2022oyu, Cordova:2022ruw, Freed:2022iao, Schafer-Nameki:2023jdn, Brennan:2023mmt, Bhardwaj:2023kri, Luo:2023ive, Shao:2023gho, Iqbal:2024pee, Costa:2024wks} and the references therein.} In the sandwich construction \cite{Gaiotto:2020iye, Freed:2022qnc, Kaidi:2022cpf,Bhardwaj:2023ayw}, the $(d+1)$-dimensional SymTFT is placed on a slab consisting of the $d$-dimensional spacetime of the underlying QFT cross a compact interval with two boundaries. The gapless degrees of freedom of the $d$-dimensional QFT reside on the non-topological physical boundary. On the other end of the interval we have a topological boundary condition, known as the symmetry boundary, that determines the global form and the generalized global symmetries of the $d$-dimensional QFT. 

Operationally, the topological operators of the SymTFT encode the symmetry generators and the charged operators of the QFT.
Specifically, charged operators of the absolute theory arise from bulk topological operators anchored between the two boundaries.
In the standard SymTFT construction, the physical boundary allows the termination of all bulk topological operators \cite{Freed:2022qnc, Bhardwaj:2023ayw} and, consequently, can give rise to charged operators if they can consistently end on the symmetry boundary.
This situation is realized in most string constructions studied in the literature \cite{Bergman:2020ifi, Apruzzi:2021nmk, Apruzzi:2022rei, Bergman:2022otk, Heckman:2022muc, vanBeest:2022fss, Bashmakov:2022uek, Heckman:2022xgu, Chen:2023qnv, Bah:2023ymy, Baume:2023kkf, Yu:2023nyn, Gould:2023wgl, Najjar:2023hee, Heckman:2024oot, DelZotto:2024tae, GarciaEtxebarria:2024fuk, Braeger:2024jcj, Franco:2024mxa, Heckman:2024zdo, Bergman:2024aly, GarciaEtxebarria:2024jfv, Tian:2024dgl, Apruzzi:2024cty, Yu:2024jtk, Najjar:2024vmm, Bergman:2024its, Bonetti:2024etn, Najjar:2025rgt, Cvetic:2025kdn, Heckman:2025isn, DeMarco:2025pza, Braeger:2025rov, Heckman:2025lmw, Bergman:2025isp, Bah:2025vfu, Dramburg:2025tlb, Chakrabhavi:2026bbi}, in particular in the context of QFTs engineered geometrically on conical singularities \cite{Apruzzi:2021nmk, Heckman:2022muc, vanBeest:2022fss, Bashmakov:2022uek, Heckman:2022xgu, Chen:2023qnv, Bah:2023ymy, Yu:2023nyn, Gould:2023wgl, Najjar:2023hee, DelZotto:2024tae, Braeger:2024jcj, Franco:2024mxa, Tian:2024dgl, Apruzzi:2024cty, Najjar:2024vmm, Bonetti:2024etn, Najjar:2025rgt, DeMarco:2025pza, Braeger:2025rov, Dramburg:2025tlb}.

An interesting exception to this (empirical) pattern are gauge theories obtained from so-called frozen singularities \cite{Witten:1997bs, deBoer:2001wca, Atiyah:2001qf, Tachikawa:2015wka, Bhardwaj:2018jgp}:
If we identify the SymTFT operators from the topology on the asymptotic boundary, then not all of them can terminate on the physical boundary \cite{Cvetic:2022uuu,Cvetic:2024mtt}.
The absence of the associated charged operators then leads to a set of ``trivially-acting'' symmetry generators, originating from the topological operators in the bulk that link with those non-endable defects.
This non-effective symmetry action was crucial to explain the apparent mismatch between the electric and magnetic charged operators obtained from wrapped M2- vs.~M5-branes in the seven dimensional super-Yang--Mills (SYM) theories, which is the low-energy effective field theory (EFT) of M-theory compactified on frozen ADE singularities~\cite{Cvetic:2022uuu,Cvetic:2024mtt}.

In ``bottom-up'' QFT models, the non-effective action of a symmetry is usually observed in the deep infrared limit, when the objects charged under the symmetry are often absent in the IR description of the QFT. 
The SymTFT realization of non-effective symmetry actions and the absence of charged objects have already appeared in the literature \cite{Bhardwaj:2023bbf,Bhardwaj:2024qrf}, and involve \emph{topological interfaces} in the SymTFT bulk.\footnote{An alternative approach to trivially acting symmetries starts from the idea that a $p$-dimensional topological operator which can terminate on a $(p-1)$-dimensional topological operator will link trivially with any $(d-p-1)$-dimensional operator, and this has also been implemented in a SymTFT context~\cite{Robbins:2022wlr, Lin:2025oml}. Trivially acting symmetries can have interesting implications; for example gauging a trivially acting $(d-2)$-form symmetry leads to a theory with $(d-1)$-form symmetry. Consequently, the gauged theory decomposes \cite{Hellerman:2006zs, Sharpe:2022ene, Sharpe:2014tca, Sharpe:2019ddn, Sharpe:2021srf, Komargodski:2020mxz, Cherman:2020cvw, Tanizaki:2019rbk,Nguyen:2021naa, Lin:2022xod, Cherman:2021nox, Robbins:2020msp }.}
Such an interface can pose non-trivial boundary conditions for some of the bulk defects $U^{(p)}$, such that these terminate on the interface rather than on the physical boundary where the gapless degrees of freedom reside.
Pushing the interface on top of the physical boundary results in a new boundary where the ``dual'' defects $W^{(d-p)}$, which link with $U^{(p)}$ in the bulk, can no longer end there, resulting in ``missing charges''.

The purpose of this work is to give a top-down origin of this structure in M-theory geometric engineering. We show that torsion in the (co)homology of the internal space produces topological degrees of freedom on the physical boundary, alongside the gauge sector. Depending on the geometry, these can be either coupled or decoupled from the gauge sector. 
In both cases the physical boundary no longer admits endpoints for all bulk SymTFT operators. We refer to such a boundary as a \textit{modified physical boundary}. 
By going to a dual geometric description, we show that frozen singularities fit this framework: said dual geometries carry torsion, and this torsion produces an additional topological sector as well as the modified physical boundary found in \cite{Cvetic:2024mtt}. We will see that in this case, the topological sector and the gauge sector turn out to be coupled non-trivially, in the sense of Kapustin and Seiberg \cite{Kapustin:2014gua}. 
Importantly, we will demonstrate how we can extract the equivalent interface-description from the geometry.
In the following, we give an overview of our approach and results.

\subsubsection*{Modified physical boundaries from the bottom-up}

From the bottom-up perspective, a modified physical boundary can be obtained by half-space gauging \cite{Choi:2022zal, Choi:2021kmx,Choi:2022jqy}. 
Starting from a SymTFT that encodes some symmetry category ${\cal S}$\footnote{For simplicity, and with the specific top-down models in mind, we will only focus on finite invertible higher-form symmetries.}, one can gauge an appropriate bulk symmetry only in a half-space within the SymTFT slab \cite{Perez-Lona:2025ncg, Robbins:2025puq}. 
This produces a topological interface ${\cal I}$ between the original SymTFT and one with \textit{reduced} symmetry ${\cal S}'$.
Fusing this interface with a physical boundary $\mathfrak{B}_{\rm phys}^{({\cal S}')}$ of the ${\cal S}'$-SymTFT gives a new physical boundary $\mathfrak{B}^{({\cal S})}_{\rm phys}$ of the ${\cal S}$-SymTFT,
\begin{equation}\label{eq:intro-bottom-up}
  \mathfrak B^{(\mathcal S)}_{\rm phys}
  =\mathcal I \otimes \mathfrak B^{(\mathcal S')}_{\rm phys}\,.
\end{equation}
Only a subset of the $\mathcal S$-SymTFT operators can terminate on this new boundary. The remaining operators see a Neumann-type boundary condition. The resulting absence of their endpoints on the physical boundary makes the symmetry generators (dual to them) act trivially on any absolute theory which is specified once we pick a topological symmetry boundary condition $\mathfrak{B}^{({\cal S})}_{\rm sym}$ for the SymTFT.

This construction is closely related to the ``club sandwich/quiche'' \cite{Bhardwaj:2023bbf,Bhardwaj:2024qrf,Robbins:2025puq}, with the difference being the type of boundary one starts with. In that construction, the interface is fused with a topological symmetry boundary to obtain a new topological symmetry boundary corresponding to the reduced SymTFT. Here we instead fuse the interface with the physical boundary. The resulting boundary supports the non-topological degrees of freedom, but no longer admits endpoints for every operator of the larger SymTFT.

\subsubsection*{Modified physical boundaries from the top-down}

We now turn to the realization of modified physical boundaries directly from M-theory compactification. In the standard geometric-engineering dictionary~\cite{Apruzzi:2021nmk}, the SymTFT follows from reducing M-theory on the asymptotic boundary $\partial X$ of an internal space $X$, while compactification on $X$ determines its physical boundary. In many such setups the homology of the internal space is torsion-free. Then, compactly supported fluxes describe the dynamical Cartan sector \cite{Morrison:2020ool,Albertini:2020mdx}, while the quotient of ordinary cohomology classes by compactly supported cohomology classes determines the defect group and its background gauge fields \cite{Albertini:2020mdx,DelZotto:2015isa,GarciaEtxebarria:2019caf}. 

The situation changes when $X$ has torsion in its homology. In particular, suppose that\footnote{Here and throughout we write $H_k(X)\equiv H_k(X;\Z)$ and $H^k(X)\equiv H^k(X;\Z)$  for homology and cohomology groups with integer coefficients respectively.}
\begin{equation}\label{eq:intro-torsion}
  \mathcal Z=\text{Tor}\, H_1(X)\neq0\, .
\end{equation}
The universal coefficient theorem (UCT) and Poincar\'e--Lefschetz duality then imply the appearance of two new inputs,
\begin{equation}
\text{Tor}\, H^2(X)\cong\mathcal Z^\vee\,,\qquad
  \text{Tor}\, H^{m-1}(X,\partial X)\cong\mathcal Z\,,
\end{equation}
where $m=\dim_{\mathbb R}X$. The former produces additional finite background fields. 
Since the corresponding cohomology classes restrict non-trivially to the boundary of the internal space, these fields are distinct from the background fields associated with the 1-form symmetries of QFT sectors localized in $X$ which act on defects that are, in contrast, associated to torsional classes collapsing along the QFT's support in $X$.
On the other hand, the torsion in $H^{m-1}(X,\partial X)$ give additional compactly supported fluxes which, upon Kaluza--Klein-reducing the 11d supergravity action, give dynamical fields that must be summed over in the path integral. These two new ingredients pair naturally through the torsion linking of the internal space. Consequently, reducing the M-theory kinetic term produces an additional finite topological sector.  
This sector is distinct from the ordinary gauge sector obtained from the free cohomology.

The dynamical fields of this torsional sector provide endpoints for a set of SymTFT operators on the physical boundary. More importantly, the operators dual to them do not admit any corresponding endpoints. The physical boundary produced by compactification with torsion therefore has precisely the desired property that certain topological operators do not terminate on it, which is the key characterization of what we refer to as a modified physical boundary. 
Upon choosing an appropriate symmetry boundary condition, the absence of the corresponding objects charged under the symmetry causes a part of the symmetry to act non-effectively or trivially. 

\subsubsection*{Topological interfaces from bordisms}

\begin{figure}[ht]
\centering
\scalebox{0.8}{
\begin{tikzpicture}
	\begin{pgfonlayer}{nodelayer}
		\node [style=none] (2) at (-2.25, 1.75) {};
		\node [style=none] (3) at (2.25, 0.75) {};
		\node [style=none] (4) at (2.25, 1.75) {};
		\node [style=none] (6) at (-2.25, 1.75) {};
		\node [style=none] (7) at (-2.25, 0.75) {};
		\node [style=none] (8) at (2.5, 0.75) {};
		\node [style=none] (9) at (2.5, 1.75) {};
		\node [style=none] (10) at (4.25, 0.75) {};
		\node [style=none] (11) at (4.25, 1.75) {};
		\node [style=none] (12) at (4.5, 0.75) {};
		\node [style=none] (13) at (4.5, 1.75) {};
		\node [style=none] (18) at (-5, 1.75) {};
		\node [style=none] (19) at (-5, 0.75) {};
		\node [style=none] (20) at (-2.5, 1.75) {};
		\node [style=none] (21) at (-2.5, 0.75) {};
		\node [style=none] (29) at (0, -2) {};
		\node [style=none] (30) at (0, -5) {};
		\node [style=none] (31) at (6.5, -2) {};
		\node [style=none] (32) at (6.5, -5) {};
		\node [style=none] (33) at (-5, -2) {};
		\node [style=none] (34) at (-5, -5) {};
		\node [style=none] (36) at (-5, -5.5) {Topological};
		\node [style=none] (37) at (6.5, -5.5) {Physical};
		\node [style=none] (38) at (0, -5.5) {Interface};
		\node [style=none] (39) at (-3.5, 0) {};
		\node [style=none] (40) at (-2.5, -1.5) {};
		\node [style=none] (41) at (3.5, 0) {};
		\node [style=none] (42) at (3.25, -1.5) {};
		\node [style=none] (43) at (0, -1.5) {};
		\node [style=none] (44) at (0, -0.25) {};
		\node [style=none] (45) at (5.75, 0) {};
		\node [style=none] (46) at (6.5, -1.5) {};
		\node [style=none] (47) at (0, 0.25) {};
		\node [style=none] (48) at (0, 2.25) {};
		\node [style=none] (49) at (6.5, 1.25) {};
		\node [style=none] (50) at (-0.5, 1.2) {};
		\node [style=none] (51) at (0.5, 1.2) {};
		\node [style=none] (52) at (0.75, 1.25) {};
		\node [style=none] (53) at (-0.75, 1.25) {};
		\node [style=Star] (54) at (6.5, 1.25) {};
        \node [style=none] (55) at (0, -6.75) {};
        \node [style=none] (56) at (-5.75, 1.25) {$X$};
		\node [style=none] (57) at (8, 1.25) {Singularity};
		\node [style=none] (58) at (2, -0.75) {Compactification};
        \node [style=none] (59) at (-5, -6) {Boundary};
        \node [style=none] (60) at (6.5, -6) {Boundary};
        \node [style=none] (61) at (-8.75, -0.75) {};
        \node [style=none] (62) at (0, 2.75) {Bordism $B$};
	\end{pgfonlayer}
	\begin{pgfonlayer}{edgelayer}
        \filldraw[fill=green!20, draw=green!20]  (0, -2) -- (0, -5) -- (6.5, -5) -- (6.5, -2) -- cycle;
        \filldraw[fill=blue!20, draw=blue!20]  (0, -2) -- (0, -5) -- (-5, -5) -- (-5, -2) -- cycle;
		\draw [style=ThickLine, bend left=15] (3.center) to (4.center);
		\draw [style=ThickLine, bend right=15] (3.center) to (4.center);
		\draw [style=ThickLine, bend left=15] (7.center) to (6.center);
		\draw [style=ThickLine, bend left=15] (6.center) to (7.center);
		\draw [style=GreenLine, bend left=15] (8.center) to (9.center);
		\draw [style=GreenLine, bend left=15] (9.center) to (8.center);
		\draw [style=GreenLine, bend left=15] (10.center) to (11.center);
		\draw [style=GreenLine, bend right=15] (10.center) to (11.center);
		\draw [style=GreenLine] (9.center) to (11.center);
		\draw [style=GreenLine] (8.center) to (10.center);
		\draw [style=ThickLine, bend left=15] (12.center) to (13.center);
		\draw [style=ThickLine, bend left=15] (13.center) to (12.center);
		\draw [style=BlueLine, bend left=15] (19.center) to (18.center);
		\draw [style=BlueLine, bend left=15] (18.center) to (19.center);
		\draw [style=BlueLine, bend left=15] (21.center) to (20.center);
		\draw [style=BlueLine, bend left=15] (20.center) to (21.center);
		\draw [style=BlueLine] (20.center) to (18.center);
		\draw [style=BlueLine] (19.center) to (21.center);
		\draw [style=ThickLine] (33.center) to (29.center);
		\draw [style=ThickLine] (29.center) to (31.center);
		\draw [style=ThickLine] (31.center) to (32.center);
		\draw [style=ThickLine] (32.center) to (30.center);
		\draw [style=ThickLine] (33.center) to (34.center);
		\draw [style=ThickLine] (34.center) to (30.center);
		\draw [style=ThickLine] (29.center) to (30.center);
		\draw [style=ArrowLineRight] (44.center) to (43.center);
		\draw [style=ThickLine, in=-180, out=0] (6.center) to (48.center);
		\draw [style=ThickLine, in=-180, out=0] (48.center) to (4.center);
		\draw [style=ThickLine, in=0, out=-180] (3.center) to (47.center);
		\draw [style=ThickLine, in=0, out=-180] (47.center) to (7.center);
		\draw [style=ThickLine] (13.center) to (49.center);
		\draw [style=ThickLine] (49.center) to (12.center);
		\draw [style=ThickLine, bend left=15] (50.center) to (51.center);
		\draw [style=ThickLine, bend left=15] (52.center) to (53.center);
	\end{pgfonlayer}
\end{tikzpicture}
}
\caption{Geometric realization of a SymTFT interface. The internal geometry $X$ decomposes into a local model containing the singular locus and an interpolating bordism $B$. Compactification on $B$ produces an interface between the SymTFTs associated with the two cylinders. }
\label{fig:Bordism}
\end{figure}
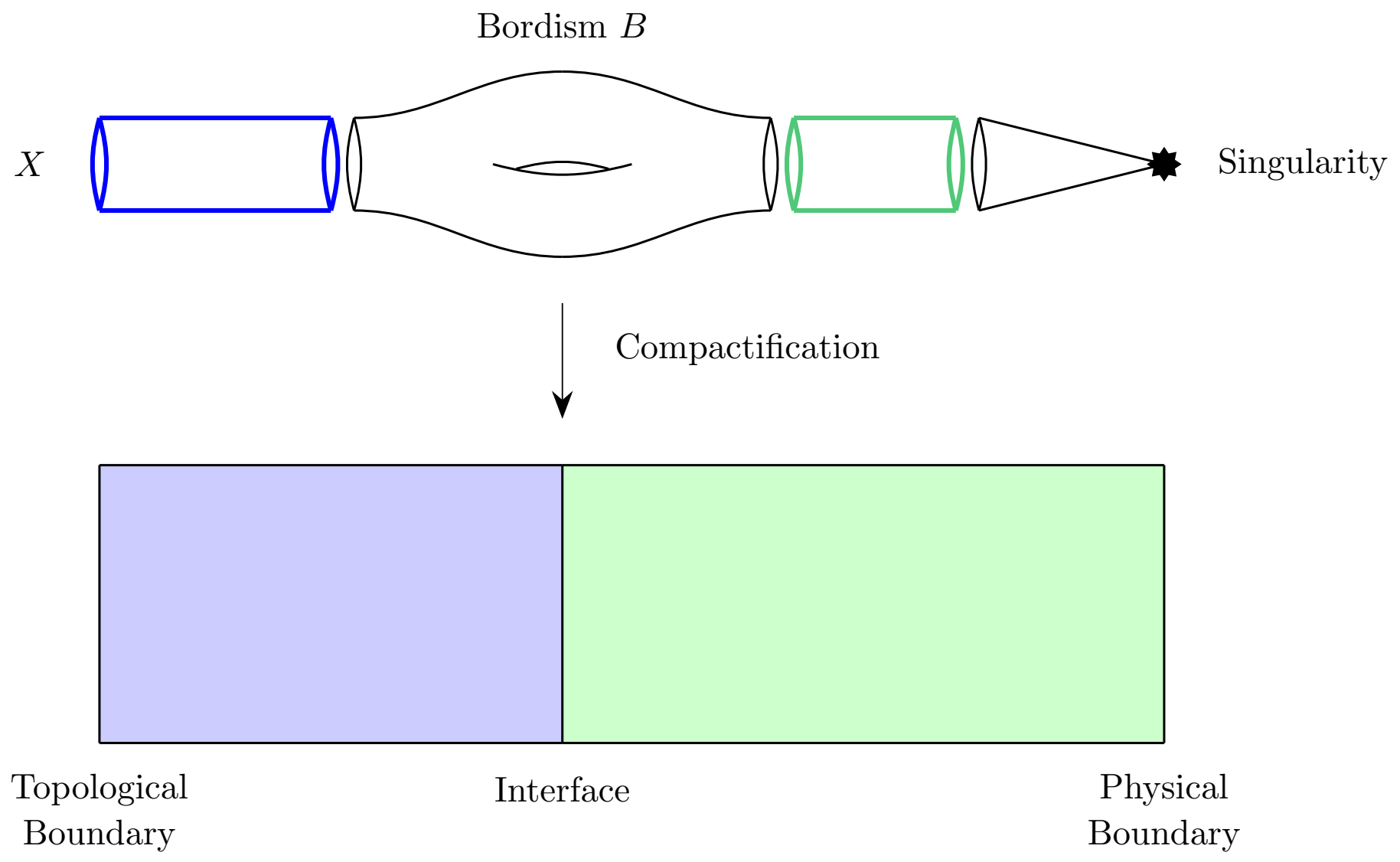

While the presence of torsion provides a top-down realization of modified physical boundaries, this description leaves no room for the interfaces in the SymTFT bulk as in~\eqref{eq:intro-bottom-up}. To uncover these interfaces, we need to move beyond conical geometries, as first suggested in \cite{Apruzzi:2021nmk}, and consider geometries that decompose into pieces, such as cylinders, where the standard extra-dimensional formalism applies, and interpolating regions, which we call bordisms; see Figure~\ref{fig:Bordism}.

Such constructions have already appeared in the literature \cite{Baume:2023kkf, Braeger:2024jcj, Heckman:2024zdo, Cvetic:2024dzu, GarciaEtxebarria:2024jfv, Heckman:2025isn, Heckman:2025lmw}. The correspondence between bordisms and interfaces, junctions, boundaries, and corners in extra-dimensional constructions of SymTFTs is now understood in considerable generality. For our purposes, however, the question is more specific: when do bordisms map to \emph{topological} SymTFT building blocks, such as topological interfaces?

We show that the interface theory associated with a bordism $B$ is obtained by reducing string or M-theory on $B$. The boundary conditions imposed on the bulk SymTFT fields at the interface are determined by how $B$ is glued to the rest of the geometry along $\partial B$. These boundary conditions can eat up all the degrees of freedom associated, via a naive compactification, to $B$. The (co)homology lift of the embedding $\partial B\hookrightarrow B$, stratified across torsional and free classes, informs whether the interface purely realizes boundary conditions for the bulk fields, supports topological degrees of freedom, or also supports non-topological degrees of freedom.

In the first two cases, the interface is a topological interface, and the geometry realizes a club-sandwich-like configuration \cite{Bhardwaj:2023bbf, Bhardwaj:2024qrf}. Fusing the interface with the physical boundary of the SymTFT associated with the local model produces exactly a modified physical boundary. If the interface instead supports non-topological degrees of freedom, the configuration is of the more general SymTree type \cite{Baume:2023kkf}. 

\subsubsection*{Summary of results}

In Section~\ref{sec:top-down}, we derive the modified physical boundary directly from M-theory compactification. We show that the torsion linking pairing of the internal geometry gives rise
to additional finite topological terms in the physical boundary action whose boundary conditions reproduce the asymmetric termination of SymTFT operators described above.

In Section~\ref{sec:bordism_general}, we develop a general prescription for determining the interface theory associated with a bordism $B$. The central ingredient is the long exact cohomology sequence of the pair $(B,\partial B)$,
\begin{equation}
\begin{aligned}
\cdots \longrightarrow H^{n-1}(\partial B)
\xrightarrow{\ d_{n-1}\ } H^n(B,\partial B)
\xrightarrow{\ \iota^n\ } H^n(B)
\xrightarrow{\ r_n\ } H^n(\partial B)
\longrightarrow\cdots \, .
\end{aligned}
\end{equation}
The maps $r_n$ and $d_{n-1}$ determine the gluing conditions for the bulk SymTFT fields at the interface, while $\mathrm{Im}\,\iota^n$ provides the criterion for the presence of independent interface degrees of freedom. We illustrate this prescription in the toy model of Section~\ref{sec:example}, where the bordism realizes a topological interface and the additional topological sector decouples from the gauge sector.

Our main application, in Section~\ref{sec:frozen-setting}, returns to the original motivation for this work: namely the symmetry structure of M-theory on frozen singularities \cite{Cvetic:2024mtt}.
More specifically, we study an M-theory compactification on a five-manifold $X$,
\begin{equation}
X=(Y_{I_2}^{\rm ell}\times S^1)/\Z_2\,,
\end{equation}
which is dual to the circle compactification of a frozen $D_5$ singularity and engineers a six-dimensional relative theory $\mathcal{T}^{\frozen}$ containing an $\mathfrak{sp}(1)\cong\mathfrak{su}(2)$ gauge sector \cite{Tachikawa:2015wka}.\footnote{We thank Shani Meynet for suggesting the notation used to denote the frozen theory.} The presence of torsion in homology,
\begin{equation}
    \mathrm{Tor}\,H_1(X)\cong\Z_2\, ,
\end{equation}
allows us to apply the general prescription of Section~\ref{sec:top-down}. The resulting additional dynamical topological sector modifies the physical boundary conditions, which we identify by studying the coupled bulk--boundary system consisting of the six-dimensional theory and its seven-dimensional SymTFT. The latter is the $\Z_4$ SymTFT determined by $\partial X$. We write it as two $\Z_2$ BF pairs with a mixed term,
\begin{equation}
\begin{aligned}
S_{\rm SymTFT}^{\mathcal{T}^{\frozen}} = 2\pi\int_{W_7} \left( \frac12 E_2\,dE_4 +\frac12 C_2\,dB_4 +\frac14 E_2\,dB_4 \right). 
\end{aligned}
\end{equation}
In particular, we determine which bulk topological operators can terminate on the physical boundary, recovering the modified physical boundary observed in \cite{Cvetic:2024mtt}.

We then apply the bordism prescription to separate this modified physical boundary into a topological interface and an ordinary physical boundary. By excising a tubular neighborhood of the twisted circle of $A_1$ singularities, we isolate the local $\mathfrak{su}(2)$ gauge sector from the remaining geometry. The resulting bordism gives rise to a topological interface between the $\Z_4$ SymTFT of the full theory and the $\Z_2$ SymTFT associated with the local model. We determine the gluing conditions across this interface and show that its fusion with the ordinary physical boundary reproduces the termination conditions described above. Importantly, the local gauge sector has the expected $\Z_2$ electric and magnetic symmetries, with no mismatch between them.

Finally, we investigate the different global forms of $\mathcal{T}^{\frozen}$ by choosing the corresponding polarizations of the SymTFT. In the electric polarization, the absolute theory $\mathcal{T}^{\frozen}_e$ has a $\Z_4^{(1)}$ one-form symmetry, containing a trivially acting $\Z_2^{(1)}$ subgroup associated with the additional topological sector. The effective symmetry is precisely the $\Z_2^{(1)}$ electric center symmetry of the $\mathfrak{su}(2)$ gauge sector. The nontrivial relation between these sectors becomes manifest in the partition function,
\begin{equation}
\begin{aligned}
Z_{\mathcal{T}^\frozen_e}[E_2,C_2] &=\int\mathcal{D}\lambda_3\int \mathcal{D}\Phi\, \exp\left(2\pi i\int_{M_6} \mathcal{L}_{\mathcal{T}^{\frozen}_e}[\lambda_3,\Phi]\right),\\ \mathcal{L}_{\mathcal{T}^\frozen_e}[\lambda_3,\Phi] &=\mathcal{L}_{\mathfrak{su}(2)}[E_2] +\frac12 C_2\,dk_3 +\lambda_3\left( \frac12 dC_2+\frac14 dE_2 \right).
\end{aligned}
\end{equation} 
Here $\Phi$ symbolizes the set of dynamical fields of the first two terms in the second line, including for example the $\mathfrak{su}(2)$ gauge theory and $k_3$. The field $\lambda_3$ is a Lagrange multiplier realizing an extension structure, relating in particular $E_2$ and $C_2$ in a non-trivial fashion.
Thus, unlike the toy model, the additional topological sector (the $(k_3, C_2)$ sector) does not decouple from the gauge theory. This is precisely the type of coupling between a QFT and a TQFT discussed by Kapustin and Seiberg~\cite{Kapustin:2014gua}, which in our construction is dictated by the topology of the internal geometry. The magnetic polarization likewise exhibits a $\Z_4^{(3)}$ symmetry, with the additional topological sector now acting non-trivially. The apparent mismatch between the electric and magnetic symmetries of frozen theories therefore originates from this coupling, rather than from the local gauge sector itself.

\subsubsection*{Plan of the paper}

The remainder of the paper is organized as follows. In Section~\ref{sec:top-down}, we study M-theory compactifications on internal spaces with torsion in their homology and argue for the existence of the additional finite topological sector and its coupling to the SymTFT; we exemplify this first for the case of non-compact Calabi--Yau two-folds. In Section~\ref{sec:bordism_general}, we develop the general prescription for constructing SymTFT interfaces from bordisms and establish the criteria for their topological nature. We illustrate this construction in Section~\ref{sec:example} through a concrete Calabi--Yau two-fold toy model with a decoupled topological sector. 
In Section~\ref{sec:frozen-setting}, we apply these results to compactifications that are dual to M-theory on a frozen $D_5$ singularity, construct the corresponding topological interface, and analyze the global forms and their interpretation. Additional background material and details of the (co)homology computations are collected in Appendices~\ref{app:bottom-up}, \ref{app: homology_tracking},  and \ref{app: decoupled_gapless}.

\section{Modified physical boundary from torsion}\label{sec:top-down}

In this section, we study the symmetries of M-theory geometric engineering on internal spaces $X$ with torsional (co)homology.
Taking the standard geometry/symmetry dictionary, we will show that not all SymTFT operators --- which are determined by the topology of $\partial X$ --- terminate on the physical boundary.
After presenting the general arguments, we will take $X$ to be a Calabi--Yau two-fold to demonstrate the features.
This will serve as a toy model to better understand the frozen theories later on.

\subsection{Review of compactification without torsion}

We consider M-theory on a spacetime of the form
\begin{equation}
    {\cal M} = M_d \times X \, , 
\end{equation} 
where the $d$-dimensional spacetime $M_d$ is closed with torsion free cohomology, and $X$ is a smooth non-compact manifold with boundary $\partial X$, of real dimension $m = 11-d$. Compactifying M-theory on such a geometry results in a field theory living on $M_d$.
We will be mostly interested in cases where there is a massless gauge sector with center 1-form and $(d-3)$-form symmetries; in practice this is guaranteed if $X$ preserves some amount of supersymmetry.

We recall from \cite{Morrison:2020ool} that the dynamical $G_{4}$- or $G_7$-flux excitations are encoded (modulo the usual subtleties concerning the appropriate generalized cohomology theory of M-theory \cite{Fiorenza:2019usl}) in \emph{compactly supported} 4- or 7-forms 
\begin{equation}
    H_\text{cpt}^{n}({\cal M}) \cong H^{n} ({\cal M} , \partial {\cal M}) \, , \qquad (n= 4,7) \, .
\end{equation}
These compactly supported cohomology classes can be mapped to ordinary cohomology via the map $J^*$ of the following long exact sequence, 
\begin{align}\label{eq:11d_cohomology}
    \ldots \rightarrow H^n({\cal M}, \partial {\cal M}) \xrightarrow{J^*} H^n ({\cal M}) \xrightarrow{I^*} H^n(\partial {\cal M}) \rightarrow ... \, .
\end{align}
Whenever this map $J^*$ is injective \cite{Morrison:2020ool, Albertini:2020mdx}, the dynamical fluxes are $\text{im}(J^*) \subset H^n({\cal M})$. All flux configurations are then captured by $H^n({\cal M})$. It also allows us to distinguish dynamical fluxes from ``background fluxes'', which are those that are mapped non-trivially into $H^n(\partial {\cal M})$.

To begin with, we assume that the homology of $X$ is torsion free, so that by the Universal Coefficient Theorem (UCT) the cohomology of $X$ is torsion free as well. Focusing on the center symmetries of the $d$-dimensional gauge sector, the relevant cohomology groups of $X$ and $\partial X$ fit into the long exact sequence of the pair,
\begin{equation}\label{eq:SES_torsion-free}
     H^1(\partial X) \rightarrow H^2(X, \partial X) \stackrel{\jmath^*}{\rightarrow} H^2(X) \stackrel{\imath^*}{\rightarrow} H^2(\partial X) \stackrel{d}{\rightarrow} H^3(X, \partial X) \, .
\end{equation}
The electric defect group is defined as \cite{ Albertini:2020mdx, DelZotto:2015isa}
\begin{equation}
    \mathbb{D}^{(1)} = \text{coker}(\jmath^*) = H^2(X)/ \text{im}(\jmath^*) \subset H^2(\partial X) \, ,
\end{equation}
and since $H^2(X)$ has no torsion this quotient is well defined. 
By Poincar\'e--Lefschetz duality and UCT, we have $H^{m-2}(X) \cong H^2(X, \partial X)$ and $H^{m-2}(X, \partial X) = H^2(X)$.
This means that the magnetic defect group $\mathbb{D}^{(d-3)}$, defined from $\jmath^* : H^{m-2}(X, \partial X) \rightarrow H^{m-2}(X)$, is the same finite group, $\mathbb{D}^{(d-3)} \cong \mathbb{D}^{(1)} \equiv \mathbb{D}$.
The two are Pontryagin-dual under the Dirac pairing (modulo integers) of electric and magnetic defects, which geometrically is computed by the linking pairing on $H^2(\partial X) \times H^{m-2}(\partial X)$.

Via the K\"unneth formula and our assumptions about $M_d$, we have
\begin{equation}\label{eq:kunneth-general_1}
    \begin{aligned}
         H^n ({\cal M}) & \cong \bigoplus_{p+q=n} H^p(M_d) \otimes H^q (X) \oplus \bigoplus_{p+q=n+1} \operatorname{Tor} (H^p(M_d), H^q(X)) \\
         &= \bigoplus_{p+q=n} H^p(M_d) \otimes H^q (X) \, , \\
         H^n ({\cal M}, \partial {\cal M}) & \cong \bigoplus_{p+q=n} H^p(M_d) \otimes H^q (X, \partial X) \oplus \bigoplus_{p+q=n+1} \operatorname{Tor} (H^p(M_d), H^q(X, \partial X))  \\
          &= \bigoplus_{p+q=n} H^p(M_d) \otimes H^q (X, \partial X) \, .
    \end{aligned}
\end{equation}
For the center symmetries of the $d$-dimensional gauge sector, the relevant summands are,
\begin{equation}\label{eq: kunneth-2}
     H^4 ({\cal M}) \supset H^{2} (M_d) \otimes H^2(X), \qquad  H^7 ({\cal M}) \supset H^{d-2}(M_d) \otimes H^{m-2}(X) \, ,
\end{equation}
and likewise for $H^n({\cal M}, \partial {\cal M})$ with $H^2(X, \partial X)$ and $H^{m-2}(X, \partial X)$.

With $J^* = \text{id}_{M_d} \otimes \jmath^*$, the dynamical ``Cartan'' flux configurations are captured by $\text{im}(J^*)$, while the background fluxes are encoded by the quotient
\begin{align}\label{eq:conventional-defect}
    \frac{H^{2}(M_d) \otimes H^2(X)}{\text{im}(J^*)} \cong H^{2}(M_d; \mathbb{D}^{(1)}) \, , \qquad \frac{H^{d-2}(M_d) \otimes H^{m-2}(X)}{\text{im}(J^*)} \cong H^{d-2}(M_d; \mathbb{D}^{(d-3)}) \, .
\end{align}
As in \cite{Morrison:2020ool, Albertini:2020mdx}, these fluxes are mapped non-trivially onto the boundary $\partial {\cal M}$, and give rise to the background gauge fields for the global electric 1-form and magnetic $(d-3)$-form symmetries of the $d$-dimensional gauge sector.
As these are mutually non-local, they form the defect group $\mathbb{D}^{(1)} \oplus \mathbb{D}^{(d-3)}$; an absolute theory is obtained by specifying suitable boundary conditions of $G_{4/7}$ on $\partial X$, which in the SymTFT description \cite{Apruzzi:2021nmk} corresponds to picking a polarization on the symmetry boundary.

\subsection{Compactification with torsion}

We now relax the assumption that the homology of $X$ is torsion free, and allow for
\begin{equation}
   {\cal Z}:=  \text{Tor}\, (H_1(X)) \neq 0  \, .
\end{equation}
First, by the universal coefficient theorem (UCT), we also have a non-zero torsional cohomology,
\begin{equation}
    \text{Tor}(H^2(X)) \cong \text{Ext} (H_1(X), \Z) \cong \mathcal{Z}^\vee \, , \qquad  \mathcal{Z}^\vee = \text{Hom}(\mathcal Z, \mathbb{Q}/\Z) \, .
\end{equation}
Second, by Poincar\'e--Lefschetz duality, also $H^{m-1}(X, \partial X) \cong H_1(X)$ is non-zero.
These cohomology groups are Pontryagin dual under the torsion linking pairing,
\begin{equation}\label{eq:torsion-pairing-general}
    \text{Tor}(H^2(X)) \times \text{Tor}(H^{m-1}(X, \partial X)) \rightarrow \mathbb{Q}/\Z \, .
\end{equation}
These ingredients give rise to a topological sector in the $d$-dimensional theory, coupled to discrete background gauge fields.

To see this, we first use the UCT to choose a non-canonical splitting of $H^2(X)$,
\begin{equation}\label{eq:spliting-cohomology}
    H^2(X) \cong H^2_\text{free}(X) \oplus H^2_{\text{tor}}(X) = H^2_\text{free}(X) \oplus \underbrace{\text{Tor}(H_1(X))^\vee}_{\cong \mathcal{Z}^\vee} \, .
\end{equation}
The K\"unneth decomposition of  \eqref{eq:kunneth-general_1} then becomes, 
\begin{equation}
\begin{aligned}
    H^n ({\cal M}) &\supset  H^{n-2}(M_d) \otimes H^2_{\text{free}}(X) \oplus H^{n-2}(M_d)\otimes  \text{Tor}(H_1(X))^\vee \, ,\\
     H^n ({\cal M}, \partial {\cal M}) &\supset H^{n-2}(M_d) \otimes H^2(X, \partial X) \oplus H^{n-m+1}(M_d) \otimes \text{Tor}(H^{m-1}(X, \partial X)) \, .
\end{aligned}
\end{equation}
The first term in $H^n(\mathcal{M})$ is the ordinary ``Cartan'' gauge sector with electric/magnetic defects encoded by \eqref{eq:conventional-defect}. 
The second term is new, and will give rise to the finite background fields. 
Similarly, in the relative cohomology, compared with the torsion-free case, there is a new contribution from $H^{n-m+1}(M_d) \otimes \text{Tor} (H^{m-1}(X, \partial X)) $. The map $J^*$ acts as $\text{id}_{M_d} \otimes \jmath^*$ on the first term, but it kills the second term,
\begin{equation}
    \text{ker}(J^*) \supset H^{n-m+1}(M_d) \otimes \underbrace{\text{Tor} (H^{m-1}(X, \partial X))}_{\cong \mathcal{Z}} \, .
\end{equation}
These classes are nevertheless non-zero compactly supported fluxes, and one must consequently sum over these in the M-theory path integral.
Being torsion-valued, they therefore represent genuinely dynamical degrees of freedom on $M_d$ that are topological.

In the class of geometries considered here, $H^2(X, \partial X) \cong H_{m-2}(X)$ is free.
In the simpler case where 
\begin{equation}
    \text{im}(\jmath^*)\subset H^2_\text{free}(X)
\end{equation}
with respect to the splitting in \eqref{eq:spliting-cohomology}\footnote{For example, consider the internal geometry $X$ obtained by removing all the singular points of $T^4/\Z_2$. For this particular example, the exact sequence $0 \rightarrow H^2(X, \partial X) \stackrel{\jmath^*}\rightarrow H^2(X) \stackrel{\imath^*}\rightarrow H^2(\partial X) \stackrel{d}\rightarrow H^3(X, \partial X) \rightarrow 0$
takes the following form \cite{Cvetic:2023pgm, Gould:2023wgl}, 
\begin{equation}
    0 \rightarrow \Z^6 \stackrel{\jmath^*}\rightarrow \Z^6 \oplus \Z^5_2 \stackrel{\imath^*}\rightarrow \Z^{16}_2 \stackrel{d}\rightarrow \Z^5_2 \rightarrow 0 \, ,
\end{equation} 
where the map $\jmath^*$ is multiplication by $2$ into the $\Z^6$ factor.}, the electric defect group of the Abelian gauge sector, derived from $H^2(X,\partial X)\cong \text{im}(\jmath^*)$, is defined by
\begin{equation}
    \mathbb{D}^{(1)} := H_\text{free}^2(X)/\text{im}(\jmath^*)\,,
\end{equation}
and the magnetic one ($\mathbb{D}^{(d-3)} \cong \mathbb{D}^{(1)}$) in the same way in degree $m-2$. More generally, when there exists no splitting \eqref{eq:spliting-cohomology} such that
the image of $\jmath^*$ sits in $H^2_\text{free}(X)$, we use the short exact sequence 
\begin{equation}\label{eq: D+T extension}
    0 \rightarrow H^2_{\text{Tor}}(X) \rightarrow H^2(X) \stackrel{q}\rightarrow H^2_{\text{free}}(X) \rightarrow 0 \, ,
\end{equation}
and define the defect group as 
\begin{equation}\label{eq: effective_defect_group}
    \mathbb{D}^{(1)} : = \text{coker}(I) = H^2_{\text{free}}(X)/ \text{im}(q \circ \jmath^*) \, ,
\end{equation}
where $I$ is the injective map $I : = q \circ \jmath^* : H^2(X, \partial X) \rightarrow H^2_{\text{free}}(X)$ (and likewise for the magnetic defects).
This will be relevant for the frozen compactification discussed in Section~\ref{sec:frozen-setting}.

Thus, the novel aspects are the fluxes associated with 
\begin{equation}
    H^n(\mathcal{M}) \vert_{\text{new}} = H^{n-2}(M_d) \otimes \text{Tor}(H_1(X))^\vee \cong  H^{n-2}(M_d; {\cal Z}^\vee) \, ,
\end{equation}
as well as the compactly supported fluxes 
\begin{equation}
    H^n({\cal M}, \partial {\cal M}) \vert_{\text{new}} = H^{n-m+1}(M_d) \otimes \text{Tor} (H^{m-1}(X, \partial X) ) \cong H^{n-m+1}(M_d; {\cal Z}) \, .
\end{equation}
These do not appear in the settings of \cite{Morrison:2020ool, Albertini:2020mdx}.
The first part, $H^{n-2}(M_d; {\cal Z}^\vee) = H^{n-2}(M_d) \otimes H^2_\text{tor}(X) \subset H^n({\cal M})$, is mapped injectively into $H^n(\partial {\cal M})$. This part is therefore entirely background flux, i.e., the (non-dynamical) background field for some global symmetry different from the center symmetries of the gauge sector. The compactly supported fluxes, on the other hand, give rise to dynamically non-trivial discrete $(n-m+1)$-form gauge fields in the $d$-dimensional theory.
The non-trivial torsion pairing \eqref{eq:torsion-pairing-general} induces a BF-type coupling between these fields.

\subsubsection*{Specialization to Calabi--Yau two-folds}

To understand what these new global symmetries are, and the role of the additional dynamical pieces, we need to reduce the 11d kinetic term $\int_{\cal M} G_4 \wedge G_7$. To keep things tractable and explicit for the remainder of this section, we take $X$ to be a non-compact Calabi--Yau two-fold, so that $m = 4$ and $d = 7$. 
A natural class of such models is the smooth complement of the union $X^\text{loc} = \bigcup_a \bbC^2/\Gamma_a$ of ADE-singularities in a K3-manifold ($\widehat{X}$), 
\begin{equation}
    X \cong \widehat{X} \setminus X^{\text{loc}} \, , \qquad \partial X = \partial X^\text{loc} \cong \bigsqcup_a S^3 / \Gamma_a \,,
\end{equation}
where $X$ is the smooth complement,
(see, e.g., \cite{Cvetic:2023pgm}). For example, if $\widehat{X}=T^4/\Z_2$, then $X^\text{loc}$ is a union of sixteen disjoint copies of $\mathbb{C}^2/\Z_2$, and $\partial X$ consists of sixteen copies of $S^3/\Z_2$. 

The torsional data sits in
\begin{equation}
   {\cal Z} = \text{Tor}(H_1(X)) = H_1(X) \, , \qquad \text{Tor}(H^2(X)) \cong \mathcal{Z}^\vee \, , \qquad H^3(X, \partial X) \cong H_1(X) \cong \mathcal{Z} \, ,
\end{equation}
and under the standard assumption $H^1(\partial X) = 0$ the sequence \eqref{eq:SES_torsion-free} reads\footnote{The final map in \eqref{eq:torsion-cohomology-seq} is zero as $H^3(X,\partial X)\cong H_1(X)$ is pure torsion and cannot map non-trivially into the free group $H^3(X)\cong H_1(X,\partial X)$. }
\begin{equation}\label{eq:torsion-cohomology-seq}
     H^1(\partial X) = 0  \rightarrow H^2(X, \partial X) \stackrel{\jmath^*}{\rightarrow} H^2(X) \stackrel{\imath^*}{\rightarrow} H^2(\partial X) \stackrel{d}{\rightarrow} H^3(X, \partial X) \rightarrow 0 \, .
\end{equation}
The new fluxes are
\begin{equation}
    H^n(\mathcal{M}) \vert_{\text{new}} \cong  H^{n-2}(M_7; {\cal Z}^\vee) \, , \qquad H^n({\cal M}, \partial {\cal M}) \vert_{\text{new}} \cong H^{n-3}(M_7; {\cal Z}) \, ,
\end{equation}
i.e., background fields for new 1-form ($n=4$) and 4-form ($n=7$) symmetries, and dynamical 1-form ($n=4$) and 4-form ($n=7$) gauge fields in the 7d theory.

Let us pause momentarily to discuss the limiting regimes in which the above construction admits a seven-dimensional effective field theory description, including its non-topological degrees of freedom. Consider M-theory compactified on a singular space $\widehat{X}$, giving rise to a seven-dimensional supergravity theory with non-abelian gauge sectors localized at the singularities, described by local models $X^{\mathrm{loc}}$, together with abelian gauge fields in the bulk. 
The complementary space $X$ is an incomplete smooth manifold with boundary associated with bulk physics. Schematically, the full partition function can be written as
\be
Z_{\widehat{X}}=\int D\varphi_{\partial X} Z_{X^\text{loc}}[\varphi_{\partial X}] Z_{X}[\varphi_{\partial X}]\,,
\ee 
where we collect all the ``gluing'' fields into $\varphi_{\partial X}$. Now, $Z_{X^{\rm loc}}[\varphi_{\partial X}]$ is the familiar piece for which there exists a local limit such that we have a geodesically complete metric on $X^{\rm loc}$. In this limit the boundary conditions set by the gluing fields $\varphi_{\partial X}$ include the background fields of various generalized global symmetries, and $Z_{X^{\rm loc}}[\phi_{\partial X}]$ therefore naturally gives rise to a physical boundary condition in a SymTFT construction.

Similar limits, on the other hand, are not generally available for $Z_X[\phi_{\partial X}]$. However, with the goal of connecting to a SymTFT picture we can do the following: we consider the large volume limit of $\widehat X$, and correspondingly also of $X$. This limit lowers the KK-scale and we must therefore simultaneously consider a low-energy limit such that the cut-off remains below the KK-scale. In this regime, the free cohomology of $X$ gives rise to a 7d effective Abelian gauge sector in its IR/free limit, subject to the gluing conditions at $\partial X$. Strictly speaking, this is a low-energy truncation rather than an autonomous decoupled QFT, in the large volume limit the KK-scale vanishes, so the energy scale must be taken to zero accordingly. Away from this regime, for example, branes wrapping relative cycles terminating on $\partial X$ can have finite mass and need not correspond to non-dynamical probes
of the effective physics.


\subsection{KK-reduction of kinetic term}\label{sec:KK-reduction-kinetic-term}

We now carry out the reduction of $\int_{\cal M} G_4 \wedge G_7$ on the two-fold $X$. This gives the 7d theory, which furnishes the physical boundary condition for the SymTFT. We perform the reduction separately on the free and torsional cohomology classes: the free classes reproduce the ordinary Cartan gauge sector, while the torsion classes produce the additional topological sector. 

\subsubsection*{Reduction on free classes}
First, we focus on the free parts,
\begin{equation}
   H^n_\text{free}({\cal M}) \cong H^{n-2}(M_7) \otimes H^2_\text{free}(X), \qquad  H_\text{free}^n({\cal M}, \partial {\cal M}) \cong H^{n-2}(M_7) \otimes H^2(X, \partial X) ,
\end{equation} 
which are well-known to give rise to the Cartan gauge sector and the center symmetries. To this end, we introduce a basis
\begin{align}\label{eq:basis_free_parts}
    \omega_2^{(i)} \in H^2(X, \partial X)  , \qquad e_2^{(j)} \in H^2_\text{free}(X) \, , \qquad i,j = 1, \cdots, \text{rank}(H^2_\text{free}) = \text{rank}(H^2(X, \partial X)) .
\end{align}
Poincar\'e--Lefschetz duality and UCT imply $ H^2(X, \partial X) \cong \text{Hom} ( H^2_\text{free}(X), \mathbb{Z})$, which means that the basis can be chosen such that 
\begin{equation}
    \int \omega_2^{(i)} \wedge e_2^{(j)} = \delta_{ij}  .
\end{equation}
Let $C_{ij} : = \int \omega_2^{(i)} \wedge \omega_2^{(j)}$ be the intersection matrix \cite{Morrison:2020ool, Albertini:2020mdx}. With our choice of bases, $ \jmath^* \omega^{(i)}_2 = C_{ik} e^{(k)}_2$ and consequently
\begin{equation}
     \int_{X} e^{(i)}_2 \wedge e^{(j)}_2 = \big( C^{-1} \big)_{ij}  .
\end{equation}
The KK-reduction of the $G_{4/7}$-fluxes on the free parts proceeds straightforwardly. We choose representatives $e^{(a)}_2 \in H^2_{\text{free}}(X)$, from the set of basis $e^{(j)}_2$, whose equivalence classes generate the electric/magnetic defect groups $\mathbb{D}^{(1)} \cong \mathbb{D}^{(4)} \equiv \mathbb{D}$. The index $a$ therefore labels the non-trivial defect group ``directions''. These are defined only modulo $\text{im}(\jmath^*)$. We expand the fluxes, 
\begin{align}\label{eq:expansion_G4_and_G7-free}
    \begin{split}
        G_4 & =\sum_i  f_2^{(i)} \wedge \jmath^*\omega_2^{(i)} + \sum_a \tilde{E}_2^{(a)} \wedge e_2^{(a)}  , \\
        G_7 & =\sum_i  f_5^{(i)} \wedge j^*\omega_2^{(i)} + \sum_a \tilde{E}_5^{(a)} \wedge e_2^{(a)}  .
    \end{split}
\end{align}
Here $f^{(i)}_{2/5}$ are the dynamical Cartan fluxes, since they are associated with the relative cohomology classes, whereas $\tilde{E}_{2/5}^{(a)}$ are the integral lifts of the background fields associated with the defect group. To see this, we define 
\begin{equation}
    \eta_a : = \imath^*(e^{(a)}_2) \in H^2_{\text{free}}(X)/ \text{im}(\jmath^*) = \mathbb{D} \subset H^2(\partial X).
\end{equation}
The exactness of the sequence $ H^2(X, \partial X) \stackrel{\jmath^*}\rightarrow H^2(X) \stackrel{\imath^*} \rightarrow H^2(\partial X)$ implies that 
\begin{equation}
    I^* G_{4/7} = \sum_a \tilde{E}^{(a)}_{2/5} \wedge \eta_a ,
\end{equation}
where $I^* = \text{id}_{M_7} \otimes \imath^*$ and $J^* = \text{id}_{M_7} \otimes \jmath^*$ as in \eqref{eq:11d_cohomology}.
Therefore, the fields $\tilde{E}^{(a)}_{2/5}$ are precisely the background fields associated with the defect group. 

Substituting the expansions \eqref{eq:expansion_G4_and_G7-free} into the 11d kinetic term, we obtain the usual gauge sector: 
\begin{align}\label{eq:kinetic-terms-free-part}
    \begin{split}
         \int_{\cal M} G_4 \wedge G_7 
        =  \int_{M_7}  \sum_{i,j}  C_{ij} \, f_2^{(i)} \wedge f_5^{(j)} + \sum_{a,b} (C^{-1})_{ab}   \tilde{E}_2^{(a)} \wedge \tilde{E}_5^{(b)} \\ + \sum_{i,a} \delta_{i,a} \left(f_2^{(i)} \wedge \tilde{E}_5^{(a)} + f_5^{(i)} \wedge \tilde{E}_2^{(a)} \right)  .
    \end{split}
\end{align}
The first term is the kinetic term for the dynamical Cartan fluxes $f_{2/5}^{(i)}$, and the second term is the ``defect group'' pairing for the 1-form/4-form symmetries. The last two terms show the familiar structure by which the magnetic symmetries, with background field $\tilde{E}_5^{(i)}$, are generated by $f_2^{(i)}$-surfaces and analogously the  electric symmetries, with background $\tilde{E}_2^{(i)}$, are generated by $f_5^{(i)}$-surfaces.

\subsubsection*{Reduction on torsion classes}

To describe this reduction appropriately, we follow the procedure of \cite{Najjar:2024vmm} and work with the differential cohomological uplift of the kinetic term\footnote{The reduction of the kinetic term on torsional elements should perhaps be done more rigorously following \cite{GarciaEtxebarria:2024fuk}, using the Hopkins--Singer formalism of differential cocycles \cite{Hopkins:2002rd}. However, we believe that the final result is the same in either approach, so we choose to follow the prescription of \cite{Najjar:2024vmm} for convenience.}.
In particular, the authors in \cite{Najjar:2024vmm} introduced a degree 8 differential character $\check{G}_8 \in \check{H}^8({\cal M})$ to model the $G_7$ flux in M-theory, which satisfies $dG_7 = -\half G_4 \wedge G_4$. This allows us to write the differential refinement of the kinetic term\footnote{Our notation follows that of \cite{Apruzzi:2021nmk}. For a physics friendly introduction to differential cohomology, see \cite{Apruzzi:2021nmk, vanBeest:2022fss, Freed:2006ya, Freed:2006yc,Hsieh:2020jpj, Freed:2006mx}.}, 
\begin{align}\label{eq:M-theory_kinetic_diff-cohom}
    \int_{\cal M} G_4 \wedge G_7 \longrightarrow \int_{\cal M} \check{G}_4 \star \check{G}_8 , \quad \check{G}_p \in \check{H}^p({\cal M})  .
\end{align}
Here $(\star)$ denotes the product of differential characters,
\begin{equation*}
    \star : \check{H}^p(\mathcal{M}) \times \check{H}^q(\mathcal{M}) \rightarrow \check{H}^{p+q}(\mathcal{M})  ,
\end{equation*}
while $\int_\mathcal{M}$ (on the right hand side) denotes the ``fiberwise integration'' (also known as the integration map) on differential characters.  The elegance of this formulation is that, for an oriented $d$-manifold $\mathcal{M}$, integration of a degree-$d$ differential-cohomology class gives a $\mathbb{Z}$-valued primary invariant, whereas integration of a degree $(d+1)$ class gives an
$\mathbb{R}/\mathbb{Z}\cong U(1)$-valued secondary invariant \cite{Apruzzi:2021nmk}.  Applied to the internal space $X$, the former reproduces the ordinary pairing of
the free degree-two classes --- i.e., yields the same result for the terms in \eqref{eq:expansion_G4_and_G7-free}, while the latter detects the torsion
linking pairing between degree-$p$ and degree-$(\text{dim}(X)-p+1)$ classes.  At the level of differential characters, we can think of integration over $X$ as a ``map'',
\begin{equation}
    \int_X : \check{H}^p(\mathcal{M}) \rightarrow \check{H}^{p-\text{dim}(X)}(M_7)  .
\end{equation}
While previous works have focused solely on the reduction of $\check{G}_p$ on a compact manifold (mostly, on $\partial X$), it is straightforwardly applicable to the case at hand, where there is a torsion pairing between $H_\text{tor}^2(X)$ and $H^3(X, \partial X)$.

To perform the reduction, we choose a basis for the torsional pieces,
\begin{align}
     \tau_2^{(k)} \in H^2_\text{tor}(X) \cong {\cal Z}^\vee  , \qquad \sigma_3^{(\ell)} \in H^3(X, \partial X) \cong {\cal Z}  .
\end{align}
Using their differential cohomological uplift, we can expand
\begin{align}\label{eq:KK-reduction-X}
    \begin{split}
        \check{G}_4 & = \sum_k \check{C}_2^{(k)} \star \check{\tau}_2^{(k)} + \sum_\ell \check{b}_1^{(\ell)} \star \check{\sigma}_3^{(\ell)} , \\
        \check{G}_8 & = \sum_k \check{C}_6^{(k)} \star \check{\tau}_2^{(k)} + \sum_\ell \check{b}_5^{(\ell)} \star \check{\sigma}_3^{(\ell)}  .
    \end{split}
\end{align}
Substituting these expansions into \eqref{eq:M-theory_kinetic_diff-cohom} yields\footnote{Strictly speaking, because $X$ has a boundary and the reduction involves both absolute and relative cohomology classes, the products and integration maps appearing here should be formulated in a relative formulation of differential cohomology. We suppress possible subtleties associated
with this extension and assume that the absolute-relative torsion pairing admits the expected differential cohomology refinement. A fully rigorous reduction would require working with relative Cheeger--Simons differential characters, or an equivalent relative version of Hopkins--Singer differential cocycles \cite{becker2013relativedifferentialcohomology, ruffino2018relativedifferentialcohomologygeneralized}.}
\begin{align}
    \int_{\cal M} \check{G}_4 \star \check{G}_8 = \sum_{k,\ell}  \underbrace{\left( \int_{X} \check{\tau}_2^{(k)} \star \check{\sigma}_3^{(\ell)}\right)}_{\equiv L(\tau^{(k)}_2, \sigma^{(\ell)}_3)}  \int_{M_7} \left( \check{C}_2^{(k)} \star \check{b}_5^{(\ell)} + \check{b}_1^{(\ell)} \star \check{C}_6^{(k)} \right)  .
\end{align}
The fiber integral over $\check{\tau}_2^{(k)} \star \check{\sigma}_3^{(\ell)}$ reproduces the cohomological linking pairing $L(\tau_2^{(k)}, \sigma_3^{(\ell)}) \equiv L^{(k\ell)} \in \mathbb{Q}/\bbZ$.
This pairing can be ``diagonalized'' by picking Pontryagin-dual bases, i.e., if $H^2_\text{tor}(X)= \bigoplus_k \bbZ_{n_k} \cong \bigoplus_k (\bbZ_{n_k})^\vee = H^3(X, \partial X)^\vee$, then we can find generators such that $L^{(k\ell)} = \frac{1}{n_k} \delta_{k\ell}$.
Then we obtain
\begin{align}
    \int_{\cal M} \check{G}_4 \star \check{G}_8 = \sum_k \frac{1}{n_k} \int_{M_7} \left( \check{C}_2^{(k)} \star \check{b}_5^{(k)} + \check{b}_1^{(k)} \star \check{C}_6^{(k)} \right)  .
\end{align}
Notice that, since the fiberwise integration kills the product of a torsional and a free character, there is no non-zero term in the 7d action between the fields appearing in \eqref{eq:expansion_G4_and_G7-free} and those in \eqref{eq:KK-reduction-X}; for this reason we have chosen to simply neglect to include the former in the expansion \eqref{eq:KK-reduction-X}.

Following the prescription of \cite{Najjar:2024vmm}, we now retain the discrete cochain data contained in the differential characters. Since $\check{G}_8$ is constructed as the differential cohomological refinement of the exact form $dG_7$, ``projecting'' back onto integral cohomology produces the following 7d action, 
\begin{align}
   2\pi \sum_k \frac{1}{n_k} \int_{M_7} \left( C_2^{(k)} \cup \delta b_4^{(k)} + b_1^{(k)} \cup \delta C_5^{(k)} \right)  ,
\end{align}
which we recognize as the action of $\mathbb{Z}_{n_k}$-BF-pairs $(C_2^{(k)}, b_4^{(k)})$ and $(b_1^{(k)}, C_5^{(k)})$.

The crucial difference to previous works is that the distinction of (non-)compactly-sup\-ported classes on $X$ leads to the distinction of background gauge fields ($C^{(k)}_2, C_5^{(k)}$) and dynamical \emph{discrete} gauge fields $(b_1^{(k)}, b_4^{(k)})$ in the 7d theory.
Hence, the BF-action describes how $b_4$-surface operators magnetically source the background fields $C_2$, and likewise for $b_1$ operators and the background fields $C_5$.
Moreover, since the $b_p^{(k)}$ are discrete gauge fields, they form a topological sector.

\subsection{Coupling to the SymTFT and modified physical boundary}\label{sec:coupling-symtft}

Let us quickly summarize what we have discussed so far. The relevant geometric data is the sequence,
\begin{align}
    0 & \rightarrow H^2(X, \partial X) \stackrel{\jmath^*}{\rightarrow} H^2(X)\cong H^2_{\text{free}}(X) \oplus H^2_{\text{Tor}}(X) \stackrel{\imath^*}{\rightarrow} H^2(\partial X) \stackrel{d}{\rightarrow} H^3(X, \partial X) \rightarrow 0  .
\end{align}
Since $H^2(X, \partial X)$ is free and $\text{im}(j^*) \subset H^2_{\text{free}}(X)$, the sequence above splits into two short exact sequences,
\begin{align}\label{eqn:split_sequence}
    & 0 \rightarrow H^2(X, \partial X) \stackrel{\jmath^*}{\rightarrow} H^2_\text{free}(X) \rightarrow \mathbb{D} \cong \frac{H^2_{\text{free}}(X)}{\text{im}(\jmath^*)} \rightarrow 0  , \\
    & 0 \rightarrow H^2_\text{tor}(X) \rightarrow \underbrace{H^2(\partial X)/\mathbb{D}}_{=: \mathbb{T}} \rightarrow H^3(X, \partial X) \rightarrow 0  ,
\end{align}
where $H^2(\partial X) \supset \mathbb{D} = \imath^*(H^2_\text{free}(X)) \cong H^2_\text{free}(X) / \text{im}(\jmath^*)$.
As we will see below, the group $\mathbb{T}$ encodes the SymTFT-bulk extensions of the background fields $C^{(k)}_p$.
By simple intersection properties on $X$,\footnote{More precisely, by the assumption on $\jmath^*$, $\mathbb{D}$ is the discriminant group for the intersection pairing on $H_2(X) \cong H^2(X, \partial X)$.
The linking pairing on $H^2(\partial X)$ restricts to the discriminant pairing on $\mathbb{D}$ which is non-degenerate.
Since the former is also non-degenerate, we are guaranteed the orthogonal splitting $H^2(\partial X) = \mathbb{D} \oplus \mathbb{D}^\perp$. The second sequence identifies $\mathbb{D}^\perp = \mathbb{T}$.
} the boundary cohomology takes the form
\begin{equation}\label{eq:orthogonal_splitting_boundary_K3}
    H^2(\partial X) \cong \mathbb{D} \oplus \mathbb{T}  ,
\end{equation}
i.e., the subgroups $\mathbb{D}$ and $\mathbb{T}$ are orthogonal with respect to the linking pairing on $H^2(\partial X)$.
This simplifies the coupling of the 7d theory to its 8d SymTFT, in that the center symmetries of the gauge sector and the background fields coupled to the topological sector are completely decoupled.
We will see later that for the geometry dual to the frozen theories, the two sectors will mix nontrivially.

To derive the SymTFT, we need to expand the fluxes on the bases of $\partial X$. We parametrize the bases of the various groups as follows. For the free sector, $e_a$ are representatives\footnote{To reduce the clutter with notations, we have dropped the degree subscript from $e^{(a)}_2$.} of the generators in $H^2_\text{free}/H^2(X, \partial X)$ and $\eta_a (= \imath^* (e_a)) \in \mathbb{D}$. For the torsional sector, $\tau_k \in H^2_\text{tor}(X)$, $\sigma_\ell \in H^3(X, \partial X)$, and $\{t_k, s_\ell\} \subset \mathbb{T}$ with
\begin{equation}
    t_k = \imath^*(\tau_k) , \qquad d(s_\ell) = \sigma_\ell  .
\end{equation}
Poincar\'e duality and UCT imply that $H_\text{tor}^2(X)$ and $H^3(X, \partial X)$ are Pontryagin-dual to each other with respect to linking on $X$, so the indices $k$ and $\ell$ run over the same range. 
By the orthogonal splitting \eqref{eq:orthogonal_splitting_boundary_K3}, the $\mathbb{Q}/\mathbb{Z}$-valued linking pairing on $H^2(\partial X)$ is non-zero only for the combinations 
\begin{equation}\label{eq:linking_boundary}
\begin{aligned}
    D_{ab} &:= L_{\partial X} (\eta_a, \eta_b) \equiv L_X(e_a, e_b) \mod \Z , \\
    \kappa_{k\ell} &: = L_{\partial X} (t_k , s_\ell ) \equiv L_X(\tau_k, \sigma_\ell)  .
\end{aligned}
\end{equation}
With this information, we can expand the M-theory fields, 
\begin{equation}\label{eq:KK-reduction_flux_on_boundary}
\begin{aligned}
    \check{G}_4 = \check{E}^{(a)}_2 \star \check{\eta}_a + \check{B}^{(\ell)}_2 \star \check{s}_\ell + \check{C}^{(k)}_2 \star \check{t}_k  , \\
    \check{G}_8 = \check{E}^{(a)}_6 \star \check{\eta}_a + \check{B}^{(\ell)}_6 \star \check{s}_\ell + \check{C}^{(k)}_6 \star \check{t}_k  .
\end{aligned}
\end{equation}
Plugging these back into the kinetic term gives the 8d SymTFT action
\begin{align}
    S_\text{8d} = 2\pi \int_{W_8} \left( D_{ab} \, E_2^{(a)} \cdot \delta E_5^{(b)} + \kappa_{k\ell} \, B_2^{(\ell)} \cdot \delta C_5^{(k)} + \kappa_{k\ell} \, C_2^{(k)} \cdot \delta B_5^{(\ell)} \right)  ,
\end{align}
where we have once again used the exactness of the differential character $\check{G}_8$ while projecting back to integral cohomology \cite{Najjar:2024vmm}. 

Combining with the results from Section \ref{sec:KK-reduction-kinetic-term}, the total action for the physical boundary is the following\footnote{Here we have omitted the $\widetilde{E}_2 \cdot \widetilde{E}_5$ coupling from the $7d$ action. We have chosen a boundary counterterm convention in which the local counterterm $\widetilde{E}_2 \cdot \widetilde{E}_5$ is absorbed into the boundary transgression of the $8d$ bulk action. This changes the representative of the anomaly-inflow action but not its anomaly class. 
}:
\begin{align}\label{eq:7d_modified-boundary_action}
\begin{split}
    S_\text{7d} = 2\pi \int_{M_7} \Big( &C_{ij} \, f_2^{(i)} \cdot f_5^{(j)} + \delta_{ia} \, f_2^{(i)} \cdot \widetilde{E}_5^{(a)} + \delta_{ia} \, f_5^{(i)} \cdot \widetilde{E}_2^{(a)} \\
    + & \kappa_{k\ell} \, b_1^{(\ell)} \cdot \delta \widetilde{C}_5^{(k)} + \kappa_{k \ell} \widetilde{C}_2^{(k)} \cdot \delta b_4^{(\ell)} \Big)  .
\end{split}
\end{align}
Here the lower-case fields are dynamical degrees of freedom. On the other hand, the tilded capitalized fields are background fields associated with the cohomologies
\begin{align}
\begin{split}
    & \widetilde{E}_{2/5}^{(a)} \otimes e_a \in H^{2/5} (M_7) \otimes H_\text{free}^2(X)/H^2(X, \partial X) \subset H^{4/7}({\cal M}_{11})  ,\\
    & \widetilde{C}_{2/5}^{(k)} \otimes \tau_k \in H^{2/5}(M_7) \otimes H^2_\text{tor} (X) \subset H^{4/7}({\cal M}_{11})  .
\end{split}
\end{align}
In case $W_8 = M_7 \times [0,1]$, which is the standard setup for the SymTFT coupled to the physical boundary theory, we have $H^n(W_8) = H^n(M_7)$ since the interval is contractible.
Combining this with the short exact sequences above, we have the natural identifications
\begin{align}
    & H^{n}(W_8) \otimes \underbrace{\langle \eta_a \rangle}_{\mathbb{D} \subset H^2(\partial X)} \cong H^n(M_7) \otimes H^2_\text{free}(X) / H^2(X,\partial X)  , \\
    & H^n(W_8) \otimes \underbrace{\langle t_k \rangle}_{ \mathbb{T} \subset H^2(\partial X)} \cong H^n(M_7) \otimes \langle \tau_k\rangle = H^n(M_7) \otimes H^2_\text{tor}(X)  .
\end{align}
We use this to identify the tilded fields in the 7d action with the boundary values of the corresponding 8d fields in the SymTFT\footnote{As we emphasized earlier, the combination $\tilde{E}^{(a)}_{2/5} \otimes e_a$ are integral lifts of classes in $H^{2/5}(M_7, \mathbb{D})$. Under the boundary map $(i^*)$, $e_a$ is mapped to $\eta_a$, and therefore these backgrounds are naturally identified with the boundary values of the corresponding SymTFT fields.},
\begin{align}
    \widetilde{E}_{2/5}^{(a)} = E^{(a)}_{2/5} \Big|_{M_7}  , \quad \widetilde{C}_{2/5}^{(k)} = C_{2/5}^{(k)} \Big|_{M_7}  .
\end{align}
Meanwhile, there is no explicit appearance of the boundary values of the $B_{2/5}^{(\ell)}$ fields in the 7d action.

Taking the combined action $S_\text{8d} + S_\text{7d}$ to describe the coupled bulk-(physical) boundary system, we arrive at the following boundary equations of motion 
\begin{align}
\begin{split}
    E_{5/2} \ \text{ variations } \quad \Rightarrow \quad E_{2/5}^{(a)} \Big|_{M_7} & = (D^{-1})^{ab} \, \delta_{ib} \,f_{2/5}^i = (D^{-1})^{ab} \, f_{2/5}^b  , \\
    C_{5/2} \ \text{ variations } \quad \Rightarrow \quad B_{2/5}^{(\ell)} \Big|_{M_7} & = \delta b^{(\ell)}_{1/4}  .
\end{split}    
\end{align}
The first equation has the standard interpretation for a $U(1)^\#$ gauge theory, whereby the SymTFT operators built from $E^{(a)}_{2/5}$-surfaces can terminate on Wilson lines (holonomy associated with the electric field strength $f_2$) and 't Hooft surfaces (holonomy associated with the magnetic field strength $f_5$).
In other words, the $E_{2/5}$-surfaces can terminate on the physical boundary on charged objects, i.e., have Dirichlet boundary conditions here.
As is well-known, to obtain an absolute 7d theory requires further specifying the boundary conditions on the (gapped) symmetry boundary, and only those operators with Dirichlet boundary conditions on both sides will give rise to genuine Wilson lines or 't Hooft surfaces.

By the same token, the second equation then specifies Dirichlet boundary conditions for the $B_{2/5}$ operators, by letting them terminate on $b_{1/4}$ holonomy lines/surfaces.
However, in contrast, there are no corresponding equations that allow the $C_{2/5}$ operators to terminate on the physical boundary. Consequently, the corresponding $C_{2/5}$ operators cannot terminate on the physical boundary. By choosing an appropriate symmetry boundary condition, we can now realize trivially acting symmetries. For example, choose Dirichlet boundary condition on $C_{2/5}$-surfaces on the symmetry boundary. The $C_{2/5}$-surfaces then end on the symmetry boundary but not on the physical boundary, so the objects charged under the dual $B_{2/5}$-surfaces do not exist in the 7d QFT. As a consequence, the symmetry generated by $B_{2/5}$-surface act trivially. 

This asymmetric termination of the bulk operators on the physical boundary is precisely the defining feature of a modified physical boundary. Thus, the compactification with torsion provides a top-down construction of a modified physical boundary.

\section{Bordisms and SymTFT interfaces}\label{sec:bordism_general}

In the previous section, we saw that the modified physical boundary followed from reducing M-theory on the internal space $X$ with torsion. As emphasized in the introduction (and Appendix~\ref{app:bottom-up}), from the bottom-up perspective, a modified physical boundary consists of two ingredients: a topological interface and an ordinary physical boundary. This raises the question of whether the same two ingredients can be identified in the top-down construction. In this section, we show that it can, using the correspondence between extra-dimensional bordisms and interfaces in the SymTFT \cite{Baume:2023kkf,Cvetic:2024dzu}. We also derive a criterion for when such an interface is topological.

\subsection{Decomposition of the geometry}

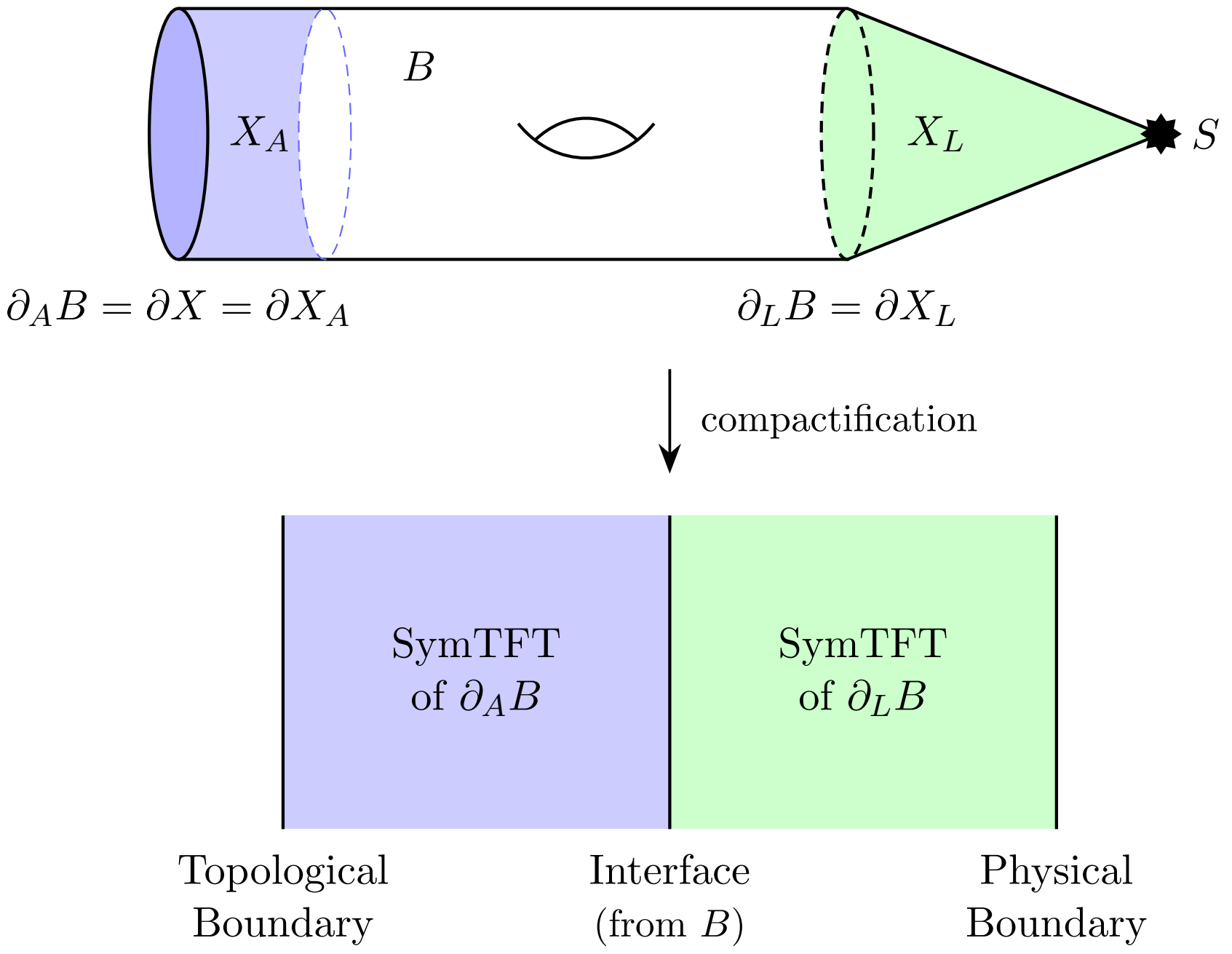
\begin{figure}[ht]
\centering
\begin{tikzpicture}[scale=0.95]

\fill[blue!20] (0,1.2) -- (1.4,1.2)
      arc[start angle=90, end angle=270, x radius=0.25, y radius=1.2]
      -- (0,-1.2)
      arc[start angle=-90, end angle=90, x radius=0.28, y radius=1.2] -- cycle;
\fill[blue!30] (0,0) ellipse (0.28 and 1.2);
\fill[green!20] (6.4,1.2) -- (9.4,0) -- (6.4,-1.2)
      arc[start angle=270, end angle=90, x radius=0.25, y radius=1.2] -- cycle;

\draw[thick] (0,1.2) -- (6.4,1.2) -- (9.4,0) -- (6.4,-1.2) -- (0,-1.2);
\draw[thick] (0,0) ellipse (0.28 and 1.2);

\draw[thick, densely dashed] (6.4,0) ellipse (0.25 and 1.2);

\draw[thin, densely dashed, blue!60] (1.4,0) ellipse (0.25 and 1.2);

\draw[thick] (3.25,0.10) .. controls (3.60,-0.34) and (4.20,-0.34) .. (4.55,0.10);
\draw[thick] (3.41,-0.06) .. controls (3.70,0.22) and (4.10,0.22) .. (4.39,-0.06);

\node[style=Star] at (9.4,0) {};
\node[right=4pt] at (9.4,0) {$S$};

\node at (0.78,0.0) {$X_A$};
\node at (2.3,0.65) {$B$};
\node at (7.25,0.0) {$X_L$};

\node[below] at (0,-1.35) {$\partial_A B = \partial X = \partial X_A$};
\node[below] at (6.4,-1.35) {$\partial_L B = \partial X_L$};

\draw[-{Stealth[scale=1.25]}, line width=0.25mm] (4.7,-2.25) -- (4.7,-3.25)
      node[midway, right=4pt] {\small compactification};

\begin{scope}[yshift=-5.15cm]
\fill[blue!20]  (1.0,-1.5) rectangle (4.7,1.5);
\fill[green!20] (4.7,-1.5) rectangle (8.4,1.5);
\draw[thick] (1.0,-1.5) -- (1.0,1.5);   
\draw[thick] (4.7,-1.5) -- (4.7,1.5);   
\draw[thick] (8.4,-1.5) -- (8.4,1.5);   
\node[align=center] at (2.85,0) {SymTFT\\ of $\partial_A B$};
\node[align=center] at (6.55,0) {SymTFT\\ of $\partial_L B$};
\node[below, align=center] at (1.0,-1.6) {Topological\\ Boundary};
\node[below, align=center] at (4.7,-1.6) {Interface\\ \small (from $B$)};
\node[below, align=center] at (8.4,-1.6) {Physical\\ Boundary};
\end{scope}
\end{tikzpicture}
\caption{Top: decomposition of the internal geometry $X = X_A \cup B \cup X_L$, where $S$ is the singular locus. Bottom: The correspondence with the SymTFT with an interface. The larger SymTFT is obtained from the asymptotic boundary. The reduced SymTFT data is encoded in the boundary of the local model and the bordism itself compactifies to the interface.   }
\label{fig:bordism-setup}
\end{figure}

We consider M-theory on a non-compact geometry $X$ with a compact locus $S\subset X$, on which all relevant QFT degrees of freedom are localized. 
In cases where the latter contains a non-Abelian gauge sector, $S$ will typically be a singular locus that can be resolved into a collection of compact curves and (real) codimension-2 cycles.
We decompose $X$ as
\be 
X = X_L \cup B \cup X_A .
\ee  
Here $X_L$ is a tubular neighborhood of $S$, which we call the local model. Excising $X_L$ from $X$ produces a geometry which has an asymptotic end $X_A$ but is otherwise a bordism $B$; that is, $B \cong B \cup X_A$, see top of Figure~\ref{fig:bordism-setup}.
If $X$ is conical, $B$ is a cylinder with cross-section $\partial X_L=\partial X_A$. If $X$ is not conical, all topology of $X$ not present in $S$ is supported by $B$. By construction, $X_L\setminus S$ and $X_A$ are cylinders with cross-sections given by the connected components of $\partial B$. We will therefore write
\be
\partial B =   \partial_L B \sqcup \partial_A B \,, \qquad \partial X_L   = \partial_L B\,, \quad \partial X_A  = \partial_A B=\partial X .
\ee

Let the theory engineered by $X$ be $d$-dimensional. Following \cite{Apruzzi:2021nmk}, each connected component of $\partial B$ gives a $(d+1)$-dimensional SymTFT. The SymTFT associated to $\partial_L B$ has as physical boundary the degrees of freedom living on $S$. The SymTFT associated to $\partial_A B$ has topological boundary conditions informed by the boundary conditions of the string background set at $\partial X$. The bordism $B$ connects the two SymTFTs. It compactifies to a $d$-dimensional interface, see Figure~\ref{fig:bordism-setup}. These are the two ingredients: $X_L$ supplies the physical boundary, and $B$ supplies the interface.

We now determine the topological structures we must in general track to determine properties of the interface. The relevant results concerning how to compute SymTFT building blocks from manifolds with boundary are spread throughout \cite{Baume:2023kkf,Cvetic:2024dzu}. The starting point is to simply attempt to apply the general compactification framework \cite{Apruzzi:2021nmk}, which produces SymTFTs from closed manifolds, to the manifold with boundary $B$. Immediately we are confronted with the question with respect to which cohomology groups we should expand supergravity fields. The two natural options are: $H^n(B,\partial B)$ and $H^n(B)$. But of course these are not independent, any cocycle which vanishes on $\partial B$ is a (possibly trivial) cocycle on $B$ and so we should reduce modulo this identification, i.e., we should expand with respect to
\be\label{eq:Internal}
H^n(B, \partial_L B \sqcup \partial_A B ) \xrightarrow[]{~\iota^n~}  H^{n}(B ) \,,
\ee
with mapping induced by inclusion and which in general has non-trivial kernel and co-kernel. These mappings sit in the long exact sequences in relative (co)homology associated to $(B,\partial B)$:
\be\label{eq:LESRH}\ba
\scalebox{0.9}{\text{$\dots \rightarrow H_n( \partial_L B)\oplus H_n(\partial_A B ) \rightarrow H_n(B)\xrightarrow[]{~\iota_n~}  H_n(B, \partial_L B \sqcup \partial_A B ) \rightarrow H_{n-1}( \partial_L B)\oplus H_{n-1}(\partial_A B ) \rightarrow \dots $}  } \\
\scalebox{0.9}{\text{$ \dots \leftarrow H^{n}( \partial_L B)\oplus H^{n}(\partial_A B ) \leftarrow H^n(B) \xleftarrow[]{~\iota^n~}  H^n(B, \partial_L B \sqcup \partial_A B ) \leftarrow H^{n-1}( \partial_L B)\oplus H^{n-1}(\partial_A B )  \leftarrow \dots  $}} \, .\\
\ea \ee
The relevance of this is as follows. Recall that $H^{n}( \partial_L B)\oplus H^{n}(\partial_A B )$ specifies the two expansion bases relevant to the SymTFTs to either side of the interface (see Figure~\ref{fig:bordism-setup} for an illustration). The mappings
\be\ba
H^{n-1}(\partial_L B) \oplus H^{n-1}(\partial_A B)  &\xrightarrow[]{~d_{n-1}~}  H^n(B, \partial_L B \sqcup \partial_A B )\,, \\  H^{n}(B )  &\xrightarrow[]{~r_n~} H^{n}(\partial_L B) \oplus H^{n}(\partial_A B) \,,
\ea \ee
then simply relate the expansion bases and are associated with boundary conditions of the bulk SymTFT fields to either side of the interface along the interface. These mappings in turn can locally be understood straightforwardly. The boundary map $d_{n-1}$ is, locally near the boundaries $\partial_L B$ and $\partial_A B$, a derivative together with wedge corresponding to the directions normal to the boundary.\footnote{More precisely, the boundary map $d_{n-1}:H^{n-1}(\partial B)\to H^n(B,\partial B)$ sends a closed boundary form $\alpha$ to the exterior derivative of an extension into $B$.
Extend $\alpha$ constantly along a collar with inward normal coordinate $r$, and choose a cutoff $\chi(r)$ equal to $1$ near the boundary and $0$ outside the collar.
Then
\be \label{eq: d-map}
d_{n-1}[\alpha]=[d(\chi\alpha)]=[\chi'(r)\,dr\wedge\alpha]\,.
\ee
This cocycle has vanishing boundary pullback, and its relative cohomology class is independent of the extension and cutoff.} The mapping $r_n$ is a restriction of the cocycle in the bordism to its boundary. They respectively correspond to Neumann and Dirichlet boundary conditions.

\subsection{Classification of interfaces}

Let us expand on the $(d+1)$-dimensional SymTFT perspective. Consider some $p$-form $11d$ supergravity field strength $G_p$ and expand it as
\be 
G_p = C_{p-n} \wedge \omega_{n}\,,
\ee 
with $\omega_{n}\in H^{n}(\partial B)$ and SymTFT bulk field $C_{p-n}$. Now, simply follow a local radial direction in $X$ from $X_L,\,X_A$ into the bordism $B$. In both cases, before entering the bordism the direction parametrizes the auxiliary SymTFT direction orthogonal to the quantum field theory spacetime, i.e., with respect to the projection of $X$ onto the SymTFT slab it is horizontal. When it enters the bordism it becomes vertical with respect to the projection of $X$. We therefore have 
\be 
\iota_{\partial_r} C_{p-n}\Big|_{\text{Interface}} = C_{p-n-1}'\,,
\ee 
where $C_{p-n-1}'$ is a SymTFT coefficient field supported on the interface and obtained by expanding $G_p$ with respect to $d_{n}\omega_n$. Of course it can also be that $\omega_n$ simply restricts to a cocycle $\omega_n'$ of the bordism. In that case we have 
\be 
 C_{p-n}\Big|_{\text{Interface}} = C_{p-n}'\,,
\ee 
where $C_{p-n}'$  is a SymTFT coefficient field supported on the interface and obtained by expanding $G_p$ with respect to $\omega_n'$ (such that $r_n(\omega_n) = \omega'_n$). In both cases, generally, the left hand side is a sum of bulk fields (one summand for every boundary component of the bordism, i.e., here two) restricted to the interface.

Returning to \eqref{eq:Internal} we see that the above two boundary conditions will fix all degrees of freedom on the interface except those in $\text{Im}\,\iota^n$, which follows by exactness of \eqref{eq:LESRH}. We therefore arrive at a first crude classification of SymTFT interfaces constructed from extra-dimensional smooth bordisms
\be\label{eq:ClassificationOfInterfaces}
\text{Im}\,\iota^n=\begin{cases} 0 \quad \qquad ~~ \qquad \qquad (\text{Pure Boundary Condition Interface}) \\ \text{Torsion}\: ~~\quad \quad  \qquad (\text{Boundary Condition Interface $+$ Topological d.o.f.})   \\ \text{Free} \oplus \text{Torsion}  \quad \! \quad (\text{Boundary Condition Interface $+$ (Non-)Topological d.o.f.})   \end{cases}
\ee
In the first two cases the interface is topological. It either purely imposes boundary conditions on the bulk fields, or additionally carries a topological sector. As emphasized in the introduction, this is very much like the club sandwich type configuration \cite{Bhardwaj:2023bbf, Bhardwaj:2024qrf, Robbins:2025puq}. 

In many cases supergravity will associate in part simply a theory of abelian gauge fields to the last case, and the configuration is instead that of a SymTree type \cite{Baume:2023kkf}, which generalizes the club sandwich. If we allow for singular $B$ then, upon appropriately generalizing the underlying cohomology theory, the interacting theories corresponding to the last case generalize further beyond this abelian class. 

The above classification is crude as it does not account for which subset of interface degrees of freedom are retained in geometric engineering limits or associated with symmetry structures relevant to degrees of freedom retained in such limits\footnote{Compared to the bottom up constructions, where SymTFTs are extracted from an underlying quantum field theory \cite{Freed:2022qnc}, we have that in top down constructions the SymTFT precedes the quantum field theory \cite{Apruzzi:2021nmk,Baume:2023kkf,Cvetic:2024dzu}. The quantum field theory is only realized in certain decoupling limits and as such general compactification procedures may over-produce SymTFT sectors and structure. Any top down approach therefore generally includes a step wherein we have to determine which structures are relevant for the degrees of freedom retained in any geometric engineering limits.}. For example, there could be a KK-symmetry associated with a free cycle which plays no conventional role in many cases where the KK-tower is decoupled. 

This said, it may be interesting to retain these symmetries which act trivially on the QFT at all scales and across its full parameter space as these in principle retain some information on the string theoretic origin of the SymTFT and a putative UV completion of the theory under consideration once coupled to gravity. However, here we will take a more pragmatic approach. We will take our starting point as the QFT and the SymTFT associated with $X_L$. There we will identify particular symmetries of interest (e.g., center symmetries), identify the relevant geometric structures and see how they extend through the bordism to the boundary $\partial X=\partial X_A$.

For this reason, we may also be interested in attempting to separate the interface structures associated with torsional and free classes in \eqref{eq:ClassificationOfInterfaces}. Formally, this may be done by further cutting out neighborhoods of representatives of free class from $B$. Geometrically, the resulting bordism $B^\circ$ then connects to geometries associated with the non-topological degrees of freedom of $B$, and the original $X_L,\,X_A$. We say formally, since in many cases the corresponding free cycles in the bordism $B$ have trivial normal bundle directions and are as such mobile, whereby it is physically unmotivated to isolate any one representative into a fixed local model.

Finally, there is also the question whether there exist QFT parameters which can be turned on to separate out the interface. With respect to the decomposition $X=X_L\cup B\cup X_A$ this amounts to asking whether a good local limit exists which zooms in on $X_L$, a question which is beyond topology. 
When such a limit exists, $X_L$ provides an ordinary physical boundary for the SymTFT of $\partial_L B$. If such a limit does not exist then the modified physical boundary condition cannot be separated into a SymTFT interface and a new physical boundary condition.

\section{A toy model example: toroidal orbifold}\label{sec:example}

We now illustrate the construction of Section~\ref{sec:bordism_general} in a simple example. We take an internal space $X = \widehat{X} \setminus X^{\text{loc}}$, obtained by excising the local models $X^{\text{loc}}$ of ADE singularities from a compact space $\widehat{X}$, so that $X$ carries torsion in its homology. The purpose of the example is one thing only: to construct the bordism $B$ explicitly and to identify the interface it gives rise to, together with the SymTFTs on its two sides. 

Note that the example is for demonstration purposes. Whether these configurations admit a limit yielding a local QFT is not clear. A candidate is the large volume and low energy limit of Section~\ref{sec:top-down}, with energies kept below the KK scale. The interpretation of the result as the SymTFT of a local QFT is conditional on such a limit.

\subsection{Bordism and the interface}

\begin{figure}[ht]
\centering
\scalebox{0.7}{
\begin{tikzpicture}
	\begin{pgfonlayer}{nodelayer}
		\node [style=none] (0) at (-2, 3.25) {};
		\node [style=none] (1) at (-2, 3.25) {};
		\node [style=none] (2) at (-2, 2.25) {};
		\node [style=none] (3) at (-2, 1.75) {};
		\node [style=none] (4) at (-2, 1.75) {};
		\node [style=none] (5) at (-2, 0.75) {};
		\node [style=none] (6) at (-2, -0.75) {};
		\node [style=none] (7) at (-2, -0.75) {};
		\node [style=none] (8) at (-2, -1.75) {};
		\node [style=none] (9) at (-2, -2.25) {};
		\node [style=none] (10) at (-2, -2.25) {};
		\node [style=none] (11) at (-2, -3.25) {};
		\node [style=none] (12) at (4, 2) {};
		\node [style=none] (13) at (4, 2) {};
		\node [style=none] (14) at (4, 1) {};
		\node [style=none] (15) at (4, 0.5) {};
		\node [style=none] (16) at (4, 0.5) {};
		\node [style=none] (17) at (4, -0.5) {};
		\node [style=none] (18) at (4, -1) {};
		\node [style=none] (19) at (4, -1) {};
		\node [style=none] (20) at (4, -2) {};
		\node [style=none] (21) at (-1, 0.25) {};
		\node [style=none] (22) at (-1, -0.25) {};
		\node [style=none] (23) at (1, 0.95) {};
		\node [style=none] (24) at (2, 0.95) {};
		\node [style=none] (25) at (2.25, 1) {};
		\node [style=none] (26) at (0.75, 1) {};
		\node [style=none] (27) at (1, -1.05) {};
		\node [style=none] (28) at (2, -1.05) {};
		\node [style=none] (29) at (2.25, -1) {};
		\node [style=none] (30) at (0.75, -1) {};
		\node [style=none] (31) at (0.25, -0.05) {};
		\node [style=none] (32) at (1.25, -0.05) {};
		\node [style=none] (33) at (1.5, 0) {};
		\node [style=none] (34) at (0, 0) {};
		\node [style=none] (35) at (2, -3) {$B$};
		\node [style=none] (36) at (-3.25, 2.75) {$(S^3/\Z_2)_1$};
		\node [style=none] (37) at (-3.25, 1.25) {$(S^3/\Z_2)_2$};
		\node [style=none] (38) at (-3.25, -1.25) {$(S^3/\Z_2)_{15}$};
		\node [style=none] (39) at (-3.25, -2.75) {$(S^3/\Z_2)_{16}$};
		\node [style=none] (41) at (5.5, 1.5) {$\partial T(\widehat{A}_1\cup \widehat{B}_1)$};
		\node [style=none] (42) at (5.5, 0) {$\partial T(\widehat{A}_2\cup \widehat{B}_2)$};
		\node [style=none] (43) at (5.5, -1.5) {$\partial T(\widehat{A}_3\cup \widehat{B}_3)$};
		\node [style=none] (44) at (9.75, 0) {$\partial T(\widehat{A} \cup \widehat{B})$};
		\node [style=none] (45) at (11, 0.5) {};
		\node [style=none] (46) at (11, 0.5) {};
		\node [style=none] (47) at (11, -0.5) {};
		\node [style=none] (48) at (13, 0.45) {};
		\node [style=none] (49) at (14, 0.45) {};
		\node [style=none] (50) at (14.25, 0.5) {};
		\node [style=none] (51) at (12.75, 0.5) {};
		\node [style=none] (52) at (13, -0.55) {};
		\node [style=none] (53) at (14, -0.55) {};
		\node [style=none] (54) at (14.25, -0.5) {};
		\node [style=none] (55) at (12.75, -0.5) {};
		\node [style=none] (56) at (13.5, 2) {};
		\node [style=none] (57) at (15.5, 0) {};
		\node [style=none] (58) at (13.5, -2) {};
		\node [style=none] (59) at (0.625, -4) {(i) \qquad $H_n(B)\cong \{\Z,  \Z_2^{11},  \Z^{15}, \Z^{18} \}$};
		\node [style=none] (60) at (13.5, -4) {(ii)\qquad $ H_n(T(\widehat{A}\cup \widehat{B}))\cong \{\Z,\Z^5,\Z^2\}$ };
		\node [style=none] (61) at (7.5, -5) {};
		\node [style=none] (62) at (13.5, -2.75) {$X_L = T(\widehat{A}_i\cup \widehat{B}_i)$};
		\node [style=none] (63) at (1.85, -4.5) {$H_{n}(\partial B) \cong \{\Z^{19}, \Z^{15} \oplus \Z_2^{22}, \Z^{15},\Z^{19}\}$};
		\node [style=none] (64) at (14.7, -4.5) {$H_n(\partial T(\widehat{A}\cup \widehat{B}))\cong\{\Z, \Z^5\oplus \Z_2^2,\Z^5,\Z\}$};
		\node [style=none] (65) at (16.5, 0) {};
	\end{pgfonlayer}
	\begin{pgfonlayer}{edgelayer}
		\draw [style=ThickLine, bend left=15] (2.center) to (1.center);
		\draw [style=ThickLine, bend left=15] (1.center) to (2.center);
		\draw [style=ThickLine, bend left=15] (5.center) to (4.center);
		\draw [style=ThickLine, bend left=15] (4.center) to (5.center);
		\draw [style=ThickLine, bend left=15] (8.center) to (7.center);
		\draw [style=ThickLine, bend left=15] (7.center) to (8.center);
		\draw [style=ThickLine, bend left=15] (11.center) to (10.center);
		\draw [style=ThickLine, bend left=15] (10.center) to (11.center);
		\draw [style=ThickLine, bend left=15] (14.center) to (13.center);
		\draw [style=ThickLine, bend left=15] (13.center) to (14.center);
		\draw [style=ThickLine, bend left=15] (17.center) to (16.center);
		\draw [style=ThickLine, bend left=15] (16.center) to (17.center);
		\draw [style=ThickLine, bend left=15] (20.center) to (19.center);
		\draw [style=ThickLine, bend left=15] (19.center) to (20.center);
		\draw [style=ThickLine, in=180, out=0] (1.center) to (13.center);
		\draw [style=ThickLine, bend right=90, looseness=2.75] (14.center) to (16.center);
		\draw [style=ThickLine, bend right=90, looseness=2.50] (17.center) to (19.center);
		\draw [style=ThickLine, in=0, out=-180] (20.center) to (11.center);
		\draw [style=ThickLine, bend right=90, looseness=2.75] (10.center) to (8.center);
		\draw [style=ThickLine, bend left=90, looseness=3.00] (2.center) to (4.center);
		\draw [style=ThickLine, in=90, out=0, looseness=1.25] (5.center) to (21.center);
		\draw [style=ThickLine, in=0, out=-90, looseness=1.25] (22.center) to (7.center);
		\draw [style=DottedLine] (21.center) to (22.center);
		\draw [style=ThickLine, bend left=15] (23.center) to (24.center);
		\draw [style=ThickLine, bend left=15] (25.center) to (26.center);
		\draw [style=ThickLine, bend left=15] (27.center) to (28.center);
		\draw [style=ThickLine, bend left=15] (29.center) to (30.center);
		\draw [style=ThickLine, bend left=15] (31.center) to (32.center);
		\draw [style=ThickLine, bend left=15] (33.center) to (34.center);
		\draw [style=ThickLine, bend left=15] (47.center) to (46.center);
		\draw [style=ThickLine, bend left=15] (46.center) to (47.center);
		\draw [style=ThickLine, bend left=15] (48.center) to (49.center);
		\draw [style=ThickLine, bend left=15] (50.center) to (51.center);
		\draw [style=ThickLine, bend left=15] (52.center) to (53.center);
		\draw [style=ThickLine, bend left=15] (54.center) to (55.center);
		\draw [style=ThickLine, in=-180, out=0] (46.center) to (56.center);
		\draw [style=ThickLine, in=90, out=0, looseness=0.75] (56.center) to (57.center);
		\draw [style=ThickLine, in=0, out=-90, looseness=0.75] (57.center) to (58.center);
		\draw [style=ThickLine, in=0, out=180] (58.center) to (47.center);
	\end{pgfonlayer}
\end{tikzpicture}
}
\caption{We have cut the geometry $X= T^4/\Z_2 \setminus X^{loc}$ into two pieces. (i). The bordism $B$, its boundary consists of 16 copies of $S^3/\Z_2$ and three tube boundaries. (ii). The local model $X_L$. The homology groups are listed below each picture.}
\label{fig:CutCut}
\end{figure}
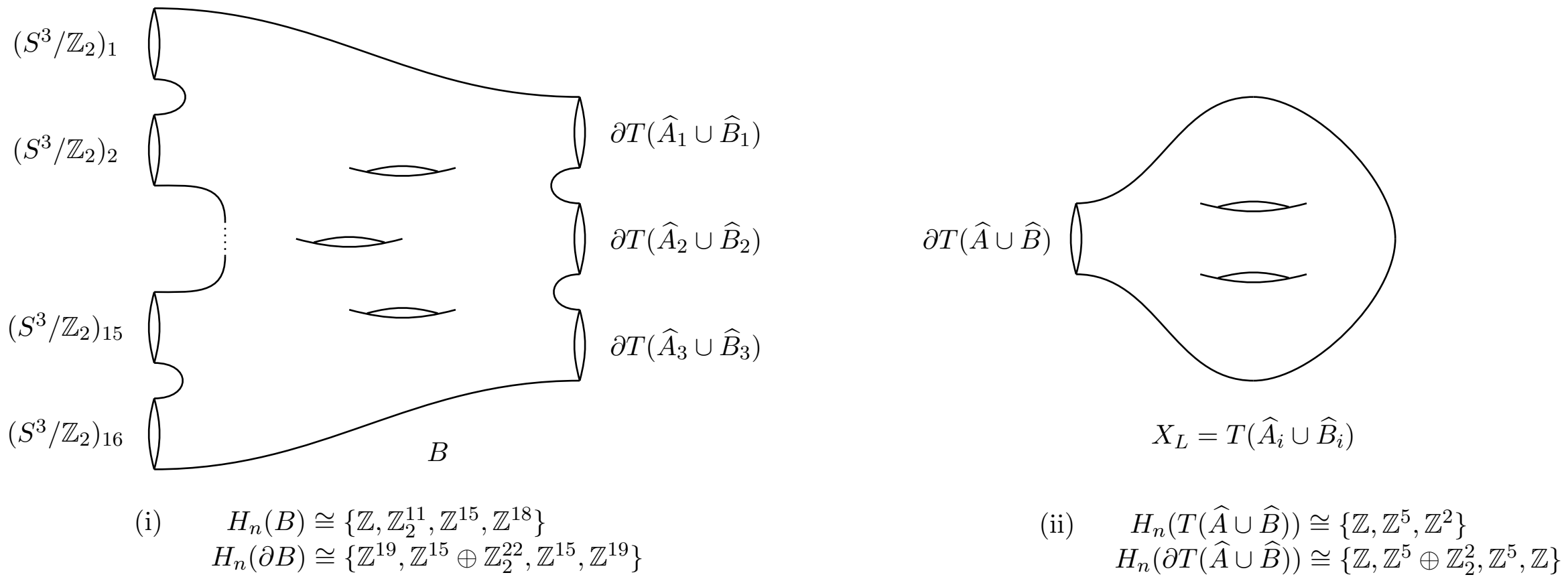

Consider the toroidal orbifold $\widehat{X}=T^4/\Z_2$. Here $X^{\text{loc}}$ is the disjoint union of 16 small balls, each modeled on a ball centered at the origin of $\mathbb{C}^2/\Z_2$. By excising them, we obtain a smooth manifold with boundary $X : = \widehat{X} \setminus X^{\text{loc}} $ whose homology groups are
\be\label{eq:MVinformant1}
H_n(X)\cong \{\Z,\Z_2^5,\Z^6,\Z^{15},0\}\,,
\ee
where left to right we list $n=0,\dots,4$. The generators are straightforward to identify. They are the point, 5 circles, 6 2-tori and $16-1=15$ boundary components $S^3/\Z_2$. The five circles originate from the four cycles of the torus (which become order two in the quotient) and one of the Hopf circles of any of the $S^3/\Z_2$ boundary components.  The 15 three-cycles come from the 16 copies of $S^3/\Z_2$ modulo one relation introduced by $X$ itself.
The 6 2-tori form 3 hyperbolic pairs and each pair has intersection matrix
\be
U(2)=\left( \begin{array}{cc} 0 & 2 \\ 2 & 0 \end{array} \right)\,,
\ee
so that, as a lattice, we have $H_2(X)\cong U(2)^{\oplus 3}$. Consider any of the three hyperbolic pairs of 2-cycles $\widehat{A},\,\widehat{B}$ represented as two 2-tori intersecting at 2 points. Then, $\widehat{A} \cup \widehat{B}$, and likewise its tubular neighborhood $T(\widehat{A}\cup \widehat{B})$, is deformation equivalent to the wedge sum $T^2\vee T^2\vee S^1$. We therefore have the homology groups
\be\label{eq:MVinformant2}
H_n(\widehat{A}\cup \widehat{B})\cong H_n(T(\widehat{A}\cup \widehat{B}))\cong \{\Z,\Z^5,\Z^2\}\,.
\ee
The tube $T(\widehat{A} \cup \widehat{B})$ is a manifold with boundary $\partial T(\widehat{A}\cup \widehat{B})$ and from the intersection form $U(2)$ informs the homology groups
\be
H_n(\partial T(\widehat{A}\cup \widehat{B}))\cong\{\Z, \Z^5\oplus \Z_2^2,\Z^5,\Z\}\,.
\ee
In total $X$ contains three such disjoint neighborhoods $T(\widehat{A}_i\cup \widehat{B}_i)$ with $i=1,2,3$ in obvious notation. We take these as our local model $X_L$ and define the bordism $B$ as, 
\begin{equation}
    X_L = \bigsqcup^3_{i=1} T(\widehat{A}_i\cup \widehat{B}_i), \qquad B = X \setminus X_L. 
\end{equation}
$B$ is a smooth manifold with boundary and the boundaries of $B$ are, in our notation, 
\begin{equation}
    \partial B = \partial_L B \, \sqcup \partial_A B \, ,
\end{equation}
where 
\begin{equation}
    \partial_L B = \partial X_L= \left( \bigsqcup_{i=1}^3 \partial T(\widehat{A}_i\cup \widehat{B}_i)\right), \qquad  \partial_A B = \partial X =  \left( \bigsqcup_{a=1}^{16} (S^3/\Z_2)_a\right)\,.
\end{equation}
It has a total of 19 components and overall homology groups of
\be
H_n(\partial B) \cong \{\Z^{19}, \Z^{15} \oplus \Z_2^{22}, \Z^{15},\Z^{19}\}\,,
\ee
where $22=16+3\times 2$, and which contain the subgroups 
\be\label{eq:MVinformant3}
H_n(\partial( T(\widehat{A}_i\cup \widehat{B}_i)))^{\oplus 3}\cong \{\Z^3,\Z^{15}\oplus \Z_2^6,  \Z^{15}, \Z^3\}\,.
\ee
The homology groups $H_n(B)$ follow via application of a Mayer--Vietoris sequence informed by \eqref{eq:MVinformant1}, \eqref{eq:MVinformant2} and \eqref{eq:MVinformant3}. Our starting point for computing  $H_n(B)$ is therefore:
\begin{equation}\label{eq:computeHY}
    \begin{array}{c||cccccc}
&   H_n(\partial X_L) &  & H_n(B)\oplus H_n(X_L) &  &   H_n(X) &      \\[0.4em] \hline \hline \\[-0.9em]
            n=3 ~&  \Z^{3} & \rightarrow &H_3(B)\oplus 0 & \rightarrow & \Z^{15} & \rightarrow  \\
         n=2~ &  \Z^{15}  &\rightarrow & H_2(B)\oplus \Z^{6}  & \rightarrow& \Z^6 &\rightarrow \\
       n=1~ &    \Z^{15}\oplus \Z_2^6  &\rightarrow &H_1(B)\oplus \Z^{15} &\rightarrow &\Z_2^5&\rightarrow  \\
        n=0 ~&   \mathbb{Z}^3  &\rightarrow & H_0(B)\oplus \Z^3 &\rightarrow & \Z  &   
    \end{array}
\end{equation}
By construction we have $ H_2(X_L) \cong H_2(X) \cong \Z^6$ which amounts to ``separating out the 2-cycles" via excision as shown in Figure~\ref{fig:CutCut}. We also have 
\begin{equation}
    \text{Free}\,H_1(\partial T(\widehat{A} \cup \widehat{B}))\cong H_1( T(\widehat{A} \cup \widehat{B}))\cong \Z^5\,,
\end{equation}
as all the 1-cycles of the tubes cut out admit representatives which lie in their boundary. Next, note that $B$ is connected by construction, i.e., $H_0(B)\cong \Z$ and it trivially realizes a chain whose boundary is $\partial B$, i.e., $H_3(B)\cong \Z^{19-1}$. We therefore extract from the above the exact split sequence
\begin{equation}\label{eq:computeHY2}
    \begin{array}{c||cccccc}
           ~&  0 & \rightarrow &0 & \rightarrow &0 & \rightarrow  \\
         ~ &  \Z^{15}  &\rightarrow & H_2(B) & \rightarrow& 0 &\rightarrow \\
       ~ &    \Z_2^6  &\rightarrow &H_1(B) &\rightarrow &\Z_2^5&\rightarrow  \\
         ~&  0  &\rightarrow & 0&\rightarrow & 0  &   
    \end{array}
\end{equation}
and deduce $H_1(B)\cong \Z_2^{11}$ and $H_2(B)\cong \Z^{15}$. Altogether
\be
H_n(B)\cong \{\Z,  \Z_2^{11},  \Z^{15}, \Z^{18} \}\,,
\ee
where left to right we list $n=0,\dots,3$. From this we obtain
\be
H^{n}(B,\partial B) \cong \{0,  \Z^{18} ,  \Z^{15},   \Z_2^{11}, \Z \}\,,
\ee
and substituting $H^{n}(\partial B) =\{\Z^{19}, \Z^{15} , \Z^{15}\oplus \Z_2^{22},\Z^{19}\}$ into the long exact sequence for relative cohomology of the pair $(B,\partial B)$ we obtain
\begin{equation}\label{eq:upslon1}
    \begin{array}{c||cccccc}
&   H^n(B,\partial B)&  & H^n(B)&  &   H^n(\partial B) &      \\[0.4em] \hline \hline \\[-0.9em]
            n=0 ~&  0& \xrightarrow[]{~\iota^0~}  &H^0(B) & \rightarrow & \Z^{19} & \rightarrow  \\
         n=1~ &  \Z^{18}  &\xrightarrow[]{~\iota^1~} & H^1(B) & \rightarrow& \Z^{15} &\rightarrow \\
       n=2~ &    \Z^{15}  &\xrightarrow[]{~\iota^2~}  &H^2(B)&\rightarrow &\Z^{15}\oplus \Z_2^{22}&\rightarrow  \\
        n=3 ~&   \mathbb{Z}_2^{11}  &\xrightarrow[]{~\iota^3~}  & H^3(B) &\rightarrow & \Z^{19}  &    \\
        n=4 ~&  \Z &\xrightarrow[]{~\iota^4~}  &H^4(B)&\rightarrow &  0&   
    \end{array}
\end{equation}
resulting in 
\begin{equation}
    H^{n}(B)\cong \{ \Z, 0, \Z^{15}\oplus \Z_2^{11}, \Z^{18},0\}\,,
\end{equation}
and the maps
\be\ba
&\iota^0\,:~H^0(B,\partial B) \cong 0 \rightarrow \Z \cong H^0(B)\,, \\ 
&\iota^1\,:~H^1(B,\partial B) \cong \Z^{18} \rightarrow 0 \cong H^1(B)\,, \\ 
&\iota^2\,:~H^2(B,\partial B) \cong \Z^{15} \rightarrow \Z^{15}\oplus \Z_2^{11} \cong H^2(B)\,, \\ 
&\iota^3\,:~H^3(B,\partial B) \cong \Z_2^{11} \rightarrow \Z^{18} \cong H^3(B)\,, \\ 
&\iota^4\,:~H^4(B,\partial B) \cong \Z \rightarrow 0 \cong H^4(B)\,. \\ 
\ea\ee
Except for $\iota^2$, all these maps are trivially zero. In fact $\iota^2$ is also zero: by exactness of \eqref{eq:upslon1}, its image is the kernel of the restriction $H^2(B) \rightarrow H^2(\partial B)$ which must vanish. From this, we conclude that the interface associated to $B$ supports no degrees of freedom, not even topological ones. It will be a pure boundary condition interface.

\subsection{SymTFTs and decoupled gapped sector}

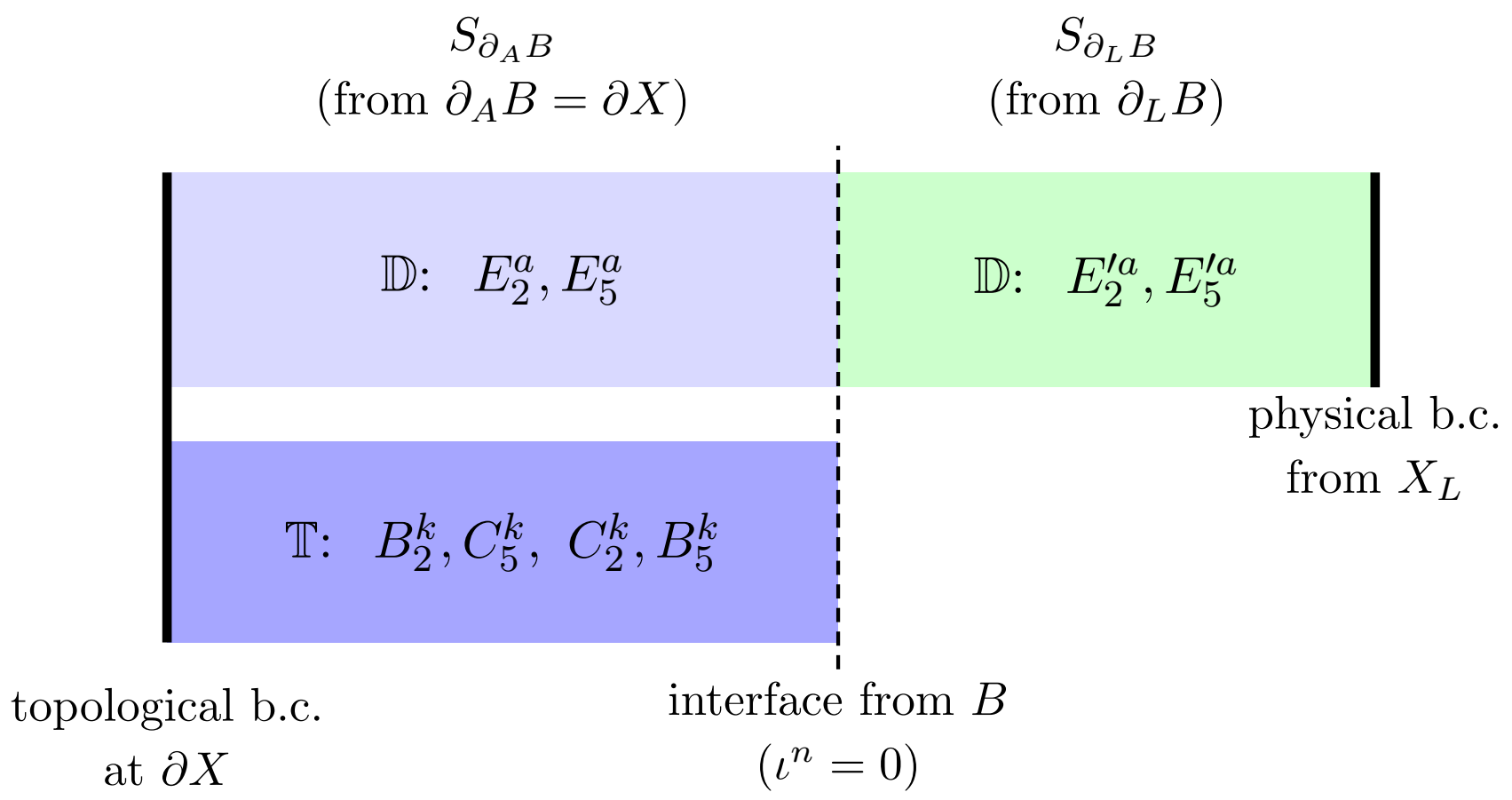
\begin{figure}[ht]
\centering
\begin{tikzpicture}[scale=1]
\fill[blue!15] (0,0.3) rectangle (5,1.9);
\fill[green!20] (5,0.3) rectangle (9,1.9);
\fill[blue!35] (0,-1.6) rectangle (5,-0.1);
\draw[line width=2pt] (0,-1.6) -- (0,1.9);
\draw[thick, densely dashed] (5,-1.8) -- (5,2.1);
\draw[line width=2pt] (9,0.3) -- (9,1.9);
\node at (2.5,1.1) {$\mathbb{D}$: \ $E_2^{a},E_5^{a}$};
\node at (7.0,1.1) {$\mathbb{D}$: \ $E_2^{\prime a},E_5^{\prime a}$};
\node at (2.5,-0.85) {$\mathbb{T}$: \ $B_2^k,C_5^k,\ C_2^k,B_5^k$};
\node[align=center] at (2.5,2.65) {\small $S_{\partial_A B}$\\ 
\small (from $\partial_A B=\partial X$)};
\node[align=center] at (7.0,2.65) {\small $S_{\partial_L B}$\\ 
\small (from $\partial_L B$)};
\node[align=center] at (0,-2.3) {\small topological b.c.\\ \small at $\partial X$};
\node[align=center] at (5,-2.3) {\small interface from $B$\\ \small ($\iota^n=0$)};
\node[align=center] at (9,-0.15) {\small physical b.c.\\ \small from $X_L$};
\end{tikzpicture}
\caption{The two SymTFTs associated with $\partial_A B=\partial X$ and $\partial_L B$. The former contains both $\mathbb D$ and $\mathbb T$, while the latter contains only $\mathbb D$. They are connected by the pure boundary-condition interface obtained from the bordism $B$.}
\label{fig:twoSymTFTs}
\end{figure}

By Poincar\'e--Lefschetz duality and the UCT, the cohomology sequence of $(X,\partial X)$ in degree two reads
\begin{equation}\label{eq:LESXT4}
    0\rightarrow \underbrace{H^2(X,\partial X)}_{\Z^6}\xrightarrow{~\jmath^*~}\underbrace{H^2(X)}_{\Z^6\oplus\Z_2^5}\xrightarrow{~\imath^*~}\underbrace{H^2(\partial X)}_{\Z_2^{16}}\xrightarrow{~d~}\underbrace{H^3(X,\partial X)}_{\Z_2^5}\rightarrow 0\,,
\end{equation}
where on the free part $\jmath^*$ is the intersection form $U(2)^{\oplus3}$. The defect group is encoded in
\begin{equation}
    \mathbb{D} = H^2_{\text{free}}(X)/\text{im}(\jmath^*)\cong \Z_2^6\,,
\end{equation}
generated by $\eta_a=\imath^*(e_a)$, with $e_a$ the free classes dual to the six 2-tori $(\widehat{A}_i, \widehat{B}_i), \, i=1,2,3$. The trivial sector, on the other hand, sits in, 
\begin{equation}
    0 \rightarrow H^2_{\text{Tor}}(X) \cong \Z^5_2 \rightarrow \mathbb{T} \cong \Z^{10}_2 \rightarrow H^3(X, \partial X) \cong \Z^5_2 \rightarrow 0 \, ,
\end{equation}
generated by $t_k = i^*(\tau_k)$, where $\tau_k$ are the generators of $H^2_{\text{Tor}}(X)$ and $s_k$ with $d(s_k) = \sigma_k$, such that $\sigma_k$ generates $H^3(X, \partial X)$. Altogether we have, 
\begin{equation}
    H^2(\partial X) \cong \mathbb{D} 
    \oplus \mathbb{T} \cong \Z^6_2 \oplus \Z^{10}_2 \cong \Z^{16}_2.
\end{equation}
The linking pairings are, 
\begin{equation}
    D_{ab}=L(\eta_a,\eta_b)\,,\qquad D=\big(U(2)^{-1}\big)^{\oplus3}= \begin{pmatrix}0 & \tfrac12 \\ \tfrac12  & 0 \end{pmatrix}^{\oplus3}, \qquad \kappa_{k\ell}=L(t_k,s_\ell)=\tfrac12\,\delta_{k\ell}\,.
\end{equation}
We expand the differential cohomology lift of $G_4$, 
\begin{equation}
    \check{G}_4 = \check{E}^{a}_2 \star \check{\eta}_a + \check{B}^{k}_2 \star \check{s}_k + \check{C}^{k}_2 \star \check{t}_k\,,
\end{equation}
and likewise for $\check{G}_8$. Reducing on $\partial_A B = \partial X = \bigsqcup^{16}_{a=1} (S^3/\Z_2)_a$ gives the SymTFT\footnote{We focus only on the electric/magnetic center symmetry. Throughout, we retain only the sector of the SymTFT obtained from the kinetic term, i.e. the BF couplings for the center symmetries and the torsional sector; contributions of the 
M-theory Chern--Simons term are not included. Moreover, in this example we do have a non-vanishing contribution from the self-linking $L(s_k,s_\ell)$. This adds terms $L(s_k,s_\ell)\, B^k_2\cup\delta B^\ell_5$ to $S_{\partial_A B}$, which we `remove' by the field redefinition $C^k_5 \to C^k_5 + 2\sum_\ell L(s_k,s_\ell)\, B^\ell_5$.}, 
\begin{equation}
    S_{\partial_A B} = 2\pi \int \Big( D_{ab} E^a_2 \cup \delta E^b_5 + \kappa_{k\ell}\,B^\ell_2 \cup \delta C^k_5 + \kappa_{k\ell}\,C^k_2 \cup \delta B^\ell_5 \Big)\, .
\end{equation}
Reducing instead on $\partial_L B =\partial X_L = \bigsqcup^3_{i=1} \partial (T (\widehat{A}_i \cup \widehat{B}_i))$ gives,
\begin{equation}
    S_{\partial_L B} = 2\pi \int D_{ab} \,E'^a_2 \cup \delta E'^b_5\, .
\end{equation}
As illustrated in Figure~\ref{fig:twoSymTFTs}, the SymTFT $S_{\partial_A B}$ associated with the asymptotic boundary $\partial_A B=\partial X$ contains both the defect group sector $\mathbb D$ and the additional sector $\mathbb T$, whereas the SymTFT $S_{\partial_L B}$ associated with the local boundary contains only $\mathbb D$.
The two SymTFTs are connected by a pure boundary-condition interface obtained by compactifying M-theory on $B$, giving a configuration analogous to the club sandwich construction \cite{Bhardwaj:2023bbf,Bhardwaj:2024qrf}. Across this interface, the fields $E^a_{2/5}$ are identified with their counterparts $E^{\prime a}_{2/5}$, while the fields belonging to the $\mathbb T$ sector do not extend to the local side.

The bordism thus isolates the defect group sector associated with the local model $X_L$. Provided a suitable large-volume and low-energy limit below the KK scale exists, compactifying M-theory on $X_L$ gives rise to a seven-dimensional QFT $\mathcal{T}^{\rm toy}$ with an abelian gauge sector. Its electric and magnetic symmetries are captured by $S_{\partial_L B}$ and act on the gauge sector of $\mathcal{T}^{\rm toy}$. The full theory, however, also contains additional degrees of freedom, which in this example form a topological sector. Their associated symmetries are encoded by $\mathbb T$ in the larger SymTFT $S_{\partial_A B}$.

To determine how these symmetries act on the full theory, we fuse the interface with the ordinary physical boundary supplied by $X_L$, recovering the modified physical boundary. On this boundary, the $B_2^k$ and $B_5^k$ operators can terminate on the corresponding holonomies, whereas the $C_2^k$ and $C_5^k$ operators have no endpoints. With that in mind, if we impose Dirichlet conditions on $E_2^a$, $C_2^k$ and $C_5^k$ at the symmetry boundary, the finite higher-form symmetry of the resulting absolute theory $\mathcal{T}^{\rm toy}$ is
\begin{equation}
    \underbrace{\bigl(\Z_2^{(1)}\bigr)^6}_{\text{symmetry of the gauge sector}}
  \times
  \underbrace{\bigl(\Z_2^{(1)}\bigr)^5 \times \bigl(\Z_2^{(4)}\bigr)^5}
             _{\text{symmetry of the TFT sector}}\,.
\end{equation}
The six $\Z_2^{(1)}$ factors associated with the gauge sector act faithfully on the Wilson lines of $X_L$. In contrast, the remaining ten factors, generated by the $B_5^k$ and $B_2^k$ operators, act trivially on the full theory $\mathcal{T}^{\rm toy}$, since the corresponding charged defects would require endpoints for the $C_2^k$ and $C_5^k$ operators, which are absent. Furthermore, $\mathbb D$ and $\mathbb T$ are orthogonal under the linking pairing on $\partial X$, so there is no mixing between the two sectors.

We will not pursue this example further, since the existence of a suitable limit yielding a local QFT remains unclear. Nevertheless, it illustrates the essential mechanism: the bordism separates the symmetries of the local gauge sector from those associated with the additional topological degrees of freedom. We now apply this construction to the frozen theories in their dual geometric description.

\section{SymTFT interface in frozen compactifications}\label{sec:frozen-setting}

While our main motivation is in the 7d SYM theories obtained from M-theory on frozen ADE singularities, the non-zero $C_3$-holonomy that defines these models make a purely geometric approach to the 7d theory difficult.
To fully utilize our strategy from Section \ref{sec:bordism_general}, we employ the duality available after a further $S^1$-reduction \cite{deBoer:2001wca, Tachikawa:2015wka, Cvetic:2024mtt}:\footnote{In fact, this originates from a 7d-duality, where M-theory on a frozen $\bbC^2/\Gamma$ is dual to F-theory on $X$.
However, the cleanest way to extract the F-theory physics is to employ M-/F-theory duality, i.e., identify M-theory on $X$ as F-theory on $X \times S^1$.
}
\begin{align}
    \text{M-theory on (frozen } \mathbb{C}^2/\Gamma_\text{ADE}) \times S^1 \qquad \Longleftrightarrow \qquad \text{M-theory on } X = (Y^\text{ell} \times S^1)/\Z_r \, .
\end{align}
On the right-hand side, the purely geometric background $X$ is a $\Z_r$-orbifold of the product of an elliptic fibration $Y^\text{ell}$ and a circle; the precise type of singular fiber in $Y^\text{ell}$ and the concrete orbifold order/action depends on $\Gamma_\text{ADE}$ and the $C_3$-holonomy on the frozen side.

In \cite{Cvetic:2024mtt} it was argued that besides ``fully frozen'' models which have no gapless sector, only frozen D-type singularities give a modified physical boundary.
Besides deriving this modification cleanly from the dual geometry, our goal will be to identify the bordism that geometrically separates the local singularity containing the gauge sector from the asymptotic boundary, and show that this bordism corresponds to a topological interface in the SymTFT.

For concreteness, we focus on the frozen $D_5$ singularity, whose low-energy EFT contains an $\mathfrak{sp}(1) \cong \mathfrak{su}(2)$ SYM sector.
The dual geometry is
\begin{align}
    X = (Y^{\rm ell}_{I_2} \times S^1)/\bbZ_2 \, ,
\end{align}
where $Y^{\rm ell}_{I_2}$ is a local elliptic surface with a central $I_2$-fiber, and the $\bbZ_2$ acts as a half-shift on the circle and an $SL(2,\bbZ)$ monodromy on the elliptic fibration \cite{Tachikawa:2015wka, Cvetic:2024mtt}.\footnote{More generally, the dual orbifold geometry for a frozen $D_{4+n}$ singularity is $(Y^\text{ell}_{I_{2n}} \times S^1)/\Z_2$. Using the partial results about the cohomology of these spaces in \cite{Cvetic:2024mtt}, the computations we present for $n=1$ here should generalize straightforwardly.}
We denote the relative 6d theory resulting from M-theory on $X$ by $\mathcal{T}^{\frozen}$.
Besides a twisted circle worth of $A_1$-singularities which supports the gauge sector, $X$ also has torsion, $\mathrm{Tor}\,H_1(X)\cong\Z_2$, which produces an additional topological sector and allows us to apply the general prescription for modified physical boundaries. We first derive this boundary and identify the SymTFT operators that cannot terminate on it, as observed in \cite{Cvetic:2024mtt}. We then use the bordism construction to isolate the local gauge sector, before examining the global forms of $\mathcal{T}^{\frozen}$.

\subsection{Identifying the modified physical boundary}\label{sec: frozen_physical_boundary}

In this subsection, we first derive the modified physical boundary from KK-reduction on $X$ (or, rather, its resolution $\wZ$).
In essence, the logic is analogous to the one outlined in Section \ref{sec:top-down}.
In particular, the non-zero torsion of $H^2(\wZ)$ and $H^4(\wZ, \partial \wZ)$ will lead to a topological sector.
The crucial difference is now that the symmetry of this sector is coupled to the center symmetry of the gauge sector through the SymTFT, since we no longer have a splitting of the form $H^2(\partial \wZ) = \mathbb{D} \oplus \mathbb{T}$.

To begin with, we resolve the exceptional curve, the non-affine $\mathbb{P}^1$ of the central $I_2$-fiber, and denote the resulting space by $\wZ$. The exceptional class gives a second generator of $H_2(Y^{\rm ell}_{I_2})\cong\Z^2$ and makes the W-bosons (and the Cartan fluxes from the dual cohomology) explicit. The resolution does not change the asymptotic boundary, $\partial\wZ=\partial X$. Following Section~\ref{sec:coupling-symtft}, we reduce the 11d kinetic term on $\wZ$, couple the resulting 6d action to the SymTFT of $\partial\wZ$, and read off the allowed endpoints from the boundary equations of motion.

The relevant homology (and, by Lefschetz duality, relative cohomology) groups of $\widehat{X}$ and $\partial \widehat{X}$ are (see the discussion in Appendix C of~\cite{Cvetic:2024mtt})
\begin{align}\label{eq:homology_resolved}
\begin{split}
    H_n(\widehat{X}) = H^{5-n}(\widehat{X}, \partial \widehat{X}) & = \begin{cases}
        \bbZ \, , & n = 0 \, , \\
        \bbZ \oplus \bbZ_2 \, , & n =1 \, , \\
        \bbZ^2 \, , & n = 2, 3 \, , \\
        0 \, , & n =4, 5 \, ;
    \end{cases} \\
    H_n(\partial \widehat{X}) = H^{4-n}(\partial \widehat{X}) & = \begin{cases}
        \bbZ \, , & n = 0\, , \\ \bbZ^2 \oplus \bbZ_4 \, , & n = 1, 2 \, , \\
        \bbZ^2 \, , & n = 3\, , \\
        \bbZ \, , & n=4\, .
    \end{cases}
\end{split}
\end{align}
To complete the long exact sequence that we will need later, we also need to determine the relative homology groups. Using UCT (which we denote by $\cong$):
\begin{align}\label{eq: relative_homology_resolved}
    \begin{split}
        H_n(\widehat{X}, \partial \widehat{X}) = H^{5-n}(\widehat{X}) \cong H_{5-n}^{\rm free}(\widehat{X}) \oplus H_{4-n}^{\rm tor}(\widehat{X}) = \begin{cases}
            0 \, , & n = 0, 1 \, , \\
            \bbZ^2 \, , & n = 2 \, , \\
            \bbZ^2 \oplus \bbZ_2 \, , & n = 3 \, , \\
            \bbZ \, , & n = 4, 5 \, .
        \end{cases}  
    \end{split}
\end{align}

\subsubsection*{The cohomology sequence}

With all the ingredients discussed above, the long exact sequence in cohomology takes the following form,
\begin{equation}\label{eq:long_seq-twisted_geometry}
\begin{aligned}
0 &\longrightarrow
\underset{\bbZ}{H^1(\wZ)}
\xrightarrow{\imath_1^*}
\underset{\bbZ^2}{H^1(\partial\wZ)}
\xrightarrow{d_1}
\underset{\bbZ^2}{H^2(\wZ,\partial\wZ)} \xrightarrow{\jmath^*_2}
\\
&\xrightarrow{\jmath_2^*}
\underset{\bbZ^2\oplus\bbZ_2}{H^2(\wZ)}
\xrightarrow{\imath_2^*}
\underset{\bbZ^2\oplus\bbZ_4}{H^2(\partial\wZ)}
\xrightarrow{d_2}
\underset{\bbZ^2}{H^3(\wZ,\partial\wZ)} \xrightarrow{\jmath^*_3}
\\
&\xrightarrow{\jmath_3^*}
\underset{\bbZ^2}{H^3(\wZ)}
\xrightarrow{\imath_3^*}
\underset{\bbZ^2\oplus\bbZ_4}{H^3(\partial\wZ)}
\xrightarrow{d_3}
\underset{\bbZ\oplus\bbZ_2}{H^4(\wZ,\partial\wZ)}
\longrightarrow 0 \, .
\end{aligned}
\end{equation}
For the calculations below, it will be important to keep track of the generators associated with different groups in the long exact sequence above. We therefore collect our basis choices and the action of the maps on these generators in Table~\ref{tab:maps-twisted-geometry} (see Appendix~\ref{app: homology_tracking} for the derivation). 

\begin{table}[ht] 
\centering
\small
\setlength{\tabcolsep}{4pt}
\renewcommand{\arraystretch}{1.4}

\begin{tabularx}{\textwidth}{
@{}
>{\raggedright\arraybackslash}p{0.29\textwidth}
>{\centering\arraybackslash}X
>{\raggedright\arraybackslash}p{0.30\textwidth}
@{}
}
\toprule
\textbf{Map} & \textbf{Generators} & \textbf{Action on generators} \\
\midrule

\(\imath_1^*:H^1(\wZ)\rightarrow H^1(\partial \wZ)\)
&
\(\displaystyle
u\longrightarrow(v_1,v_2)
\)
&
\(\imath_1^*(u)=v_2\)
\\
\midrule

\(d_1:H^1(\partial \wZ)\rightarrow H^2(\wZ, \partial \wZ)\)
&
\(\displaystyle
(v_1,v_2)\longrightarrow(\rho_2,\omega_2)
\)
&
\( d_1(v_1)=\rho_2 \)
\par
\(d_1(v_2)=0\)
\\
\midrule

\(\jmath_2^*:H^2(\wZ, \partial \wZ)\rightarrow H^2(\wZ)\)
&
\(\displaystyle
(\rho_2,\omega_2)
\longrightarrow
(z_2,\eta_2;\tau_2)
\)
\par
\(\displaystyle 2\tau_2=0\)
&
\(\jmath_2^*(\rho_2)=0\)
\par
\(\jmath_2^*(\omega_2)=2\eta_2+\tau_2\)
\\
\midrule

\(\imath_2^*:H^2(\wZ)\rightarrow H^2(\partial \wZ)\)
&
\(\displaystyle
(z_2,\eta_2;\tau_2)
\longrightarrow
(a_2,b_2;x_2)
\)
\par
\(\displaystyle
2\tau_2=0,\qquad 4x_2=0
\)
&
\(\imath_2^*(z_2)=a_2\)
\par
\(\imath_2^*(\eta_2)=x_2\)
\par
\(\imath_2^*(\tau_2)=2x_2\)
\\
\midrule

\(d_2:H^2(\partial \wZ)\rightarrow H^3(\wZ, \partial \wZ)\)
&
\(\displaystyle
(a_2,b_2;x_2)
\longrightarrow
(\omega_3,\rho_3)
\)
\par
\(\displaystyle 4x_2=0\)
&
\(d_2(a_2)=0\)
\par
\(d_2(b_2)=\rho_3\)
\par
\(d_2(x_2)=0\)
\\
\midrule

\(\jmath_3^*:H^3(\wZ, \partial \wZ)\rightarrow H^3(\wZ)\)
&
\(\displaystyle
(\omega_3,\rho_3)
\longrightarrow
(\eta_3,z_3)
\)
&
\(\jmath_3^*(\omega_3)=2\eta_3\)
\par
\(\jmath_3^*(\rho_3)=0\)
\\
\midrule

\(\imath_3^*:H^3(\wZ)\rightarrow H^3(\partial \wZ)\)
&
\(\displaystyle
(\eta_3,z_3)
\longrightarrow
(a_3,b_3;x_3)
\)
\par
\(\displaystyle 4x_3=0\)
&
\(\imath_3^*(\eta_3)=2x_3\)
\par
\(\imath_3^*(z_3)=b_3\)
\\
\midrule

\(d_3:H^3(\partial \wZ)\rightarrow H^4(\wZ, \partial \wZ)\)
&
\(\displaystyle
(a_3,b_3;x_3)
\longrightarrow
(\rho_4;\sigma_4)
\)
\par
\(\displaystyle 4x_3=0,\qquad2\sigma_4=0 \)
&
\(d_3(a_3)=\rho_4\)
\par
\(d_3(b_3)=0\)
\par
\(d_3(x_3)=\sigma_4\)
\\
\midrule

\(\jmath_4^*:H^4(\wZ, \partial \wZ)\rightarrow H^4(\wZ)=0\)
&
\(\displaystyle
(\rho_4;\sigma_4)\longrightarrow0
\)
\par
\(\displaystyle 2\sigma_4=0\)
&
\(\jmath_4^*(\rho_4)=0\)
\par
\(\jmath_4^*(\sigma_4)=0\)
\\

\bottomrule
\end{tabularx}

\caption{Generators and their images under the maps in the long
exact cohomology sequence \eqref{eq:long_seq-twisted_geometry}.}
\label{tab:maps-twisted-geometry}
\end{table}

In this sequence there are two ``types'' of free factors that appear.
In degree 2 and 3, the resolution of the singularity gives compactly supported class $\omega_{2,3} \in H^{2,3}(\widehat{X}, \partial \wZ)$ that are mapped non-trivially by $\jmath^*$.
Their images contain the non-compactly supported classes $\eta_{2,3} \in H^{2,3}(\widehat{X})$, which eventually will give rise to background fields of the gauge sector's center symmetry.

The other free factors are carried through the sequence ``trivially'', 
\begin{equation}\label{eq:passed-through-free-classes}
\begin{aligned}
    \imath_1^*(u)=v_2 \, , \qquad d_1(v_1)=\rho_2 \, , \qquad  \imath_2^*(z_2)=a_2 \, , \\
    d_2(b_2)=\rho_3 \, , \qquad \imath_3^*(z_3)=b_3 \, , \qquad d_3(a_3)=\rho_4 \, .
\end{aligned}    
\end{equation}
Physically, these eventually give rise to global $U(1)$ symmetries for KK states as well as the gauge instantons.
We will argue in Appendix~\ref{app: decoupled_gapless} that these symmetries can be consistently disregarded when it comes to the discussion of the modified physical boundary, its coupling to the SymTFT, and topological interface from the bordism.

In geometry, the maps \eqref{eq:passed-through-free-classes} splits the long exact sequence \eqref{eq:long_seq-twisted_geometry} into two pieces; the first is a short exact sequence in degree 2,
\begin{align}\label{eq:twisted_geom_seq-for-H2}
    0 \rightarrow \underbrace{\bbZ \langle\omega_2\rangle}_{\subset H^2(\wZ, \partial \wZ) = H_3(\wZ)}  \stackrel{\jmath_2^*}{\longrightarrow} \underbrace{\bbZ\langle\eta_2\rangle \oplus \bbZ_2\langle\tau_2\rangle}_{\subset H^2(\wZ)=H_3(\wZ, \partial \wZ)} \stackrel{\imath_2^*}{\longrightarrow} \underbrace{\bbZ_4 \langle x_2 \rangle}_{\subset H^2(\partial \wZ) = H_2(\partial \wZ)} \stackrel{d_2}{\longrightarrow} 0 \, .
\end{align}
The other part is
\begin{align}\label{eq:twisted_geom_seq-for-H3H4}
    0 \rightarrow \underbrace{\bbZ\langle\omega_3\rangle}_{\subset H^3(\wZ, \partial \wZ) = H_2(\wZ)} \stackrel{\jmath_3^*}{\longrightarrow} \underbrace{\bbZ\langle\eta_3\rangle}_{\subset H^3(\wZ)=H_2(\wZ, \partial \wZ)} \stackrel{\imath_3^*}{\longrightarrow} \underbrace{\bbZ_4 \langle x_3 \rangle }_{\subset H^3(\partial \wZ) = H_1(\partial \wZ)} \stackrel{d_3}{\longrightarrow} \underbrace{\bbZ_2\langle\sigma_4\rangle}_{\subset H^4(\wZ, \partial \wZ) = H_1(\wZ)} \rightarrow 0 \, ,
\end{align}
where we have denoted the generators next to the groups, in the same notation as in Table~\ref{tab:maps-twisted-geometry}.

This example differs from the toy model of Section~\ref{sec:example} in that the simplifying properties, $\text{im}(\jmath^*)\subset H^2_{\rm free}(X)$ and $H^2(\partial X)\cong\mathbb{D}\oplus\mathbb{T}$, no longer holds. 
Both fail due to the map
\begin{equation}
    \jmath_2^*(\omega_2)=2\eta_2+\tau_2\,,
\end{equation}
which has a torsional component. 
By exactness of \eqref{eq:twisted_geom_seq-for-H2}, the cokernel of $\jmath_2^*$ is,
\begin{equation}\label{eq:frozen-coker}
    \text{coker}(\jmath_2^*) = \frac{\Z\langle\eta_2\rangle\oplus\Z_2\langle\tau_2\rangle}{\langle 2\eta_2+\tau_2\rangle}\cong\Z_4\,.
\end{equation}
We can still isolate the electric defects of the gauge sector as in \eqref{eq: effective_defect_group},
\begin{equation}\label{eq:frozen-defect-group}
    \mathbb{D}^{(1)} = \frac{\Z \langle \eta_2 \rangle}{\text{im}(q\circ\jmath_2^*)} = \frac{\Z\langle\eta_2\rangle}{\langle 2\eta_2\rangle}\cong\Z_2\, ,
\end{equation}
but this is no longer canonically a subgroup of $H^2(\partial \wZ)$.
On the magnetic side, we see from Table \ref{tab:maps-twisted-geometry} that $\text{im}(\jmath_3^*) \subset \Z \langle \eta_3 \rangle = H^3_\text{free}(\wZ)$, so that
\begin{align}
    \mathbb{D}^{(3)} = \frac{\Z \langle \eta_3 \rangle}{\text{im}(\jmath_3^*)} = \frac{\Z \langle \eta_3 \rangle}{\langle 2 \eta_3\rangle} \cong \Z_2 \langle 2x_3\rangle \subset H^3(\partial \widehat{X})
\end{align}
is a subgroup of the boundary cohomology $H^3(\partial \wZ)$.

\subsubsection*{Center symmetries from cohomology}

Unlike in M-theory compactifications to 7d, the electric/magnetic objects/fields in the present case come from reduction/wrapping on different (co-)homology groups. We can expand the fluxes on these bases,\footnote{As a reminder, we have neglected the free classes in \eqref{eq:passed-through-free-classes}.}
\begin{equation}
\begin{aligned}
    G_4 &\supset f_2 \, \omega_2 + \widetilde{E}_2 \, \eta_2 \, ,
    \\
    G_7 &\supset f_4 \, \omega_3 + \widetilde{E}_4 \, \eta_3 \, .
\end{aligned}    
\end{equation} 
By the UCT and Poincar\'e--Lefschetz duality, we have 
\begin{equation}
    {\rm Free}(H^n(\wZ)) \cong {\rm Hom}({\rm Free}(H_n(\wZ)), \bbZ) \cong {\rm Hom}({\rm Free}(H^{5-n}(\wZ, \partial \wZ)), \bbZ) \, .
\end{equation}
The free factors are dual to each other under this Hom-functor, while the discarded free factors pair among themselves. 
The retained generators pair as
\begin{align}
    \bbZ \langle \eta_3 \rangle  = {\rm Hom}(\bbZ\langle \omega_2 \rangle , \bbZ) \, , \quad \bbZ\langle \eta_2\rangle  = {\rm Hom}(\bbZ\langle \omega_3 \rangle , \bbZ) \, ,
\end{align}
i.e., we have the pairings 
\begin{equation}
    \int_{\wZ} \omega_2 \, \eta_3 = \int_{\wZ} \omega_3 \, \eta_2 = 1 \, ,
\end{equation}
or, equivalently,
\begin{equation}
    \int_{\wZ} \omega_2 \omega_3 = 2, \qquad  \int_{\wZ} \eta_2 \eta_3 = 1/2\, .
\end{equation}
These will give us the expected couplings in the 6d Lagrangian for the gauge fields (on the Coulomb branch) and the background fields for the $\bbZ_2$ center symmetries of the $\mathfrak{g} = \mathfrak{su}(2)$ gauge sector: 
\begin{align}
    S_{\rm 6d} \supset 2\pi\int_{M_6} \left( 2 f_2 \, f_4 + \widetilde{E}_2 \, f_4 + \widetilde{E}_4 \, f_2 + \frac12 \widetilde{E}_2 \, \widetilde{E}_4 \right) \, .
\end{align}
In addition to this, we will have a non-compactly supported 2-form $\tau_2 \in {\rm Tor}(H^2(\wZ))$, which leads to an additional $\bbZ_2$-background field $\widetilde{C}_2$ (from $G_4 = \widetilde{C}_2 \, \tau_2  + ...$), 
which does not couple to the gauge sector's (background) fields. Instead, UCT states that
\begin{equation}
    {\rm Tor}(H^2(\wZ)) \cong {\rm Hom}( {\rm Tor}(H_1(\wZ)), \mathbb{Q}/\bbZ) \cong {\rm Hom}( {\rm Tor}(H^4(\wZ, \partial \wZ)), \mathbb{Q}/\bbZ) \, ,
\end{equation}
so $\tau_2$ has non-trivial $\bbZ_2$-torsion pairing only with the generator $\sigma_4 \in \bbZ_2 = {\rm Tor}(H^4(\wZ, \partial \wZ))$. Expanding $dG_7 \supset dk_3 \, \sigma_4$ therefore gives a dynamical discrete 3-form gauge field $k_3$, as in Section~\ref{sec:KK-reduction-kinetic-term}. Reducing the kinetic term, the 6d action takes the form
\begin{align}\label{eq:6d_phys_bd}
    S_{\rm 6d} = 2\pi\int_{M_6} \left( 2 f_2 \, f_4 + \widetilde{E}_2 \, f_4 + \widetilde{E}_4 \, f_2 + \frac12 \widetilde{E}_2 \, \widetilde{E}_4 + \frac12 \widetilde{C}_2 \, d k_3 \right) \, ,
\end{align}
where lower-case fields are dynamical.

\subsubsection*{SymTFT and modified physical boundary}

We now return to the 7d SymTFT, which is entirely determined by the boundary cohomologies ${\rm Tor}(H^2(\partial \wZ)) = {\rm Hom}({\rm Tor}(H^3(\partial \wZ)), \mathbb{Q}/\bbZ) = \bbZ_4$. Therefore, we have
\begin{align}\label{eq:7d_symtft_Z4}
 S^{\rm 7d}_{\text{SymTFT}} = 2\pi \int_{W_7} \frac14 A_2 \, d A_4 \, .
\end{align}
In order to couple the SymTFT with the physical boundary, we rewrite this $\Z_4$ BF-theory as two $\bbZ_2$ BF-pairs with a mixed anomaly according to the sequence
\begin{equation}
    0 \rightarrow \bbZ_2\langle t_n\rangle \rightarrow \bbZ_4\langle x_n \rangle \rightarrow \bbZ_2\langle s_n \rangle \rightarrow 0\, .
\end{equation}
The generators are chosen such that the order-2 subgroup of $\Z_4$ is generated by $t_n := 2x_n$ and the quotient generator is the class\footnote{Here we have abused the notation and denoted the lift by the same letter.} $s_n :=x_n$. From the sequence \eqref{eq:twisted_geom_seq-for-H2} we have 
\begin{align}\label{eq:boundary-fields-1}
    \imath_2^*(\tau_2) = 2x_2 = t_2 \, , \qquad \imath^*_2(\eta_2) = x_2 = s_2 \, .
\end{align}
And in the sequence \eqref{eq:twisted_geom_seq-for-H3H4} we have, 
\begin{equation}\label{eq:boundary-fields-2}
    \imath^*_3 (\eta_3) = 2x_3 =t_3 \, , \qquad d_3(s_3) = \sigma_4 \, , \qquad d_3(t_3) =0 \, .
\end{equation}
We can ``decompose'' the $\bbZ_4$ linking pairing as
\begin{align}
    L_{\partial \wZ} (t_2 , s_3) = L_{\partial \wZ} (s_2 , t_3) = \frac12 \mod \bbZ \, ,
\end{align}
together with $L_{\partial \wZ} (s_2,s_3) = \frac{1}{4}$ mod $\Z$. Then, we can make the following ansatz:\footnote{We have split the $\Z_4$-valued field into two $\Z_2$ valued fields. $A_2$ becomes the pair $(E_2, C_2)$ and $A_4$ becomes the pair $(B_4, E_4)$. Explicitly, $A_2 \sim E_2 + 2 C_2$ and $A_4 \sim 2 E_4 + B_4$ mod $4$. Substituting into \eqref{eq:7d_symtft_Z4} gives \eqref{eq:7d_symtft_split}. See Appendix~\ref{app:bottom-up} for further details.}
\begin{equation}
\begin{aligned}
    G_4 &= E_2 \, s_2 + C_2 \, t_2, \\
    dG_7 &= dE_4 \, t_3 + dB_4 \, s_3,
\end{aligned}    
\end{equation}
where $E_2, B_4$ are valued in the $\Z_2$ quotient, and $C_2, E_4$ are valued in the $\Z_2$ subgroup. We have
\begin{align}\label{eq:7d_symtft_split}
    S^{\rm 7d}_{\text{SymTFT}} = 2\pi \int_{W_7} \left(  \frac12 E_2 \, d E_4 + \frac12 C_2 \, dB_4 + \frac14 E_2 \, d B_4 \right) \, .
\end{align}

To couple the six-dimensional theory to the seven-dimensional SymTFT, we make the following identifications between the boundary values of 7d SymTFT fields and the fields appearing in the 6d action, using equation \eqref{eq:boundary-fields-1} and \eqref{eq:boundary-fields-2}:
\begin{equation}
    E_2 \Big|_{M_6} = \widetilde{E}_2 , \qquad  E_4 \Big\vert_{M_6} = \widetilde{E}_4 , \qquad C_2\Big\vert_{M_6} = \widetilde{C}_2 .
\end{equation}
Now we treat the combined action $(S^{\text{7d}}_{\text{SymTFT}} + S_{\text{6d}})$ as a coupled bulk-boundary system\footnote{As in Section~\ref{sec:coupling-symtft}, the local counterterm $\tfrac12 \widetilde{E}_2\widetilde{E}_4$ appearing in \eqref{eq:6d_phys_bd} is absorbed into the boundary transgression of the bulk action. }. The boundary equations of motion are, 
\begin{equation}\label{eq: modified_phys_boundary_frozen}
\begin{aligned}
    E_4~ &\text{variation} \implies  E_2 \Big|_{M_6} = - 2f_2, \\
    E_2~&\text{variation} \implies E_4\Big|_{M_6} + \half B_4\Big|_{M_6} = 2f_4, \\
    C_2~&\text{variation} \implies B_4 \Big|_{M_6} = dk_3.
\end{aligned}
\end{equation}
We can see that the bulk field $B_4$ sees a Dirichlet boundary condition on the physical boundary. In contrast, there is no similar equation for $C_2$, implying that the corresponding bulk operators cannot terminate on the physical boundary. This is the desired modified physical boundary:
while the $E_{2,4}$ surfaces can end, giving rise to Wilson-/'t Hooft-operators of the gauge sector, the topological sector only has three-dimensional charged operators from $B_4$.
Unlike the 7d toy models discussed in Section \ref{sec:top-down}, the two sectors are coupled through the last term in the SymTFT action in \eqref{eq:7d_symtft_split}.

\subsection{Bordism and the SymTFT interface}\label{sec:frozen-bordism}

We now show how the modified physical boundary can be separated into a topological interface and an ordinary physical boundary. Following the general prescription of Section~\ref{sec:bordism_general}, we cut $X$ along the boundary of a tubular neighborhood $X_L$ of the $\mathfrak{su}(2)$ locus and its complement $B$, so that $X = X_L \cup B$\,\footnote{\label{fn:XADrop}In Section~\ref{sec:bordism_general} we decomposed $X$ into $X_L,B$ and $X_A$. Note that $X_A$ is simply a cylinder and $X_A\cup B$ is homotopic to $B$, so to streamline the presentation in this section we have dropped the extra factor of $X_A$. }.  The local model $X_L$ supplies the physical boundary of the gauge sector alone. The bordism $B$ compactifies to an interface between the SymTFT associated to the local model and the $\Z_4$ SymTFT of $\partial X$.

For this computation, we contract the exceptional curve again and work with $X=(Y_{I_2}^{\text{ell}}\times S^1)/\Z_2$, where the $\mathfrak{su}(2)$ gauge symmetry is manifest. Consider $X$ as an orbi-circle bundle via projection onto the first factor
\be 
X \rightarrow Y_{I_2}^{\text{ell}}/\Z_2 \,.
\ee 
Away from the exceptional elliptic fiber, the orbi-circle bundle is an ordinary circle bundle with $S^1$ fibers. The exceptional fiber is the affine $\mathbb{P}^1$. On it, there are three marked points, locally modeled on
\be
(\mathbb{C}\times S^1)/\Z_2 \rightarrow \mathbb{C}/\Z_2\,.
\ee 
Over these points the orbi-circle bundle fiber is $S^1/\Z_2$, i.e., a circle of half-radius. Two of these circles have a smooth local model in the total five-dimensional manifold $X$. The other circle has a local model of
\be 
(\mathbb{C}^2/\Z_2 \times S^1)/\Z_2 \rightarrow S^1/\Z_2\,,
\ee 
which is a twisted circle worth of $A_1$ singularities, on which the $\mathfrak{su}(2)$ degrees of freedom are localized. We take its tubular neighborhood as the local model and define the bordism as its complement. 
We define
\be \label{eq:Defs}
X := X_L \cup B, \qquad X_L := T(S^1/\Z_2),  \qquad B := X \setminus T(S^1/\Z_2)\, ,
\ee 
with boundaries, 
\begin{equation}
    \partial_L B  = (S^3/\Z_2 \times S^1)/\Z_2 \, , \qquad \partial_A B = \partial X \, .
\end{equation}

The relevant homology groups are (see Appendix~\ref{app: homology_tracking} for the derivation) 
\begin{equation}\label{eq:rel_hom_froz}
\begin{aligned}
    H_n(X) &= \{\Z, \Z\oplus \Z_2, \Z, \Z, 0, 0\} \, , \qquad H_n(\partial_A B) = \{\Z,\Z^2 \oplus \Z_4, \Z^2 \oplus \Z_4, \Z^2, \Z \} \, ,\\
    H_n(\partial_L B) &= \{\Z, \Z \oplus \Z_2, \Z_2, \Z, \Z \} \, , \qquad H_n(X_L)= \{\Z, \Z, 0, 0, 0\} \, .
\end{aligned}    
\end{equation}
They fit into the Mayer--Vietoris sequence for $X = X_L \cup B$,
\begin{equation}\label{eq:computeHY_froz}
    \begin{array}{c||cccccc}
&   H_n(\partial_L B) &  & H_n(B)\oplus H_n(X_L) &  &   H_n(X) &      \\[0.4em] \hline \hline \\[-0.9em]
            n=4 ~&  \Z & \rightarrow &H_4(B ) \oplus 0 & \rightarrow & 0 & \rightarrow \\
            n=3 ~&  \Z & \rightarrow &H_3(B ) \oplus 0 & \rightarrow & \Z & \rightarrow  \\
         n=2~ &  \Z_2  &\rightarrow & H_2(B)\oplus 0  & \rightarrow& \Z &\rightarrow \\
       n=1~ &    \Z\oplus \Z_2  &\rightarrow &H_1(B)\oplus \Z &\rightarrow &\Z\oplus \Z_2&\rightarrow  \\
        n=0 ~&   \mathbb{Z}  &\rightarrow & H_0(B ) \oplus \Z &\rightarrow & \Z  &   
    \end{array}
\end{equation}
from which we can identify
\be 
H_1(B) \cong \Z \oplus \Z_4 \,, \qquad H_2(B) \cong \Z \oplus  \Z_2 \, .
\ee 
Overall we have, 
\begin{equation}
    H_n(B) = \{ \Z, \Z \oplus \Z_4, \Z \oplus \Z_2, \Z^2, \Z \} \, .
\end{equation}
Applying Poincar\'e--Lefschetz duality, $H^n(B, \partial B) \cong H_{5-n} (B)$, yields
\begin{equation}\label{eq: relative-cohomology}
    H^n(B, \partial B) = \{0, \Z, \Z^2, \Z \oplus \Z_2, \Z \oplus \Z_4, \Z \} \, .
\end{equation}

The boundary of $B$ has two disconnected components. Therefore, 
\begin{equation}
    H_n(\partial B) = H_n(\partial_A B ) \oplus H_n(\partial_L B) = \{\Z^2, \Z^3\oplus \Z_2 \oplus \Z_4, \Z^2 \oplus \Z_2 \oplus \Z_4, \Z^3, \Z^2\} \, .
\end{equation}
By applying UCT, $H^n(\partial B) \cong \text{Free}\, H_n(\partial B) \oplus \text{Tor}(H_{n-1}(\partial B))$, we obtain
\begin{equation}
    H^n(\partial B) = \{ \Z^2, \Z^3, \Z^2 \oplus \Z_2 \oplus \Z_4, \Z^3 \oplus \Z_2 \oplus \Z_4, \Z^2  \} \, .
\end{equation}
Combining all of this into the long exact sequence of cohomology, 
\begin{equation}\label{eq:computeHY_froz_2}
    \begin{array}{c||cccccc}
&   H^n(B,\partial B) &  & H^n(B) &  &   H^n(\partial B) &      \\[0.4em] \hline \hline \\[-0.9em]
n=0 ~&  0 & \xrightarrow{\iota^0} &H^0(B) \ & \rightarrow & \Z\oplus \Z & \rightarrow \\
            n=1 ~&  \Z & \xrightarrow{\iota^1} & H^1(B) & \rightarrow & \Z^2\oplus \Z & \rightarrow \\
            n=2 ~&  \Z^2 & \xrightarrow{\iota^2} &H^2(B) & \rightarrow & \Z^2\oplus \Z_4 \oplus \Z_2 & \rightarrow  \\
         n=3~ &   \Z \oplus \Z_2 & \xrightarrow{\iota^3} & H^3(B)   & \rightarrow& \Z^3 \oplus\Z_4 \oplus \Z_2 &\rightarrow \\
       n=4~ & \Z \oplus  \Z_4 &\xrightarrow{\iota^4} & H^4(B) &\rightarrow & \Z\oplus \Z &\rightarrow  \\
        n=5 ~&   \Z &\xrightarrow{\iota^5} & H^5(B)  &\rightarrow & 0  &   
    \end{array}
\end{equation}
we deduce
\begin{equation}\label{eq: bordism cohomology}
    H^n(B) = \{\Z,\Z, \Z \oplus \Z_4, \Z^2 \oplus \Z_2, \Z, 0 \} \, ,
\end{equation}
and the maps, 
\begin{equation}\label{eq: iota_maps}
\begin{aligned}
&\iota^0:~H^0(B,\partial B) \cong 0\rightarrow \Z\cong H^0(B),\\
&\iota^1:~H^1(B,\partial B) \cong \Z \rightarrow \Z \cong H^1(B),\\
&\iota^2:~H^2(B,\partial B)\cong\Z^2 \rightarrow \Z \oplus\Z_4 \cong H^2(B),\\
&\iota^3:~H^3(B,\partial B)\cong \Z \oplus \Z_2 \rightarrow \Z^2 \oplus \Z_2 \cong H^3(B),\\
&\iota^4:~H^4(B,\partial B)\cong\Z \oplus \Z_4 \rightarrow \Z \cong H^4(B), \\
&\iota^5:~H^5(B,\partial B)\cong\Z\rightarrow 0 \cong H^5(B).
\end{aligned}
\end{equation}
The maps above satisfy
\begin{equation}
  \iota^n=0 \qquad \text{for all}\,\, n  ,
\end{equation}
which is verified by tracking geometrically how the generators in the dual homology are mapped onto each other after excising $B$, see Appendix~\ref{app: homology_tracking} for details. Thus, in the classification of Section~\ref{sec:bordism_general}, the bordism implements boundary conditions for the fields of the SymTFT \eqref{eq:7d_symtft_split}.
In particular, it is a topological interface within this SymTFT.

\subsubsection*{SymTFTs and the interface}

\begin{figure}[H]
\centering
\begin{tikzpicture}[font=\small]
\fill[blue!10] (-0.5,0) rectangle (6.1,3.0);
\fill[orange!15] (6.1,0) rectangle (12.1,3.0);
\node[align=center] at (2.8,2.35) {$S_{\dA}$\\ (from $\dA=\partial X$)};
\node[align=center] at (2.8,0.95) {$\Z_4$:\ \ $A_2,\ A_4$};
\node[align=center] at (9.1,2.35) {$S_{\dL}$\\ (from $\dL$)};
\node[align=center] at (9.1,0.95) {$\Z_2$:\ \ $A'_2,\ A'_4$};
\draw[very thick,dashed] (6.1,-0.35) -- (6.1,3.35);
\draw[very thick] (-0.5,-0.35) -- (-0.5,3.35);
\draw[very thick] (12.1,-0.35) -- (12.1,3.35);
\node[align=center,font=\footnotesize] at (-0.5,-0.9) {topological b.c.\\ at $\partial X$};
\node[align=center,font=\footnotesize] at (6.1,-1.0) {interface $\mathcal I$ from $B$ ($\iota^n=0$)\\ $B^\mathcal{I}_2,B^\mathcal{I}_3 \in\Z_4$,\ \ $B^\mathcal{I}_1,B^\mathcal{I}_4 \in\Z_2$ };
\node[align=center,font=\footnotesize] at (12.1,-0.9) {physical b.c.\\ from $X_L$};
\end{tikzpicture}
\caption{The two SymTFTs --- the $\Z_4$-BF theory (blue) of $\dA=\partial X$ and the $\Z_2$-BF theory of $\dL$ (orange) --- connected by the interface  $\mathcal I$.}
\label{fig: interface-sandwich}
\end{figure}
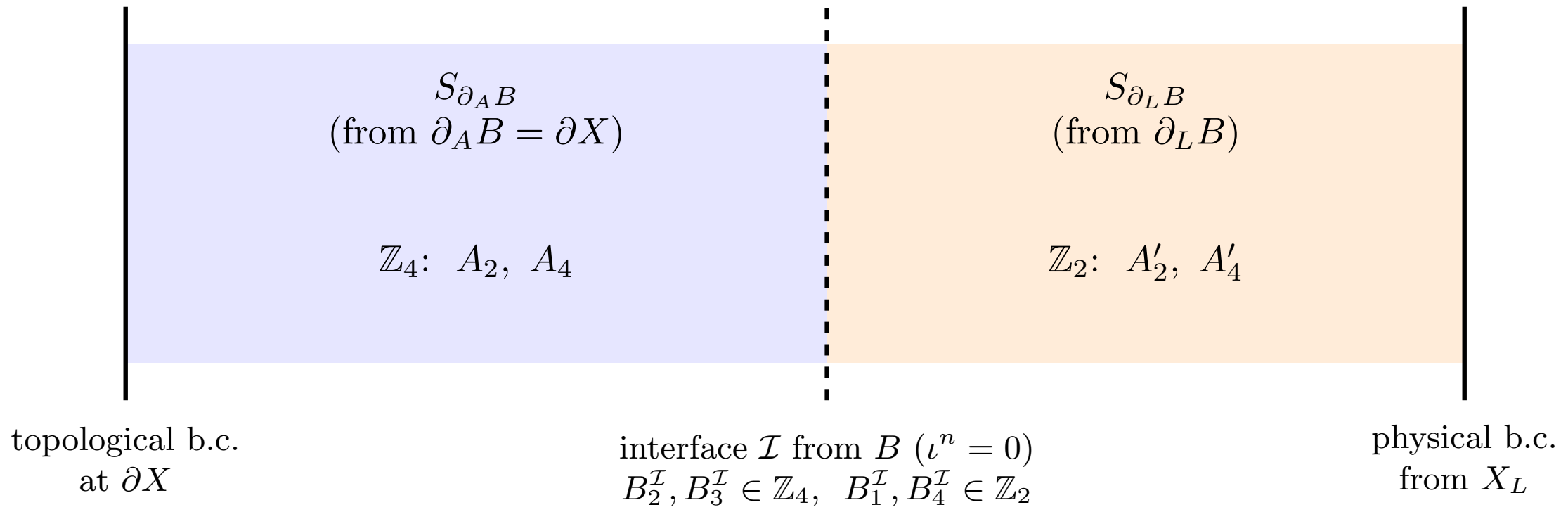

Next, we identify the SymTFTs on the two sides of the interface, the six-dimensional interface action, and the gluing conditions across it. All three follow from reducing the kinetic term of the eleven dimensional supergravity on $\partial_A B$, $\partial_L B$ and $B$. From \eqref{eq:rel_hom_froz} and \eqref{eq: bordism cohomology}, we collect the relevant cohomology groups and introduce their generators,
\begin{equation}\label{eq: generators}
\begin{aligned}
\Tor H^2(B) &= \Z_4 \langle w_2\rangle, &\qquad \Tor H^3(B)&= \Z_2\langle\theta_3\rangle, \\
\Tor H^3(B,\partial B) &=\Z_2\langle\nu_3\rangle, &\qquad \Tor H^4(B,\partial B)&=\Z_4\langle\mu_4\rangle,\\
\Tor H^2(\dA)&=\Z_4\langle x_2\rangle, &\qquad \Tor H^3(\dA)&=\Z_4\langle x_3\rangle,\\
\Tor H^2(\dL)&=\Z_2\langle x'_2\rangle, &\qquad \Tor H^3(\dL)&=\Z_2\langle x'_3\rangle .
\end{aligned}
\end{equation}
We expand the differential cohomology lifts of $G_4$ and $G_7$ as, 
\begin{equation}\label{eq: bulk expansion}
\begin{aligned}
\check{G}_4 &= \check{B}^\mathcal{I}_2 \star \check{w}_2 + \check{B}^\mathcal{I}_1 \star   \check{\nu}_3, \\ 
\check{G}_8 &= \check{F}^\mathcal{I}_5 \star \check{\theta}_3 + \check{F}^\mathcal{I}_4\star\check{\mu}_4 \, ,
\end{aligned}
\end{equation}
where $\check G_8$ refines $dG_7$, therefore, $\check{F}^\mathcal{I}_5 = \delta B^\mathcal{I}_4$ and $\check{F}^\mathcal{I}_4 = \delta B^\mathcal{I}_3$. Reducing $\check G_4\star\check G_8$ therefore gives
\begin{equation}\label{eq:interface_action}
S_{\mathcal{I}} = 2\pi \int_{\mathcal{I}} \left( \frac14 B^\mathcal{I}_2 \cup\delta B^\mathcal{I}_3 + \frac12  B^\mathcal{I}_1  \cup \delta B^\mathcal{I}_4 \right) ,
\end{equation}
where we have fixed the coefficients using the perfect pairing $\text{Tor}\, H^k(B)\times \text{Tor}\,H^{6-k} (B,\partial B)\longrightarrow\mathbb Q/\mathbb Z$. To be precise, we have used $L_B(w_2,\mu_4)=\frac14, L_B(\theta_3,\nu_3)=\frac12$. This is the interface theory. The fields $ B^\mathcal{I}_2,\,B^\mathcal{I}_3$ are valued in $\mathbb Z_4$, whereas $B^\mathcal{I}_1,\, B^\mathcal{I}_4$ are valued in $\mathbb Z_2$. Similarly, expanding $G_{4/7}$ fluxes, with unprimed fields on $\dA$ and primed fields on $\dL$, yields
\begin{equation}\label{eq: boundary_expansions}
\begin{aligned}
\partial_A B: & \qquad \check G_4 = \check{A}_2\star \check x_2, & \qquad \check G_8 & = \check F_5 \star \check{x}_3,\\
\partial_L B: & \qquad \check G_4 = \check A'_2 \star \check x'_2, &\qquad \check G_8 & = \check F'_5 \star \check x'_3,
\end{aligned}
\end{equation}
and using the linking pairings on the two boundary components, $L_{\dA}(x_2,x_3)=\frac14$ and $L_{\dL}(x'_2,x'_3)=\frac12$, we get the SymTFTs on each side of the interface: 
\begin{equation} \label{eq: two_symtfts}
S_{\dA} = 2 \pi \int_{W_7} \frac14 \, A_2 \cup \delta A_4 , \qquad
S_{\dL} = 2 \pi \int_{W_7} \frac12 \, A'_2 \cup \delta A'_4 .
\end{equation}
Here, $A_2,\,A_4$ are valued in $\Z_4$ and $A'_2,\,A'_4$ are valued in $\Z_2$. The SymTFT $S_{\dA}$ is the $\Z_4$ BF theory \eqref{eq:7d_symtft_Z4} of the full frozen theory (in its dual geometric frame), obtained from the asymptotic geometry. The SymTFT associated with the boundary of the local model is $S_{\dL}$, which is a $\Z_2$ BF theory, see Figure~\ref{fig: interface-sandwich}.

\subsubsection*{The gluing conditions}

Finally, we derive the gluing conditions. Since $\iota^n=0$ for all $n$, we have the following short exact sequence,
\begin{equation}
0\longrightarrow H^n(B)\xrightarrow{\ r\ }H^n(\dA)\oplus H^n(\dL)\xrightarrow{\ d\ }H^{n+1}(B,\partial B)\longrightarrow 0 \, .
\end{equation}
We have already identified that the interface is a pure boundary condition interface. It carries no degrees of freedom and it can be specified by gluing conditions of the bulk fields on the two sides. To identify the gluing conditions,  we need the action of $r$ and $d$ on the generators \eqref{eq: generators}. 

By the UCT, a torsion class in $H^k$ is a homomorphism on the torsion classes in $H_{k-1}$,
\begin{equation}\label{eq: homology-generators}
\begin{aligned}
\Tor H_1(\partial_A B) &= \Z_4\langle\beta_A\rangle \, , &\qquad \Tor H_2(\partial_A B) &= \Z_4\langle\tau(\alpha)_A\rangle \, , \\
\Tor H_1(\partial_L B) &= \Z_2\langle\alpha_L\rangle \, , &\qquad \Tor H_2(\partial_L B) &= \Z_2\langle\tau(\alpha)_L\rangle \, , \\
\Tor H_1(B) &= \Z_4\langle\beta\rangle \, , &\qquad \Tor H_2(B) &= \Z_2\langle\tau(\alpha)\rangle \, ,
\end{aligned}
\end{equation}
with $2\beta_A=\alpha_A$ and $2\tau(\alpha)_A=T^2_{\alpha,\gamma}$ in $\partial_A B$ (See Table~\ref{tab:bordism-relative-homology}), where the indices $A$ and $L$ specify whether the cycle is on $\partial_A B$ or $\partial_L B$. Therefore,
\begin{equation}
\begin{aligned}
x_2(\beta_A) &= \tfrac14, \quad w_2(\beta)=\tfrac14, \quad x'_2 (\alpha_L) = \tfrac12 \\
x_3(\tau(\alpha)_A) &= \tfrac14,  \quad \theta_3(\tau(\alpha))=\tfrac12, \quad x'_3(\tau(\alpha)_L) = \tfrac12 \, ,
\end{aligned}
\end{equation}
from which we can pin down the gluing conditions across the interface.  Writing elements of $H^n(\dA)\oplus H^n(\dL)$ as pairs $(y,y')$,
\begin{equation}\label{eq: r-map}
r(w_2)=(x_2,\,x'_2),\qquad r(\theta_3)=(2x_3,\,x'_3) \, .
\end{equation}
By exactness, $d$ vanishes on the image of $r$ and is onto. Table~\ref{tab: interface-maps} collects the action of the maps on the generators.

\begin{table}[H]
\centering
\renewcommand{\arraystretch}{1.4}
\begin{tabular}{|l|l|}
\hline
\textbf{Map} & \textbf{Action on generators} \\
\hline
$r:H^2(B)\to H^2(\partial_A B) \oplus H^2(\partial_L B)$
& $r(w_2)=(x_2,x'_2)$ \\
\hline
$d:H^2(\partial_A B)\oplus H^2(\partial_L B) \to H^3(B,\partial B)$
& $d(x_2,0)=\nu_3,\quad d(0,x'_2)=\nu_3$ \\
\hline
$r:H^3(B)\to H^3(\partial_A B) \oplus H^3(\partial_L B)$
& $r(\theta_3)=(2x_3,x'_3)$ \\
\hline
$d:H^3(\partial_A B) \oplus H^3(\partial_L B) \to H^4(B,\partial B)$
& $d(x_3,0)=\mu_4,\quad d(0,x'_3)=2\mu_4$ \\
\hline
\end{tabular}
\caption{Action of the maps $r$ and $d$.}
\label{tab: interface-maps}
\end{table}

Restricting the bulk expansion \eqref{eq: bulk expansion} to the boundary via $r$, using \eqref{eq: r-map}, and comparing with the boundary expansions \eqref{eq: boundary_expansions} gives the following gluing conditions, 
\begin{equation}\label{eq: gluing-1}
\begin{aligned}
 A_2\big|_{\mathcal{I}} &= B^{\mathcal{I}}_2  ,\qquad A_4\big|_{\mathcal{I}} = 2B^{\mathcal{I}}_4 \, \text{mod}\,\, 4 ,\\
 A'_2\big|_{\mathcal{I}} &=B^{\mathcal{I}}_2\,\, \text{mod}\,\, 2  , \qquad A'_4\big|_{\mathcal{I}}= B^{\mathcal{I}}_4  ,
\end{aligned}
\end{equation}
where $|_{\mathcal{I}}$ is the restriction to the interface.  
Next we are going to consider the map $d$, which takes into account the coefficient fields of the relative cohomology classes $\nu_3$ and $\mu_4$, $B^{\mathcal{I}}_1$, and $B^{\mathcal{I}}_3$ respectively. As in equation \eqref{eq: d-map}, the image of a boundary class $\alpha$ under the map $d$ can be represented as  $\chi'(r)\, dr \wedge \alpha$. From Table~\ref{tab: interface-maps}, we have
\begin{equation}\label{eq: relative-classes-collar}
\begin{aligned}
\text{near } \partial_A B: &\qquad \nu_3 = d(x_2,0) \sim dr\wedge x_2 , \qquad \mu_4 = d(x_3,0) \sim dr \wedge x_3 , \\
\text{near } \partial_L B: &\qquad \nu_3 = d(0,x'_2) \sim dr\wedge x'_2 , \qquad 2\mu_4 = d(0,x'_3) \sim dr \wedge x'_3 .
\end{aligned}
\end{equation}
We split the SymTFT fields into the part tangent to $\mathcal{I}$ and the part with a leg along $dr$,
\begin{equation}\label{eq: normal-split}
A_2 = A_2^{\parallel} + dr \wedge \iota_{\partial_r} A_2 , \qquad A_4 = A_4^{\parallel} + dr \wedge \iota_{\partial_r} A_4 ,
\end{equation}
and likewise for $A'_2$, $A'_4$. 
The restriction $|_{\mathcal{I}}$ sets $r=0$, so \eqref{eq: gluing-1} constrains the tangential parts. From \eqref{eq: bulk expansion} and \eqref{eq: relative-classes-collar}, the part of $G_4$ with a leg along $dr$ is $B^{\mathcal{I}}_1 \wedge dr \wedge x_2$. From \eqref{eq: boundary_expansions} and \eqref{eq: normal-split}, it is $dr \wedge \iota_{\partial_r} A_2 \wedge x_2$. Comparing the two, and repeating the same for $G_7$, gives the normal gluing conditions,
\begin{equation}\label{eq: gluing-2}
\begin{aligned}
B^{\mathcal{I}}_1 &= \iota_{\partial_r} A_2 \big|_{\mathcal{I}} \,\, \text{mod}\,\, 2 , \qquad B^{\mathcal{I}}_3 = \iota_{\partial_r} A_4 \big|_{\mathcal{I}} , \\
B^{\mathcal{I}}_1 &= \iota_{\partial_r} A'_2 \big|_{\mathcal{I}} , \qquad\qquad\quad\ \, B^{\mathcal{I}}_3 = 2\, \iota_{\partial_r} A'_4 \big|_{\mathcal{I}} \,\, \text{mod}\,\, 4 .
\end{aligned}
\end{equation}

In summary, equations \eqref{eq: gluing-1} and \eqref{eq: gluing-2} specify the gluing conditions across $\mathcal I$. They identify the tangential components of the bulk fields with $B^{\mathcal I}_2$ and $B^{\mathcal I}_4$, and the normal components with $B^{\mathcal I}_1$ and $B^{\mathcal I}_3$. These conditions completely determine all four interface fields, leaving no independent degrees of freedom on $\mathcal I$, consistent with $\iota^n=0$. Thus, the bordism gives rise to a pure boundary-condition interface between the $\Z_4$ SymTFT associated with $\partial X$ and the $\Z_2$ SymTFT of the local $\mathfrak{su}(2)$ gauge sector.

The gluing conditions also determine which surface operators can pass through the interface. From \eqref{eq: gluing-1}, we have $A'_2=A_2\bmod 2$ on $\mathcal{I}$. Consequently, among the $A_2$-surface operators of $S_{\partial_A B}$ (see Figure~\ref{fig: interface-sandwich}), only even multiples can pass through $\mathcal I$ as the nontrivial $A'_2$ surface and terminate on Wilson lines of the local model. Similarly, $A_4=2A'_4\bmod 4$ implies that only the odd multiples can pass through and end on a 't~Hooft defect of the local model, while the even multiples of $A_4$ operators terminate on $\mathcal I$ itself. Fusing $\mathcal I$ with the physical boundary supplied by $X_L$ gives back the modified physical boundary of Section~\ref{sec: frozen_physical_boundary}. As we found there, the $E_2$-surfaces terminate on Wilson lines, while all four magnetic surfaces can terminate on the physical boundary, since both $E_4$ and $B_4$ obey Dirichlet boundary conditions.

This construction makes explicit how the bordism separates the local $\mathfrak{su}(2)$ gauge sector from the additional degrees of freedom of $\mathcal{T}^{\frozen}$. The SymTFT $S_{\partial_L B}$ on the local side of the interface therefore captures precisely the electric and magnetic symmetries of the $\mathfrak{su}(2)$ gauge sector, both of which are $\Z_2$.

\subsection{Global forms of the frozen theory}

The bordism prescription isolated the local gauge sector, together with its ``ordinary'' physical boundary realizing all Wilson- and 't Hooft-operators, from the topological sector. 
We will now first show that fusing the interface ${\cal I}$ with this ordinary physical boundary recovers the modified physical boundary action from Section~\ref{sec: frozen_physical_boundary}.
We will then proceed with discussing the possible polarizations and the corresponding global forms of the 6d theory ${\cal T}^\frozen$.

We return to continuous notation for the fields. The gluing conditions \eqref{eq: gluing-1} and \eqref{eq: gluing-2} then read
\begin{equation}\label{eq: gluing-U1}
\begin{aligned}
A_2\big|_{\mathcal{I}} &= B^{\mathcal{I}}_2 ,\qquad\ \, A_4\big|_{\mathcal{I}}=B^{\mathcal{I}}_4 , \qquad\quad\ \ B^{\mathcal{I}}_1 = 2\, \iota_{\partial_r} A_2 \big|_{\mathcal{I}} , \qquad B^{\mathcal{I}}_3 = \iota_{\partial_r} A_4 \big|_{\mathcal{I}} , \\
A'_2\big|_{\mathcal{I}} &= 2 B^{\mathcal{I}}_2, \qquad A'_4\big|_{\mathcal{I}} = B^{\mathcal{I}}_4 , \qquad\quad\ \ B^{\mathcal{I}}_1 = \iota_{\partial_r} A'_2 \big|_{\mathcal{I}} , \qquad\ \, B^{\mathcal{I}}_3 = \iota_{\partial_r} A'_4 \big|_{\mathcal{I}} .
\end{aligned}
\end{equation}
Using the splittings of the $\Z_4$-valued fields introduced in Section~\ref{sec: frozen_physical_boundary},
\begin{equation}
    A_2 = E_2 + 2C_2\, , \qquad A_4 = 2E_4 + B_4 ,
\end{equation}
we eliminate the interface fields from the gluing conditions to obtain
\begin{equation}
A'_2\big|_{\mathcal I} = E_2\big|_{\mathcal I} \, , \qquad A'_4\big|_{\mathcal I} = E_4\big|_{\mathcal I} \, , \qquad B_4\big|_{\mathcal I} = 0  ,
\end{equation}
while $C_2$ remains unconstrained on $\mathcal I$. The first two conditions identify $E_2$ and $E_4$ with the $\Z_2$ backgrounds of the  gauge sector.  After fusion, they give rise to the gauge sector (the first four terms) of \eqref{eq:6d_phys_bd}. The third condition imposes a Dirichlet boundary condition on $B_4$. Therefore, the gauge parameters of $B_4$ that do not vanish on $\mathcal I$ become physical degrees of freedom there \cite{Hsieh:2020jpj, Hosseini:2025oka, GarciaEtxebarria:2024fuk, Banerjee:2026wop, Robbins:2026kmh}. In other words, the pure gauge configurations of the bulk are no longer trivial at $\mathcal I$. This gives rise to a $\Z_2$-valued 3-form $k_3$ on $\mathcal I$, with $B_4|_{\mathcal I} = dk_3$. The bulk term  $\frac12 C_2 \, dB_4$ in \eqref{eq:7d_symtft_split} gives the boundary term $\frac12 {C}_2 \, dk_3$. Upon fusion, this produces the additional topological term of \eqref{eq:6d_phys_bd}.

Together, these contributions give the physical boundary action for $\mathcal{T}^{\frozen}$, the six-dimensional relative theory engineered by M-theory on $\wZ$,
\begin{align}
    S_{\rm 6d} = 2\pi\int_{M_6} \left( 2 f_2 \, f_4 + \widetilde{E}_2 \, f_4 + \widetilde{E}_4 \, f_2 + \frac12 \widetilde{E}_2 \, \widetilde{E}_4 + \frac12 \widetilde{C}_2 \, d k_3 \right)  .
\end{align}
The first four terms are the Cartan fluxes of the $\mathfrak{su}(2)$ gauge sector, coupled to the backgrounds $\widetilde{E}_2$, $\widetilde{E}_4$ of its $\Z_2$ center symmetries. The last term is a topological sector descending from the torsional data. Unlike the toy model of Section~\ref{sec:example},  these two sectors are not independent. Although their coupling is not manifest in the six-dimensional action, it is encoded in the SymTFT, as we will see below.

The SymTFT associated to this frozen gauge theory is
\begin{align}
    S^{\mathcal{T}^\frozen}_{\text{SymTFT}} = 2\pi \int_{W_7} \left(  \frac12 E_2 \, d E_4 + \frac12 C_2 \, dB_4 + \frac14 E_2 \, d B_4 \right) \, ,
\end{align}
where $E_2, E_4$ are associated with gauge sector and $C_2, B_4$ to the topological sector. In particular, the mixed term $\frac14 E_2\,dB_4$ captures the coupling between the two sectors. As discussed in Section~\ref{sec: frozen_physical_boundary}, the corresponding physical boundary is the modified boundary \eqref{eq: modified_phys_boundary_frozen}, on which the $E_2$-, $E_4$-, and $B_4$-surfaces can terminate, whereas the $C_2$-surfaces cannot.

With the physical boundary fixed, different choices of boundary conditions at the symmetry boundary give rise to different absolute theories. We summarize these choices in Table~\ref{tab:frozen-polarizations}, together with the boundary conditions at the physical boundary for comparison. We denote the absolute theories obtained in the electric, magnetic, and mixed polarizations by $\mathcal{T}^\frozen_e$, $\mathcal{T}^\frozen_m$, and $\mathcal{T}^\frozen_{\rm mix}$, respectively.

\begin{table}[ht]
\centering
\renewcommand{\arraystretch}{1.2}
\begin{tabular}{|c|c|c|c||c|}
\hline
Field & electric & magnetic & mixed & physical boundary \\
\hline
$E_2$ & D & N & D & D \\
$E_4$ & N & D & N & D \\
$C_2$ & D & N & N & N \\
$B_4$ & N & D & D & D \\
\hline
\end{tabular}
\caption{List of boundary conditions for the $\Z_4$ SymTFT and for the modified physical boundary associated to $\mathcal{T}^\frozen$.}
\label{tab:frozen-polarizations}
\end{table}

In the electric polarization, we impose Dirichlet boundary conditions on $E_2$ and $C_2$ at the symmetry boundary. The resulting absolute theory $\mathcal{T}^{\frozen}_e$ has a $\Z_4^{(1)}$ one-form symmetry generated by the $E_4$- and $B_4$-surface operators. The $\Z_2^{(1)}$ part generated by $E_4$-surfaces is the electric center symmetry of the $\mathfrak{su}(2)$ gauge sector and acts faithfully on its Wilson lines. The remaining $\Z_2^{(1)}$, associated with the $B_4$-operators, arises from the additional topological sector and acts trivially on $\mathcal{T}^{\frozen}_e$, since the dual $C_2$-surfaces have no endpoints on the physical boundary. Thus, the effective one-form symmetry is precisely the electric $\Z_2^{(1)}$ center symmetry of the gauge sector. The full symmetry fits into the extension
\begin{equation}
\underbrace{\Z_2^{(1)}}_{\text{symmetry of TFT sector, acts trivially}} \rightarrow \underbrace{\Z_4^{(1)}}_{\text{ full 1-form symmetry of } \mathcal{T}^{\frozen}} \rightarrow \underbrace{\Z_2^{(1)}}_{\text{electric center symmetry of gauge sector}}.
\end{equation}
The coupling between the two sectors becomes apparent when we examine the partition function of the absolute theory. We have
\begin{equation}
\begin{aligned}
Z_{\mathcal{T}^\frozen_e}[E_2,C_2] &=\int\mathcal{D}\lambda_3\int \mathcal{D}\Phi\, \exp\left(2\pi i\int_{M_6} \mathcal{L}_{\mathcal{T}^{\frozen}_e}[\lambda_3,\Phi]\right),\\ \mathcal{L}_{\mathcal{T}^\frozen_e}[\lambda_3,\Phi] &=\mathcal{L}_{\mathfrak{su}(2)}[E_2] +\frac12 C_2\,dk_3 +\lambda_3\left( \frac12 dC_2+\frac14 dE_2 \right).
\end{aligned}
\end{equation} 
Here $\Phi$ symbolizes the set of dynamical fields of the first two terms in the second line, including for example the $\mathfrak{su}(2)$ gauge theory and $k_3$. The field $\lambda_3$ is a Lagrange multiplier edge mode introduced at the symmetry boundary. Integrating over $\lambda_3$ imposes
\begin{equation}
\frac12 dC_2+\frac14 dE_2=0,
\end{equation}
which encodes the extension of the background fields into a $\Z_4$ two-form gauge field. The backgrounds associated with the gauge and topological sectors are therefore not independent. This is in contrast to the toy model of Section~\ref{sec:example}, where the corresponding sectors were orthogonal under the linking pairing and no such mixing arose. Here, the coupling is enforced by the geometry and is analogous to the type of coupling between a TQFT and a QFT discussed in \cite{Kapustin:2014gua}.

The magnetic polarization is obtained by gauging the full electric $\Z_4^{(1)}$ symmetry,
$\mathcal{T}^\frozen_m= \mathcal{T}^\frozen_e/\Z^{(1)}_4$. The resulting theory carries a dual $\Z_4^{(3)}$ three-form symmetry generated by the $E_2$- and $C_2$-surface operators. Like before, the $\Z^{(3)}_2$ part generated by the $E_2$-surfaces belongs to gauge sector. It is the magnetic center symmetry of the gauge sector. In addition, we have the $\Z^{(3)}_2$ generated by the additional topological sector which mixes non-trivially to form full $\Z_4^{(3)}$, giving 
\begin{equation}
\underbrace{\Z_2^{(3)}}_{\text{magnetic center symmetry of gauge sector}} \rightarrow \underbrace{\Z_4^{(3)}}_{\text{full 3-form symmetry of } \mathcal{T}^{\frozen}} \rightarrow \underbrace{\Z_2^{(3)}}_{\text{ symmetry of the TFT sector}}.
\end{equation}
Unlike the electric polarization, the additional symmetry arising from the topological sector is not trivially acting: the full $\Z_4^{(3)}$ symmetry couples to the gauge sector. This mixing can be seen directly from the partition function of the magnetic theory,
\begin{equation}\label{eq:L-magnetic}
    Z_{\mathcal{T}^\frozen_m}[E_4, B_4] =  \int {\cal D}[E_2,C_2] \, Z_{\mathcal{T}^\frozen_e}[E_2, C_2] \; \exp\Big( 2\pi i \int \Big( \frac12 E_2 \, E_4 + \frac14 E_2 \, B_4 + \frac12 C_2 \, B_4 \Big) \Big) \, .
\end{equation}
The terms in the exponent form the gauging kernel. We integrate over $(E_2,C_2)$ as a single flat $\Z_4$ two-form gauge field, modulo gauge transformations, with $(E_4,B_4)$ held fixed. In particular, the cross term $\frac14 E_2 B_4$ makes explicit the coupling between the gauge and topological sectors, showing why the additional magnetic symmetry cannot be treated independently of the gauge sector.

Therefore, from the perspective of the gauge sector, we see a $\Z^{(1)}_2$ symmetry in one polarization and a $\Z^{(3)}_4$ symmetry in its dual polarization. This is indeed puzzling at first sight. However, this apparent mismatch between the electric and magnetic symmetries of the frozen theory observed in \cite{Cvetic:2024mtt} is an artifact of the additional topological degrees of freedom forced upon us by the geometry itself. In fact, when we isolated the local model by cutting the geometry in the previous section, we saw that the $\mathfrak{su}(2)$ gauge sector itself carries a $\Z_2$ worth of electric and magnetic symmetry, as expected, with no mismatch whatsoever. 

Beyond these two, we can also gauge the trivially acting $\Z_2$ subgroup of the electric one-form symmetry. This gives the mixed polarization, $\mathcal{T}^\frozen_{\rm mix} = \mathcal{T}^\frozen_e/\Z^{(1)}_2$. The symmetry structure of this theory is
\begin{equation}
    \underbrace{\Z_2^{(1)}}_{\text{acts on the gauge sector}}
  \tilde{\times}
  \underbrace{\Z_2^{(3)}}_{\text{acts on the TFT sector}}\,.
\end{equation}
The first factor is generated by the $E_4$ operators and the second by the $C_2$ surfaces. Both act effectively, and there is a mixed anomaly between them, which we have denoted by a $\tilde{\times}$. This mixed anomaly is another manifestation of the coupling between the two sectors discussed at the beginning of this subsection. 
Consequently, the symmetries of the gauge sector and the topological sector do not factorize.

\section{Conclusion and outlook}

In this work we demonstrated how bulk topology in local string backgrounds can result in a topological field theory sector coupled to a localized QFT sector. We studied bordisms which connect the boundary of a tubular neighborhood centered on the extra-dimensional support of the localized QFT sector to the asymptotic boundary of the local string background. We found that in the SymTFT framework such bordisms map onto interfaces running parallel to the physical and topological boundary conditions. The interface theories, building on \cite{Baume:2023kkf,GarciaEtxebarria:2024jfv, Cvetic:2024dzu}, then simply followed by compactifying string or M-theory on the bordism. Our primary goal lay with isolating the part of the interface theory which interacted with the local QFT sector and extended its categorical symmetry structures, resulting in well-known field theory structures \cite{Bhardwaj:2023bbf,Cordova:2024jlk, Cordova:2024mqg}. We demonstrated our considerations using a 6d gauge theory example, engineered by a Calabi--Yau twofold in M-theory twisted over a circle.

Open questions remain. The first concerns equivalence relations induced by local limits. Consider a local limit from a local string background as above onto the tubular neighborhood centered on the extra-dimensional support of a localized QFT sector within. The bulk topology is pushed with reference to the QFT support to infinity, bulk modes decouple and we are left with a local relative QFT possibly coupled to a topological sector and, in principle, other decoupled topological sectors. Our construction explicitly identified one particular bordism representative which interpolates between the old and new local model asymptotics and computations proceeded with respect to this geometry. The natural question is then which properties of the bordism are retained by the localized QFT, and which are forgotten in the local limit. More precisely, different bordisms interpolating between the same asymptotic geometries may give rise to distinct interface theories, yet become indistinguishable upon fusion with the physical boundary supporting the localized QFT, up to completely decoupled topological sectors. This suggests an equivalence relation on bordisms induced by their coupling to the localized QFT, whose characterization would be interesting.

In the examples studied here, the relevant information is captured by the maps between ordinary, relative and boundary cohomology, together with their torsion linking pairings. These determine which defects can pass through the bordism, which can terminate on the associated interface, and which remain absent from the localized QFT. More generally, additional geometric and topological data may be required. What is the minimal bordism data which determines its coupling to a given localized QFT, and thereby characterize the equivalence relation induced by local limits?

Another open question is what types of topological extension of local QFT sectors can arise. This question is best asked with reference to a fixed higher-dimensional theory, e.g., M-theory, and restrictions on the bordism considered, e.g., purely geometric or of particular tangential structure or even special holonomy. We considered examples of purely geometric smooth bordisms in M-theory and found BF-sectors (and some continuous abelian degrees of freedom) which resulted in alterations of the defect group. Then, to what class of interface theories can this be extended by decorating the bordism with other stringy ingredients such as fluxes, branes and curvature singularities? Further, one may allow for additional boundary components of the bordism to realize changes of global form and discrete $\theta$-parameters \cite{Kapustin:2014gua,Baume:2023kkf}.

Moreover, we should be cautious when ``extrapolating'' the results for the 6d theory obtained from M-theory on the orbifold $(Y^\text{ell}_{I_2} \times S^1)/\Z_2$ to the 7d frozen compactification.
There may be subtleties arising from the \emph{free} (co)homology classes \eqref{eq:passed-through-free-classes} we have discarded, since our primary goal was to understand the discrete gauge fields of the SymTFT which in part couple to the center symmetries of the SYM sector.
While we expect this splitting to be natural for the 6d theory (see Appendix \ref{app: decoupled_gapless}), it is a priori not guaranteed that they have a proper 7d uplift, i.e., have a well-defined 7d origin which \emph{also} does not mix with the discrete sector.
From the abridged analysis in Appendix \ref{app: decoupled_gapless}, it seems that the additional $U(1)$ symmetries associated with those discarded free classes mostly have a well-behaved uplift, in that they give rise to a decoupled KK-sector and the 6d analogue of the 7d instanton symmetry.
However, the fields associated with a set of free classes in $H^2(\wZ, \partial \wZ)$ and $H^3(\wZ)$ evade such a clean interpretation.
We hope to address their role in the parent 7d theory --- M-theory on a frozen $D_5$ singularity --- which was one of the main motivation for this work, in future works.

Finally, let us mention that our extra-dimensional approach naturally connects, in holographic contexts, to the non-effective action of symmetries in the deep infrared limit. Here, the naturally corresponding setup are those for which the extra-dimensional geometry (and possible structures carried by it, such as flux) varies as a function of the bulk radius which itself may be identified with the RG scale of the boundary theory \cite{deBoer:1999tgo}. Evolving the extra-dimensional geometry, which determines the SymTFT, along the bulk radius it may interpolate between two different structures and indicate how symmetries evolve as a function of scale, see for example \cite{GarciaEtxebarria:2024jfv}. Of course, as these structures are often discrete, such evolution is in many cases not continuous and characterized by localized interfaces attached to the bordisms connecting geometries relevant at different scale. Given these similarities to the cases discussed here, such setups offer naturally adjacent settings for future investigation.

\subsection*{Acknowledgements}

We thank Vivek Chakrabhavi, Jonathan Heckman, Shani Meynet, and Ethan Torres for valuable discussions and feedback. We used GPT 6.1 Sol to generate figures. The authors would like to gratefully acknowledge support from the Simons Center for Geometry and Physics, Stony Brook University, at which some of the research for this paper was performed. S.R. and L.L. thank the organizers of Lotus \& Swamplandia 2026 for the hospitality during part of this work. S.R. and M.H. thank the organizers of Global Categorical Symmetries 2026 for the hospitality during part of this work. S.R. thanks IHP Paris, the Department of Physics at Virginia Tech and the Theory group at UPenn for the hospitality during part of this work.

\appendix

\section{Bottom up approach to modified physical boundaries}\label{app:bottom-up}

In this appendix, we engineer SymTFTs with modified physical boundary conditions from a bottom--up perspective. These are physical boundaries on which certain bulk topological operators do not terminate. They can be constructed via a two-step process: a half-space gauging \cite{Choi:2021kmx} within the SymTFT, followed by fusing the resulting half space gauging interface with the physical boundary. 

\subsubsection*{General construction}

We consider an extension of finite Abelian groups 
\begin{equation}\label{eq:ses_ZN}
    0 \rightarrow \Z_N^{\text{(sub)}} \rightarrow \Z_{N^2} \rightarrow \Z_N^{\text{(quotient)}} \rightarrow 0 \, .
\end{equation}
Based on this extension, a BF-type action which describes the SymTFT for a pair of $(p-1)$-/$(d-p-1)$-form $\Z_{N^2}$ symmetry is given by \cite{Robbins:2025puq, Apruzzi:2023uma, Bergman:2026lnz},
\begin{equation}\label{eqn:general_symtft}
    S^{(d+1)}_{\text{SymTFT}} := \frac{2\pi}{N}  \int_{\mathcal{M}_{d+1}}  \left( a_p \cup \delta \hat{a}_{d-p} +  b_p \cup \delta \hat{b}_{d-p}  - a_p \cup \text{Bock}(\hat{b}_{d-p})  \right) ,
\end{equation}
where $b_p$ and $\hat{a}_{d-p}$ are $\Z^{\text{sub}}_N$-valued gauge fields, and $a_p$ and $\hat{b}_{d-p}$ are valued in the quotient $\Z^{\text{quotient}}_N$. Here $\text{Bock}: H^i (X, \Z_N^{\text{quotient}}) \rightarrow H^{i+1} (X, \Z_N^{\text{sub}})$ is the Bockstein map of the extension. On cochains it acts as
\begin{equation}
    \text{Bock}(x) = \frac{1}{N}\, \delta x \mod N \, .
\end{equation}
The last term is therefore $\frac{2\pi}{N^2}\, a_p \cup \delta \hat{b}_{d-p}$. It is the mixed anomaly of the two $\Z_N$ symmetries with background fields $a_p$ and $\hat{b}_{d-p}$. With that information, the $a_p$-equation of motion takes the following form, 
\begin{equation}
    \delta \hat{a}_{d-p} = \text{Bock} (\hat{b}_{d-p}) = \frac{1}{N} \delta \hat{b}_{d-p}\, ,
\end{equation}
which encodes the group extension condition \cite{Tachikawa:2017gyf}. Hence, the pair $(\hat{a}_{d-p}, \hat{b}_{d-p})$ encodes together the background gauge field for the $\Z_{N^2}$ $(d-p-1)$-form symmetry. In the same way, $(a_p, b_p)$ can be viewed as the background gauge field of the quantum dual $(p-1)$-form symmetry.

We can now gauge the symmetry generated by the surface operator, $\exp\Big(\frac{2\pi i}{N}\int \hat{{b}}_{d-p} \Big)$, which is a $p$-form symmetry of the bulk, in one half of $\mathcal{M}_{d+1}$. To do so we turn on background gauge field of this symmetry on that half space (call it $\mathcal{M}^+_{d+1}$),
\begin{equation}
    S \supset \frac{2\pi}{N} \int_{\mathcal{M}^+_{d+1}} c_{p+1}\cup \hat{{b}}_{d-p} \, ,
\end{equation}
where $c_{p+1}$ is the background gauge field of the $p$-form symmetry. We can treat $c_{p+1}$ as Lagrange multiplier and set $\hat{b}_{d-p}$ to zero on $\mathcal{M}^+_{d+1}$. After the half-space gauging operation, we are left with a $\Z_N$ BF-action,
\begin{equation}
    S'^{(d+1)}_{\text{SymTFT}} := \frac{2\pi}{N} \int_{\mathcal{M}^+_{d+1}}  a_p \cup \delta \hat{a}_{d-p} \, .
\end{equation}
In this setting, the SymTFTs of $S^{(d+1)}_{\text{SymTFT}}$ and $S'^{(d+1)}_{\text{SymTFT}}$ are now separated by a topological interface. The $\hat{b}_{d-p}$ surface operators terminate on this interface. This is the club sandwich construction of \cite{Bhardwaj:2023bbf, Bhardwaj:2024qrf}. In its language, $S'^{(d+1)}_{\text{SymTFT}}$ is the \textit{reduced topological order}. 

The physical boundary we will consider allows every bulk operator of $S'^{(d+1)}_{\text{SymTFT}}$ to terminate on it.  So, $a_p$ and $\hat{a}_{d-p}$ have Dirichlet boundary condition on the physical boundary. Meanwhile, $\hat{b}_{d-p}$ has Dirichlet boundary condition on the interface. Fusing the interface and the physical boundary gives new physical boundary with Dirichlet boundary condition for $a_p$, $\hat{a}_{d-p}$ and $\hat{{b}}_{d-p}$. However, the field ${b}_{p}$ remains unconstrained. This is the feature observed in the SymTFT analysis of the frozen gauge theories, where certain operators of the SymTFT bulk do not terminate on the physical boundary \cite[Section 5]{Cvetic:2024mtt}. 

The consequence of such a modified physical boundary is that the symmetry generated by $\hat{{b}}_{d-p}$-surfaces acts trivially in the corresponding absolute theory (where $\hat{{b}}_{d-p}$ has Neumann boundary condition on the symmetry boundary), because the operators charged under them (which would allow the $b_p$-surfaces to end) do not exist in the absolute quantum field theory.
In other words, the absolute theory has a trivially acting $\Z_N^{(p-1)}$ symmetry\footnote{Trivially acting symmetries were introduced in a completely different point of view in \cite{Robbins:2022wlr}, where it was also shown that gauging a trivially acting $(d-2)$-form symmetry leads to decomposition.}.

\subsubsection*{Example: SymTFT for a 6d QFT with trivially acting $\Z_2$ 1-form symmetry}

We can now take this general prescription and apply it to a six dimensional quantum field theory with a $\Z^{(1)}_4$ symmetry, with a $\Z^{(1)}_2 \subset \Z^{(1)}_4$ acting trivially. Therefore, we set $d=6, p=2$ and $N=2$. The extension \eqref{eq:ses_ZN} becomes
\begin{equation}\label{eq:sesz4}
    0 \rightarrow \Z^{\text{sub}}_2 \xrightarrow{\times 2} \Z_4 \xrightarrow {\text{mod}\, 2} \Z^{\text{quotient}}_2 \rightarrow 0 \, .
\end{equation}
The 7d SymTFT action is then
\begin{equation}\label{eqn:symtft_mixed}
    S^\text{7d}_{\text{SymTFT}} = \pi \int_{\mathcal{M}_\text{7d}} \left( a_2 \delta \hat{a}_4 + b_2 \delta \hat{b}_4 -\half a_2 \delta \hat{b}_4 \right) \, ,
\end{equation}
here, $b_2$ and $\hat{a}_4$ are valued in the $\Z^{\text{sub}}_2$, while $a_2$ and $\hat{b}_4$ are valued in the $\Z^{\text{quotient}}_2$. The last term  captures the extension data. We can also write the 7d SymTFT action in terms of $\Z_4$ valued fields, instead of a pair of $\Z_2$-valued fields. To see this, we define the $\Z_4$-valued fields
\begin{equation}\label{eq: z4_redef}
    A_2 := a_2 - 2 b_2 \, , \qquad B_4 := 2 \hat{a}_4 - \hat{b}_4 \, .
\end{equation}
In terms of them,
\begin{equation}
    \frac{\pi}{2} \int_{\mathcal{M}_\text{7d}} A_2 \cup \delta B_4 = \pi \int_{\mathcal{M}_\text{7d}} \left( a_2 \cup \delta \hat{a}_4 + b_2 \cup \delta \hat{b}_4 -\half\, a_2 \cup \delta \hat{b}_4 \right) - 2\pi \int_{\mathcal{M}_\text{7d}} b_2 \cup \delta \hat{a}_4 \, .
\end{equation}
The last term is a multiple of $2\pi$ and drops out. So \eqref{eqn:symtft_mixed} is the same as the $\Z_4$ BF theory\footnote{The same holds for any $N$, $p$ and $d$. The fields $A_p := a_p - N b_p$ and $B_{d-p} := N \hat{a}_{d-p} - \hat{b}_{d-p}$ turn \eqref{eqn:general_symtft} into $\frac{2\pi}{N^2} \int A_p \cup \delta B_{d-p}$. In 3d with $N=2$ this is the statement that $\text{Vec}(\Z_4)$ and $\text{Vec}(\Z_2 \times \Z_2, \omega)$ are Morita equivalent.}
\begin{equation}\label{eqn:z4_bf}
    S^\text{7d}_{\text{SymTFT}} = \frac{\pi}{2} \int_{\mathcal{M}_\text{7d}} A_2 \cup \delta B_4 \, .
\end{equation}
However, splitting the $\Z_4$ fields in two $\Z_2$ pairs allows us to apply the half gauging procedure much more conveniently. 

To realize a trivially acting symmetry, only a subset of the bulk topological operators may terminate on the physical boundary. 
In particular, to realize a trivially acting $\Z_2^\text{sub}$ $1$-form symmetry, $b_2$ must have Neumann boundary condition on the physical boundary. Consequently, the $\hat{b}_4$-surfaces would generate a trivially acting 1-form symmetry. 
Starting from \eqref{eqn:symtft_mixed}, such a physical boundary is constructed by half-gauging the symmetry generated by $\hat{b}_4$ surfaces and then fusing the gauging interface (where $\hat{b}_4$-surfaces terminate) with the physical boundary. 

\begin{figure}[ht]
    \centering
    \begin{tikzpicture}[scale=1.35]
    
    \fill[blue!15] (0,0) rectangle (2,3); 
    \fill[green!15] (2,0) rectangle (4,3); 
    
    \draw[thick, black] (2,0) -- (2,3);
    
    \draw[thick] (0,0) -- (0,3);
    \draw[thick] (4,0) -- (4,3);
    
    \node at (0.0,-0.3) {$\mathfrak{B}_\text{sym}$}; 
    \node at (4.5,-0.3) {$\mathfrak{B}_\text{phys}$}; 
    \node at (2.1,3.3) {$\mathcal{I}$}; 
    \node at (1.0, 1.65) {\footnotesize{$S^\text{7d}_{\text{SymTFT}}$}};
    \node at (3.0, 1.65) {$S'$};
    \node at (3.0, -0.3) {${\cal M}^+_\text{7d}$};

    \node at (5,1.5) {$\longrightarrow$}; 

    \begin{scope}[shift={(5.8,0)}] 
        \fill[blue!15] (0,0) rectangle (2,3); 
        
        \draw[thick, black] (2,0) -- (2,3);
        
        \draw[thick] (0,0) -- (0,3);
        \draw[thick] (2,0) -- (2,3); 

        \node at (2.6, 1.7) {$\mathfrak{B}'_\text{phys}$}; 
        \node at (3.4, 1.2) {$\equiv a_2, \hat{b}_4, \hat{a}_4: \text{Dirichlet}$};
        \node at (0,-0.3) {$\mathfrak{B}_\text{sym}$}; 
        \node at (1.0, 1.5) {\footnotesize{$S^\text{7d}_{\text{SymTFT}}$}};
    \end{scope}
    
    \end{tikzpicture}
    
    \caption{ On the left, the half space gauging procedure leads us to a scenario where we have two SymTFTs separated by a topological interface, $\mathcal{I}$. 
    The topological operators of the right half of the sandwich (on the $S'$-side) terminates on the physical boundary. We can now slide the interface and fuse it with the physical boundary ($\mathfrak{B}'_{\text{phys}} = \mathfrak{B}_{\text{phys}} \otimes \mathcal{I}$), resulting in the figure on the right. In this SymTFT, the operator $\exp(i\pi \int b_2)$ does not terminate on $\mathfrak{B}'_{\text{phys}}$. Consequently, the corresponding symmetry generated by the dual object acts trivially, as the operator charged under this symmetry will be missing from the absolute theory.   }
    \label{fig:half_gauging_}
\end{figure}
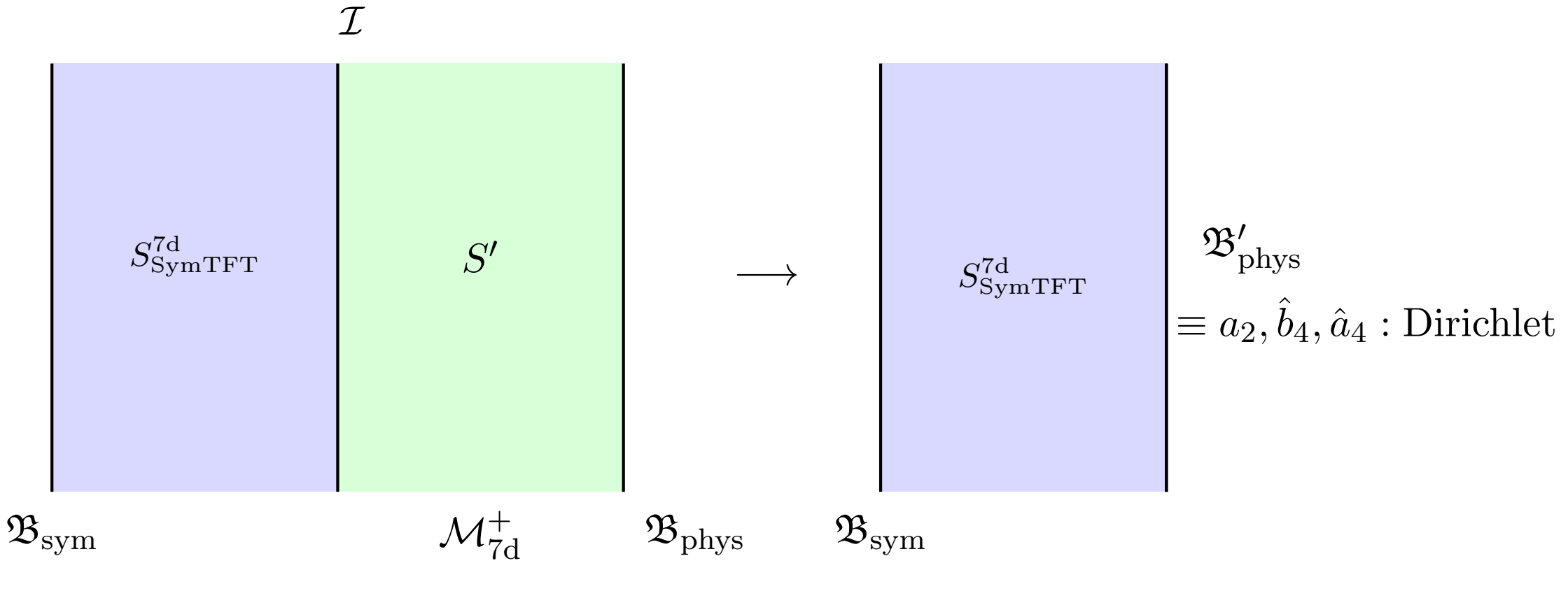

The $\hat{b}_4$-surfaces generate a $2$-form symmetry in the SymTFT bulk, whose charged objects are the $b_2$-surfaces. To gauge it, we add the following term to the action (only in the half space), 
\begin{equation*}
    \pi \int_{\mathcal{M}^+_\text{7d}} c_3 \cup \hat{b}_4 \, ,
\end{equation*}
and sum over $c_3$. As a Lagrange multiplier, $c_3$ enforces $\hat{b}_4 =0$ on the half space.
This leaves us with a ``reduced'' SymTFT (in the half space) captured by the following action:
\begin{equation*}
    S' = \pi \int_{\mathcal{M}^+_\text{7d}} a_2 \delta \hat{a}_4 \, .
\end{equation*}
Fusing the interface between the SymTFT $S^\text{7d}_{\text{SymTFT}}$ and $S'$ with the physical boundary gives a new physical boundary on which the $b_2$-surfaces can not terminate, whereas the $\hat{b}_4$-surfaces do.
Finally, we can also end the $a_2$- and $\hat{a}_4$-surfaces on this physical boundary by introducing suitable degrees of freedom; a simple choice would be an $\mathfrak{su}(2)$ gauge sector with Wilson and 't Hooft operators charged under a $\Z_2^{(1)} \times \Z_2^{(3)}$ defect group.
The situation is summarized in Figure \ref{fig:half_gauging_}.

The possible absolute theories and their symmetries are now determined by the possible choices of symmetry boundary conditions of the action \eqref{eqn:symtft_mixed}.
Let us discuss them one by one.

To begin with, imposing Dirichlet boundary condition on $\hat{a}_4$ and $\hat{b}_4$ gives rise to a $\Z^{(3)}_4$ symmetry that acts effectively.
The corresponding charges are an extension of 't Hooft surfaces from the $\hat{a}_4$-endpoints by the $\hat{b}_4$-endpoints \cite{Tachikawa:2017gyf}. The boundary conditions are summarized in \eqref{tab:bndrycond_1}, where ``N''/``D'' stand for Neumann/Dirichlet boundary conditions:
\begin{align}\label{tab:bndrycond_1}
\renewcommand{\arraystretch}{1.1}
    \begin{array}{|c|c|c|}
    \hline
        \text{Fields} & \text{Symmetry Boundary} & \text{Physical Boundary} \\
        \hline
         a_2 & \text{N} & \text{D}  \\
         \hline
         b_2 & \text{N} & \text{N} \\
         \hline
         \hat{a}_4 & \text{D} & \text{D}  \\
          \hline
         \hat{b}_4 & \text{D} & \text{D}  \\
         \hline
    \end{array}
\end{align}
Changing the boundary conditions on both $\hat{a}_4$ and $\hat{b}_4$ corresponds to gauging the full 3-form symmetry and gives a $\Z^{(1)}_4$ symmetry.
The corresponding boundary conditions are listed in \eqref{tab:bndrycond_2}.
Since $b_2$ has Neumann boundary condition on the physical boundary, there are no genuine line operators in the absolute theory which are charged under the $\hat{b}_4$-surface operators.
Hence the $\Z^{(1)}_2$ subgroup of the $\Z^{(1)}_4$ acts trivially.
\begin{align}\label{tab:bndrycond_2}
\renewcommand{\arraystretch}{1.1}
    \begin{array}{|c|c|c|}
    \hline
        \text{Fields} & \text{Symmetry Boundary}  & \text{Physical Boundary} \\
        \hline
         a_2 & \text{D} & \text{D}  \\
         \hline
         b_2 & \text{D} & \text{N} \\
         \hline
         \hat{a}_4 & \text{N} & \text{D}  \\
         \hline
         \hat{b}_4 & \text{N} & \text{D}  \\
         \hline
    \end{array}
\end{align}
Finally, we may gauge $\Z^{(3)}_2 \subset \Z^{(3)}_4$ (cf.~\eqref{tab:bndrycond_3}). This gives rise to a scenario where we have $\Z^{(1)}_2 \times \Z^{(3)}_2$ symmetry with a mixed anomaly. Both factors act effectively: while the 1-form symmetry acts on the Wilson lines of the gauge sector, the 3-form symmetry acts on the $\hat{b}_4$-endpoints that originate from the interface.
\begin{align}\label{tab:bndrycond_3}
\renewcommand{\arraystretch}{1.1}
    \begin{array}{|c|c|c|}
    \hline
        \text{Fields} & \text{Symmetry Boundary}  & \text{Physical Boundary} \\
        \hline
         a_2 & \text{D} & \text{D}  \\
         \hline
         b_2 & \text{N} & \text{N} \\
         \hline
         \hat{a}_4 & \text{N} & \text{D}  \\
         \hline
         \hat{b}_4 & \text{D} & \text{D}  \\
         \hline
    \end{array}
\end{align}

\section{Generator tracking and sequence chasing in (co)homology}\label{app: homology_tracking}

In this appendix, we first track homology generators in the contracted geometry to establish that the maps in \eqref{eq: iota_maps} vanish, giving a pure boundary condition interface. We then extend the analysis to the resolved geometry $\wZ$ and apply Lefschetz duality to determine the cohomology maps in Table~\ref{tab:maps-twisted-geometry}.

\subsection{Tracking homology generators}\label{app: B.1}

We denote the cycles as follows: $\alpha,\beta,\gamma,F,c$, where 
\begin{itemize}
    \item $F \cong T^2$ is the generic elliptic fiber of $Y^{\rm ell}_{I_2}$,
    \item  the 1-cycles $\alpha$ and $\beta$ generate $H_1(F)$, with $\alpha$ the cycle that shrinks at the central fiber,
    \item $\gamma$ is the circle in the base of $Y^{\rm ell}_{I_2}$ that surrounds the central fiber, and 
    \item $c = S^1/\Z_2\cong S^1$ is the circle factor of $X = (Y^{\rm ell}_{I_2} \times S^1)/\Z_2$. 
\end{itemize}
Beyond that, for a cycle $x$, $\tau(x)$ denotes $x$ swept around $c$. Note that homology computations done below are for the contracted geometry $X$. 

To compute the homology group for a fibration $V\hookrightarrow X\to S^1$ with monodromy $M$, we use the following sequence \cite[Section 4]{Cvetic:2022imb}, 
\begin{equation}
0 \longrightarrow\text{coker}(M_k-1) \longrightarrow H_k(X) \longrightarrow\text{ker}(M_{k-1}-1) \longrightarrow 0,
\end{equation}
where $M_k$ is the representation of the monodromy action $M$ on $H_k(V)$. For the geometry $X$ at hand, the monodromy action on $\alpha$ and $\beta$ around $\gamma$ and $c$ are, respectively,
\begin{equation}
M= \begin{pmatrix} 1 & 2 \\ 0 & 1 \end{pmatrix}, \qquad M^{\mathrm{out}} =  \begin{pmatrix} -1 & 1 \\ 0 &-1 \end{pmatrix}.
\end{equation}
The induced actions on the fiber classes are
\begin{center}
\small
\renewcommand{\arraystretch}{1.25}
\begin{tabular}{c|cccccc}
 & $\mathrm{pt}$ & $\alpha$ & $\beta$ & $F$ & $\gamma$
 & $T^2_{\alpha,\gamma}$ \\ 
\hline
$M$ & $\mathrm{pt}$ & $\alpha$ & $\beta+2\alpha$ & $F$ & --- & --- \\
$M^{\mathrm{out}}$ & $\mathrm{pt}$ & $-\alpha$ & $\alpha-\beta$
& $F$ & $\gamma$ & $-T^2_{\alpha,\gamma}$
\end{tabular}
\end{center}
where $T^2_{\alpha,\gamma}$ is $\alpha$ swept around $\gamma$.

\subsubsection*{The elliptic boundary $\partial Y_{I_2}^{\mathrm{ell}} $}

For the fibration $F\hookrightarrow\partial Y_{I_2}^{\mathrm{ell}}\to S^1_\gamma$, we obtain the classes summarized in Table~\ref{tab: boundary-Y}.
\begin{table}[H]
\centering
\small
\renewcommand{\arraystretch}{1.5}
\setlength{\tabcolsep}{4pt}
\begin{tabularx}{\textwidth}{|c|
>{\centering\arraybackslash}X|
>{\centering\arraybackslash}X|
>{\centering\arraybackslash}X|}
\hline
$k$ & Fiber classes ($\text{coker}$)
& Invariant classes  ($\text{ker}$)
& $H_k(\partial Y_{I_2}^{\mathrm{ell}})$ \\ \hline
0 & $\mathrm{pt}$ & --- & $\Z\langle pt \rangle $ \\ 
\hline
1 & $\alpha,\beta;\quad 2\alpha=0$
& $\mathrm{pt} \leadsto \gamma$
& $\Z \langle\gamma\rangle \oplus \Z \langle \beta\rangle \oplus \Z_2\langle\alpha\rangle $ \\ 
\hline
2 & $F$ & $\alpha\leadsto T^2_{\alpha,\gamma}$
& $ \Z \langle F\rangle \oplus\Z \langle T^2_{\alpha,\gamma}\rangle $ \\ 
\hline
3 & --- & $F\leadsto [\partial Y_{I_2}^{\mathrm{ell}}] $
& $\Z\langle[\partial Y_{I_2}^{\mathrm{ell}}]\rangle$ \\ 
\hline
\end{tabularx}
\caption{The cycle $\alpha$ has order two.}
\label{tab: boundary-Y}
\end{table}

\subsubsection*{The asymptotic boundary $\partial X$}

For the fibration $\partial Y_{I_2}^{\mathrm{ell}}\hookrightarrow\partial X \to S^1_c$, we obtain the results of Table~\ref{tab: boundary-Z}.

\begin{table}[H]
\centering
\small
\renewcommand{\arraystretch}{1.5}
\setlength{\tabcolsep}{4pt}
\begin{tabularx}{\textwidth}{|c|
>{\centering\arraybackslash}X|
>{\centering\arraybackslash}X|
>{\centering\arraybackslash}X|}
\hline
$k$ & Fiber classes ($\text{coker}$)
& Invariant classes ($\text{ker}$)
& $H_k(\partial X)$ \\ \hline
0 & $\mathrm{pt}$ & --- & $\Z$ \\ 
\hline
1 & $\gamma,\alpha,\beta;\, \alpha = 2\beta$
& $\mathrm{pt}\leadsto c$
& $ \Z\langle \gamma \rangle \oplus \Z \langle c\rangle \oplus\Z_4\langle\beta\rangle$ \\ 
\hline
2 & $F,T^2_{\alpha,\gamma}; \, 2T^2_{\alpha,\gamma}=0$ & $\gamma\leadsto\tau(\gamma), 
\alpha\leadsto\tau(\alpha)$  & $\Z \langle F\rangle \oplus \Z\langle \tau(\gamma)\rangle \oplus \Z_4\langle\tau(\alpha)\rangle$ \\ 
\hline
3 & $[\partial Y_{I_2}^{\mathrm{ell}}]$ & $F\leadsto\tau(F)$
& $ \Z\langle[\partial Y_{I_2}^{\mathrm{ell}}]\rangle \oplus \Z\langle\tau(F)\rangle$ \\ 
\hline
4 & --- & $[\partial Y_{I_2}^{\mathrm{ell}}] \leadsto[\partial X]$
& $\Z \langle[\partial X]\rangle$ \\ 
\hline
\end{tabularx}
\caption{The torsion relations are $\alpha=2\beta$ and $2\tau(\alpha)=T^2_{\alpha,\gamma}$, with $\alpha$ and $T^2_{\alpha,\gamma}$ having order two.}
\label{tab: boundary-Z}
\end{table}

\subsubsection*{The contracted internal space $X$}

For $Y_{I_2}^{\mathrm{ell}} \hookrightarrow X\to S^1_c$, the contracted surface has $H_1(Y_{I_2}^{\mathrm{ell}}) = \Z \langle\beta\rangle$ and $H_2(Y_{I_2}^{\mathrm{ell}})=\Z\langle F\rangle$.
This leads to the classes listed in Table~\ref{tab: contracted-Z}.

\begin{table}[H]
\centering
\small
\renewcommand{\arraystretch}{1.5}
\setlength{\tabcolsep}{4pt}
\begin{tabularx}{\textwidth}{|c|
>{\centering\arraybackslash}X|
>{\centering\arraybackslash}X|
>{\centering\arraybackslash}X|}
\hline
$k$ & Fiber classes ($\text{coker}$)
& Invariant classes  ($\text{ker}$)
& $H_k(X)$ \\ 
\hline
0 & $\mathrm{pt}$ & --- & $\Z$ \\ 
\hline
1 & $\beta;\quad 2\beta=0$ & $\mathrm{pt}\leadsto c$ & $\Z \langle c\rangle \oplus \Z_2 \langle\beta \rangle$ \\ 
\hline
2 & $F$ & --- & $\Z \langle F\rangle$ \\ \hline
3 & --- & $F \leadsto\tau(F)$ & $\Z\langle\tau(F)\rangle$ \\ 
\hline
$4,5$ & --- & --- & $0$ \\ 
\hline
\end{tabularx}
\caption{The image of $\beta$ has order two in $X$.}
\label{tab: contracted-Z}
\end{table}

\subsubsection*{Boundary of the local model $\partial X_L$}

For $S^3/\Z_2\hookrightarrow\partial X_L\to S^1_c$, the nonzero fiber homology groups are $H_0=\Z$, $H_1=\ \Z_2\langle\alpha\rangle$, and $H_3=\Z\langle[S^3/\Z_2]\rangle$. The monodromy acts as
\begin{center}
\small
\renewcommand{\arraystretch}{1.25}
\begin{tabular}{c|ccc}
 & $\mathrm{pt}$ & $\alpha$ (order two) & $[S^3/\Z_2]$ \\ 
\hline
$M^{\mathrm{out}}$ & $\mathrm{pt}$ & $-\alpha=\alpha$ & $[S^3/\Z_2]$
\end{tabular} \, .
\end{center}
This leads to the classes listed in Table~\ref{tab: local-boundary}.

\begin{table}[H]
\centering
\small
\renewcommand{\arraystretch}{1.5}
\setlength{\tabcolsep}{4pt}
\begin{tabularx}{\textwidth}{|c|
>{\centering\arraybackslash}X|
>{\centering\arraybackslash}X|
>{\centering\arraybackslash}X|}
\hline
$k$ & Fiber classes ($\text{coker}$)
& Invariant classes ($\text{ker}$)
& $H_k(\partial X_L)$ \\ 
\hline
0 & $\mathrm{pt}$ & --- & $\Z$ \\ 
\hline
1 & $\alpha; \quad 2\alpha=0$ & $\mathrm{pt}\leadsto c$ & $ \Z \langle c\rangle \oplus \Z_2 \langle \alpha \rangle $ \\
\hline
2 & --- & $\alpha\leadsto\tau(\alpha)$ & $\Z_2 \langle \tau(\alpha)\rangle$ \\ 
\hline
3 & $[S^3/\Z_2]$ & --- & $\Z \langle[S^3/\Z_2] \rangle$ \\
\hline
4 & --- & $[S^3/\Z_2] \leadsto[\partial X_L]$ & $\Z \langle[\partial X_L] \rangle$ \\ 
\hline
\end{tabularx}
\caption{The class $\tau(\alpha)$ has order two on $\partial X_L$,
compared with order four on $\partial X$, while $\alpha$ still has order two.}
\label{tab: local-boundary}
\end{table}

\subsubsection*{Homology of the bordism, $B$}

For the decomposition $X=B\cup X_L$, the Mayer--Vietoris sequence gives 
\begin{equation}
\cdots \longrightarrow H_k(\partial X_L) \longrightarrow H_k(B) \oplus H_k(X_L) \longrightarrow H_k(X) \longrightarrow H_{k-1}(\partial X_L) \longrightarrow \cdots.
\end{equation}
Using the homology groups computed in previous tables, we compute the homology of $B$ listed in Table~\ref{tab:bordism-homology}. 

\begin{table}[H]
\centering
\small
\renewcommand{\arraystretch}{1.5}
\setlength{\tabcolsep}{4pt}
\begin{tabularx}{\textwidth}{|c|
>{\centering\arraybackslash}X|
>{\centering\arraybackslash}X|
>{\centering\arraybackslash}X|}
\hline
$k$ & $H_k(\partial X_L)$ & $H_k(X)$ & $H_k(B)$ \\ 
\hline
0 & $\Z$ & $\Z$ & $\Z$ \\ 
\hline
1 & $ \Z \langle c\rangle \oplus \Z_2\langle \alpha \rangle $ & $ \Z\langle c\rangle  \oplus\Z_2 \langle \beta \rangle$ & $ \Z\langle c\rangle \oplus \Z_4\langle\beta\rangle$ \\ 
\hline
2 & $\Z_2\langle\tau(\alpha)\rangle$ & $\Z\langle F\rangle$ & $\Z\langle F\rangle \oplus\Z_2\langle\tau(\alpha)\rangle$ \\ 
\hline
3 & $\Z\langle[S^3/\Z_2]\rangle$ & $\Z\langle\tau(F)\rangle$ & $ \Z\langle[S^3/\Z_2]\rangle
\oplus\Z\langle\tau(F)\rangle$ \\
\hline
4 & $\Z \langle[\partial X_L] \rangle$ & $0$
& $\Z\langle[\partial X_L]\rangle$ \\ 
\hline
\end{tabularx}
\caption{Mayer--Vietoris data and the homology generators of $B$.}
\label{tab:bordism-homology}
\end{table}
\subsubsection*{Relative homology of the bordism}

Denote $\partial_A B=\partial X$ and $\partial_L B=\partial X_L$. The long exact sequence of the pair $(B,\partial B)$ reads
\begin{equation}
\cdots\longrightarrow H_k(\partial X_L \sqcup \partial X )\xrightarrow{i_*}H_k(B)
\longrightarrow H_k(B,\partial X_L \sqcup \partial X )
\xrightarrow{\partial}H_{k-1}(\partial X_L \sqcup \partial X )\longrightarrow\cdots.
\end{equation}
Let $I$ be a chain joining a point in $\partial X_L$ to a point in $\partial X$. We will write $x\times I$ for a tube joining boundary classes
and $\text{cone}(x)$ for a chain bounding $x$ on $\partial X$.
Then the relative homology groups of $B$ are as shown in Table~\ref{tab:bordism-relative-homology}.

\begin{figure}[H]
\centering
\begin{tikzpicture}[scale=1.0, every node/.style={font=\small}]
\colorlet{nullcol}{blue!65!black}
\colorlet{partcol}{orange!80!black}


\fill[gray!6] (0,1.6) -- (7,1.6) arc (90:-90:0.5 and 1.6) -- (0,-1.6) arc (-90:90:0.5 and 1.6) -- cycle;
\fill[gray!6] (0,1.6) -- (7,1.6) arc (90:270:0.5 and 1.6) -- (0,-1.6) arc (270:90:0.5 and 1.6) -- cycle;
\draw[thick] (0,1.6) -- (7,1.6);
\draw[thick] (0,-1.6) -- (7,-1.6);
\draw[thick, dotted] (0,1.6) -- (-1.3,1.6);
\draw[thick, dotted] (0,-1.6) -- (-1.3,-1.6);
\draw[-{Stealth[length=2mm]}] (-0.6,0) -- (-1.25,0);
\node[left] at (-1.3,0) {$r\to\infty$};

\fill[gray!22] (10.9,0) -- (8.4,1.6) arc (90:-90:0.5 and 1.6) -- cycle;
\fill[gray!22] (10.9,0) -- (8.4,1.6) arc (90:270:0.5 and 1.6) -- cycle;
\draw[gray!60!black, thick] (8.4,1.6) -- (10.9,0);
\draw[gray!60!black, thick] (8.4,-1.6) -- (10.9,0);
\draw[gray!60!black, thick, dashed] (8.4,1.6) arc (90:270:0.5 and 1.6);
\draw[gray!60!black, thick]         (8.4,1.6) arc (90:-90:0.5 and 1.6);
\fill[gray!60!black] (10.9,0) circle (0.05);
\node[right, gray!40!black] at (10.95,0) {singular locus};
\node[gray!40!black] at (9.5,0) {$X_L$};
\node[below] at (8.4,-1.7) {$\partial X_L$};

\draw[thick, dashed] (0,1.6) arc (90:270:0.5 and 1.6);
\draw[thick]         (0,1.6) arc (90:-90:0.5 and 1.6);
\draw[thick, dashed] (7,1.6) arc (90:270:0.5 and 1.6);
\draw[thick]         (7,1.6) arc (90:-90:0.5 and 1.6);

\node[below] at (0,-1.7) {$\partial_A B=\partial X$};
\node[below] at (6.9,-1.7) {$\partial_L B$};
\node at (5.8,-0.1) {\large $B$};

\fill[partcol!15] (0.43,1.1) -- (7.43,1.1) arc (90:-90:0.12 and 0.3) -- (0.43,0.5) arc (-90:90:0.12 and 0.3) -- cycle;
\draw[partcol, thick] (0.43,1.1) -- (7.43,1.1);
\draw[partcol, thick] (0.43,0.5) -- (7.43,0.5);
\draw[partcol, thick] (0.43,0.8) ellipse (0.12 and 0.3);
\draw[partcol, thick] (7.43,0.8) ellipse (0.12 and 0.3);
\node[partcol, anchor=north west] at (0.55,0.48) {$\alpha_A=2\beta_A$};
\node[partcol, anchor=north east] at (7.35,0.48) {$\alpha_L$};
\node[partcol] at (3.9,0.8) {$\alpha\times I$};
\node[partcol, above] at (3.9,1.12) {\textsf{partial bordism}};

\fill[nullcol!10]
   (0.43,-0.5) .. controls (1.4,-0.5) and (2.6,-0.62) .. (3.4,-0.8)
   .. controls (2.6,-0.98) and (1.4,-1.1) .. (0.43,-1.1)
   arc (-90:90:0.12 and 0.3) -- cycle;
\draw[nullcol, thick, dashed] (0.43,-0.5)  .. controls (1.4,-0.5)  and (2.6,-0.62) .. (3.4,-0.8);
\draw[nullcol, thick, dashed] (0.43,-1.1)  .. controls (1.4,-1.1)  and (2.6,-0.98) .. (3.4,-0.8);
\draw[nullcol, thick] (0.43,-0.8) ellipse (0.12 and 0.3);
\fill[nullcol] (3.4,-0.8) circle (0.04);
\node[nullcol, anchor=north] at (0.8,-1.12) {$\gamma_A$};
\node[nullcol, right] at (3.5,-0.62) {$\mathrm{cone}(\gamma)$};
\node[nullcol, right] at (3.5,-1.0) {\textit{null bordism}};

\end{tikzpicture}
\caption{The bordism $B$ between $\partial_A B=\partial X$ and $\partial_L B=\partial X_L$.
The local model $X_L$ is drawn cut off from $B$. The relative classes of Table~\ref{tab:bordism-relative-homology} are of two kinds. A partial bordism has one boundary component on each end of $B$. For example, $\alpha\times I$, with $\partial(\alpha\times I)=2\beta_A-\alpha_L$. A null bordism ($\mathrm{cone}(\gamma)$) has its boundary on $\partial_A B$ only and caps off inside $B$.}
\label{fig: bordism_classes}
\end{figure}
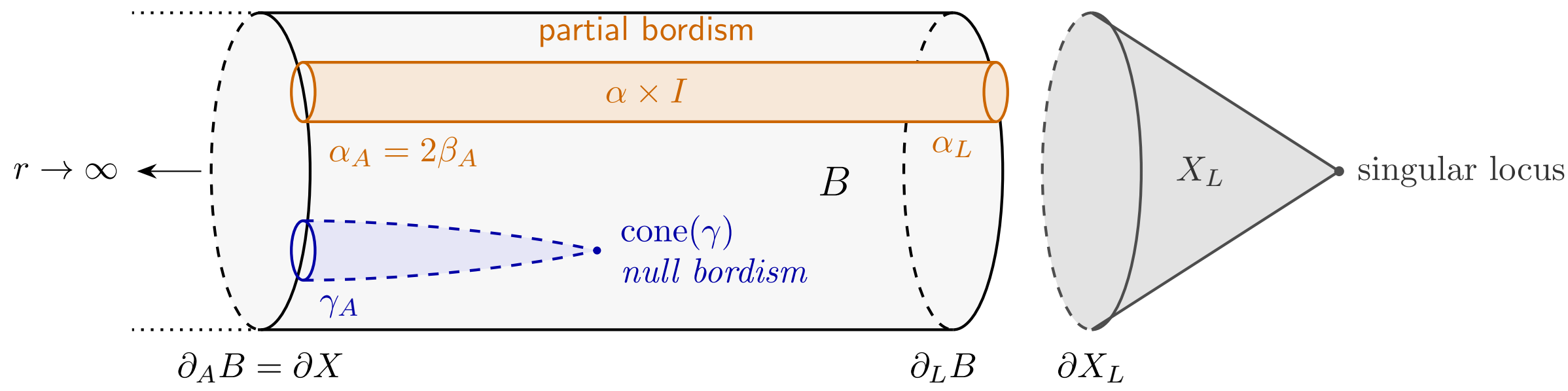

\begin{table}[H]
\centering
\small
\renewcommand{\arraystretch}{1.55}
\setlength{\tabcolsep}{5pt}
\begin{tabularx}{\textwidth}{|c|
>{\centering\arraybackslash}X|
>{\centering\arraybackslash}X|}
\hline
$k$ & $H_k(B,\partial B)$ & Boundaries of the listed generators \\ 
\hline
0 & $0$ & --- \\ 
\hline
1 &
$\Z\langle I\rangle$ & $[\mathrm{pt}]_A-[\mathrm{pt}]_L$ \\ 
\hline
2 & $\begin{gathered} \Z\langle\text{cone}(\gamma)\rangle \\ \oplus\Z\langle c\times I\rangle \\
\oplus \Z_2\langle \alpha\times I\rangle \end{gathered} $
& 
$\begin{gathered}  \gamma_A \, (\text{Null Bordism}),\\ 
c_A-c_L, \\ 
2\beta_A-\alpha_L \, (\text{Partial Bordism}) 
\end{gathered}$ \\ 
\hline
3 & $\Z \langle\text{cone}(\tau(\gamma))\rangle  \oplus\Z_4\langle\tau(\alpha)\times I\rangle $ & $\begin{gathered} \tau(\gamma)_A\, (\text{Null bordism}),  \\ 
\tau(\alpha)_A-\tau(\alpha)_L \, (\text{Partial Bordism}) \end{gathered}
$ \\ 
\hline
4 & $\Z \langle[Y_{I_2}^{\mathrm{ell}}\setminus X_L]\rangle$
& $[\partial Y_{I_2}^{\mathrm{ell}}]_A-[S^3/\Z_2]_L$ \\ 
\hline
5 & $\Z\langle[B]\rangle$ & $[\partial_A B]-[\partial_L B]$ \\ 
\hline
\end{tabularx}
\caption{Relative homology generators, specified by their boundaries on $A$ and $L$.}
\label{tab:bordism-relative-homology}
\end{table}
Figure~\ref{fig: bordism_classes} summarizes the scenario, 

\subsection{Cohomology sequence chasing}

We now use the homology generators introduced above to derive the cohomology maps in Table~\ref{tab:maps-twisted-geometry}. We obtain the bulk homology of the resolved geometry $\wZ$ by adding the generators of the additional free cycles to those of $X$, while the boundary homology is unchanged. Combining these groups with the relative homology of $\wZ$ from \cite[Appendix C]{Cvetic:2024mtt}, we track the generators through the long exact sequence in homology and apply Lefschetz duality.

\subsubsection*{Relevant homology groups and generators}

The boundary $\partial \wZ = \partial X$ is unchanged by the resolution. From Table~\ref{tab: boundary-Z} we have
\begin{align}\label{eq:res_boundary}
    H_n(\partial \wZ) = \begin{cases}
        \Z \langle \gamma \rangle \oplus \Z \langle c \rangle \oplus \Z_4 \langle \beta \rangle \,, & n = 1 \,, \\
        \Z \langle F \rangle \oplus \Z \langle \tau(\gamma) \rangle \oplus \Z_4 \langle \tau(\alpha) \rangle \,, & n = 2 \,, \\
        \Z \langle [\partial Y^{\rm ell}_{I_2}] \rangle \oplus \Z \langle \tau(F) \rangle \,, & n = 3 \,, \\
        \Z \langle [\partial \wZ] \rangle \,, & n = 4 \,,
    \end{cases}
\end{align}
with the relations $\alpha = 2 \beta$ and $T^2_{\alpha,\gamma} = 2 \tau(\alpha)$. 

Next, we turn to $\wZ$.  The homology groups of the contracted geometry $X$ are specified in Table~\ref{tab: contracted-Z}. The resolution adds one free generator in degree two and one in degree three~\eqref{eq:homology_resolved},
\begin{align}\label{eq:res_bulk}
    H_n(\wZ) = \begin{cases}
        \Z \langle c \rangle \oplus \Z_2 \langle \beta \rangle \,, & n = 1 \,, \\
        \Z \langle F \rangle \oplus \Z \langle E \rangle \,, & n = 2 \,, \\
        \Z \langle \tau(F) \rangle \oplus \Z \langle \tau(E) \rangle \,, & n = 3 \,, \\
        0 \,, & n = 4, 5 \,.
    \end{cases}
\end{align}
Here $E$ is the exceptional $\mathbb{P}^1$ of the resolved central fiber and $\tau(E)$ is $E$ swept around $c$. They generate the additional free factors.

The relative homology groups of the resolved geometry are those of \eqref{eq: relative_homology_resolved}. We write them with generators as
\begin{align}\label{eq:res_relative}
    H_n(\wZ, \partial \wZ) = \begin{cases}
        0 \,, & n = 0, 1 \,, \\
        \Z \langle D \rangle \oplus \Z \langle \mathcal{C} \rangle \,, & n = 2 \,, \\
        \Z \langle \tau(D) \rangle \oplus \Z \langle \tau(\mathcal{C}) \rangle \oplus \Z_2 \langle W \rangle \,, & n = 3 \,, \\
        \Z \langle [Y^{\rm ell}_{I_2}] \rangle \,, & n = 4 \,, \\
        \Z \langle [\wZ] \rangle \,, & n = 5 \, ,
    \end{cases}
\end{align}
where we have introduced the relative cycles $D, \mathcal{C},\tau(D), \tau(\mathcal{C})$ and $W$.
\begin{itemize}
    \item $D$ is the base of the elliptic fibration in $Y^{\text{ell}}_{I_2}$ touching the central fiber at a point.
    \item  In $\wZ$ the central fiber $F_0$ is resolved into two $\mathbb{P}^1$'s, $E$ and $E'$, meeting transversally at two points, and the generic fiber degenerates onto it, $ F = E + E' \quad \text{in } H_2(\wZ)$. This is why $F$ and $E$ can be taken as the basis of $H_2(\wZ)$ in \eqref{eq:res_bulk}, and $\tau(F)$ and $\tau(E)$ as the basis of $H_3(\wZ)$. The cycle $\alpha$ of the generic fiber is the one that becomes contractible at the central fiber, as in Appendix~\ref{app: B.1}. Then $\mathcal{C}$ is the relative cycle swept out by pulling $\alpha$ into the interior and shrinking it to zero. We transport $\alpha$ along an annulus from $\partial Y^{\rm ell}_{I_2}$ to the central fiber. There $\alpha$ becomes the equator of $E$. We glue the annulus onto a hemisphere $h \subset E$ to complete $\mathcal{C}$. The boundary of $\mathcal{C}$ is $\alpha$.
    \item $W$ is the relative cycle obtained by fibering the normal (inside of $Y^{\text{ell}}_{I_2}$) circle bundle over the hemisphere of $E$.  $[W]$ has order two in $H_3(\widehat{X},\partial \wZ)$ since it is odd under the action of the involution.
    \item Finally, $\tau(\mathcal{C})$ and $\tau(D)$ are obtained by sweeping $\mathcal{C}$ and $D$ around $c$.
    \end{itemize}
In addition to this, the boundaries of the relative generators are
\begin{equation}\label{eq:boundaries_resolved}
\begin{aligned}
    \partial D &= \gamma \,, & \partial \mathcal{C} &= \alpha = 2 \beta \,, & \partial [Y^{\rm ell}_{I_2}] &= [\partial Y^{\rm ell}_{I_2}] \,, \\
    \partial \tau(D) &= \tau(\gamma) \,, & \partial \tau(\mathcal{C}) &= \tau(\alpha) \,, & \partial W &= T^2_{\alpha,\gamma} = 2 \tau(\alpha) \,.
\end{aligned}
\end{equation}
We can collect all these data from equations~\eqref{eq:res_boundary}, \eqref{eq:res_bulk}, \eqref{eq:res_relative} into the long exact sequence of homology, 
\begin{equation}\label{eq:LES_homology_resolved}
    \scalebox{0.9}{$
    \begin{array}{c||ccccccc}
    & H_n(\partial \wZ) & & H_n(\wZ) & & H_n(\wZ, \partial \wZ) & \\[0.4em] \hline \hline \\[-0.9em]
    n=4 ~& \Z \langle [\partial \wZ] \rangle & \xrightarrow{\imath_*} & 0 & \xrightarrow{\jmath_*} & \Z \langle [Y^{\rm ell}_{I_2}] \rangle & \xrightarrow{\partial} \\
    n=3 ~& \Z \langle [\partial Y^{\rm ell}_{I_2}] \rangle \oplus \Z \langle \tau(F) \rangle & \xrightarrow{\imath_*} & \Z \langle \tau(F) \rangle \oplus \Z \langle \tau(E) \rangle & \xrightarrow{\jmath_*} & \Z \langle \tau(D) \rangle \oplus \Z \langle \tau(\mathcal{C}) \rangle \oplus \Z_2 \langle W \rangle & \xrightarrow{\partial} \\
    n=2 ~& \Z \langle F \rangle \oplus \Z \langle \tau(\gamma) \rangle \oplus \Z_4 \langle \tau(\alpha) \rangle & \xrightarrow{\imath_*} & \Z \langle F \rangle \oplus \Z \langle E \rangle & \xrightarrow{\jmath_*} & \Z \langle D \rangle \oplus \Z \langle \mathcal{C} \rangle & \xrightarrow{\partial} \\
    n=1 ~& \Z \langle \gamma \rangle \oplus \Z \langle c \rangle \oplus \Z_4 \langle \beta \rangle & \xrightarrow{\imath_*} & \Z \langle c \rangle \oplus \Z_2 \langle \beta \rangle & \xrightarrow{\jmath_*} & 0. &
    \end{array}
    $}
\end{equation}
Under Lefschetz duality this long exact sequence produces the cohomology sequence in \eqref{eq:long_seq-twisted_geometry}. The generators of the homology groups get mapped according to the Table \ref{tab:homology-maps-resolved}.

\begin{table}[H]
\centering
\renewcommand{\arraystretch}{1.5}
\setlength{\tabcolsep}{6pt}
\newcommand{\gcell}[1]{\raisebox{0pt}[\dimexpr\height+3pt\relax][\dimexpr\depth+3pt\relax]{$\begin{gathered}\relax#1\end{gathered}$}}
\begin{tabular}{|c||c|c|c|}
    \hline
    & $\imath_* : H_n(\partial \wZ) \to H_n(\wZ)$ & $\jmath_* : H_n(\wZ) \to H_n(\wZ, \partial \wZ)$ & $\partial : H_n(\wZ, \partial \wZ) \to H_{n-1}(\partial \wZ)$ \\
    \hline\hline
    $n = 4$ & $[\partial \wZ] \mapsto 0$ & $0$ & $[Y^{\rm ell}_{I_2}] \mapsto [\partial Y^{\rm ell}_{I_2}]$ \\
    \hline
    $n = 3$ & \gcell{[\partial Y^{\rm ell}_{I_2}] \mapsto 0 \\ \tau(F) \mapsto \tau(F)}
            & \gcell{\tau(F) \mapsto 0 \\ \tau(E) \mapsto 2 \tau(\mathcal{C}) + W}
            & \gcell{\tau(D) \mapsto \tau(\gamma) \\ \tau(\mathcal{C}) \mapsto \tau(\alpha) \\ W \mapsto 2 \tau(\alpha)} \\
    \hline
    $n = 2$ & \gcell{F \mapsto F \\ \tau(\gamma) \mapsto 0 \\ \tau(\alpha) \mapsto 0}
            & \gcell{F \mapsto 0 \\ E \mapsto 2 \mathcal{C}}
            & \gcell{D \mapsto \gamma \\ \mathcal{C} \mapsto \alpha = 2 \beta} \\
    \hline
    $n = 1$ & \gcell{\gamma \mapsto 0 \\ c \mapsto c \\ \beta \mapsto \beta}
            & $0$ & $0$ \\
    \hline
\end{tabular}
\caption{Action of the maps of the long exact sequence \eqref{eq:LES_homology_resolved} on the homology generators.}
\label{tab:homology-maps-resolved}
\end{table}
The maps $\partial$ are fixed by \eqref{eq:boundaries_resolved}. By exactness, $\imath_*$ kills $[\partial Y^\text{ell}_{I_2}], \tau(\gamma), \tau(\alpha)$ and $\gamma$. It maps the remaining generators $\tau(F), F, c$ and $\beta$.

Exactness at $H_n(\wZ)$ and at $H_n(\wZ, \partial \wZ)$ gives $\text{im}(\imath_*)= \ker(\jmath_*) $ and $\text{im}(\jmath_*) = \ker(\partial)$. Focusing on degree 3,  $\ker(\jmath_*) = \langle \tau(F) \rangle$, and $\jmath_* \tau(E)$ generates $\ker(\partial)$.
\begin{equation}
    \partial \big( a \, \tau(D) + b \, \tau(\mathcal{C}) + L \, W \big) = a \, \tau(\gamma) + (b + 2L) \, \tau(\alpha) \,, \qquad a, b \in \Z \,, \quad L \in \{0, 1\} \,.
\end{equation}
This vanishes if and only if 
\begin{equation}
    a = 0, \qquad b + 2L \equiv 0 \mod 4\,.
\end{equation}
Therefore, 
\begin{equation}
    \jmath_* \tau(E) = 2 \tau(\mathcal{C}) + W \,.
\end{equation}
In degree two, $\ker(\jmath_*) = \langle F \rangle$ and $\jmath_* E$ generates $\ker(\partial)$. But 
\begin{equation}
    \partial ( a \, D + b \, \mathcal{C} ) = a \, \gamma + 2 b \, \beta 
\end{equation}
vanishes if and only if $a = 0$ and $b$ is even, since $\beta$ has order four on $\partial \wZ$. Therefore, 
\begin{equation}
    \jmath_* E = 2 \mathcal{C}\,.
\end{equation}

\subsubsection*{Dualizing to cohomology}

Now, we dualize to cohomology. Lefschetz duality for the compact five-manifold $\wZ$ and Poincar\'e duality for its closed boundary,  
\begin{equation}
    H^k(\wZ, \partial \wZ) \cong H_{5-k}(\wZ) \,, \qquad H^k(\wZ) \cong H_{5-k}(\wZ, \partial \wZ) \,, \qquad H^k(\partial \wZ) \cong H_{4-k}(\partial \wZ) \,,
\end{equation}
allow us to write down the dual cohomology sequence,  
\begin{equation}\label{eq:dual_sequences}
\renewcommand{\arraystretch}{1.3}
\begin{array}{c @{\;} c @{\;} c @{\;} c @{\;} c @{\;} c @{\;} c}
    H^k (\wZ, \partial \wZ) & \xrightarrow{\;\jmath^*_k\;} & H^k(\wZ) & \xrightarrow{\;\imath^*_k\;} & H^k(\partial \wZ) & \xrightarrow{\;d_k\;} & H^{k+1} (\wZ, \partial \wZ) \\
    \rotatebox[origin=c]{90}{$\cong$} & & \rotatebox[origin=c]{90}{$\cong$} & & \rotatebox[origin=c]{90}{$\cong$} & & \rotatebox[origin=c]{90}{$\cong$} \\
    H_{5-k}(\wZ) & \xrightarrow{\;\jmath_*\;} & H_{5-k}(\wZ, \partial \wZ) & \xrightarrow{\;\partial\;} & H_{4-k}(\partial \wZ) & \xrightarrow{\;\imath_*\;} & H_{4-k}(\wZ)\,.
\end{array}
\end{equation}
We list the dual generators in Table~\ref{tab:duality_dictionary}.  
\begin{table}[H]
\centering
\small
\renewcommand{\arraystretch}{1.5}
\setlength{\tabcolsep}{5pt}
\begin{tabular}{|c|c||c|c|}
\hline
Cohomology groups in \eqref{eq:long_seq-twisted_geometry} & Generators & Dual Homology group & Generators \\
\hline
$H^1(\wZ)$ & $u$ & $H_4(\wZ, \partial \wZ)$ & $[Y^{\rm ell}_{I_2}]$ \\
\hline
$H^1(\partial \wZ)$ & $(v_1, v_2)$ & $H_3(\partial \wZ)$ & $(\tau(F), [\partial Y^{\rm ell}_{I_2}])$ \\
\hline
$H^2(\wZ,\partial \wZ)$ & $(\rho_2, \omega_2)$ & $H_3(\wZ)$ & $(\tau(F), \tau(E))$ \\
\hline
$H^2(\wZ)$ & $(z_2, \eta_2; \tau_2)$ & $H_3(\wZ, \partial \wZ)$ & $(\tau(D), \tau(\mathcal{C}); W)$ \\
\hline
$H^2(\partial \wZ)$ & $(a_2, b_2; x_2)$ & $H_2(\partial \wZ)$ & $(\tau(\gamma), F; \tau(\alpha))$ \\
\hline
$H^3(\wZ,\partial \wZ)$ & $(\omega_3, \rho_3)$ & $H_2(\wZ)$ & $(E, F)$ \\
\hline
$H^3(\wZ)$ & $(\eta_3, z_3)$ & $H_2(\wZ, \partial \wZ)$ & $(\mathcal{C}, D)$ \\
\hline
$H^3(\partial \wZ)$ & $(a_3, b_3; x_3)$ & $H_1(\partial \wZ)$ & $(c, \gamma; \beta)$ \\
\hline
$H^4(\wZ,\partial \wZ)$ & $(\rho_4; \sigma_4)$ & $H_1(\wZ)$ & $(c; \beta)$ \\
\hline
$H^4(\wZ)$ & $0$ & $H_1(\wZ, \partial \wZ)$ & $0$ \\
\hline
\end{tabular}
\caption{The generators of Table~\ref{tab:maps-twisted-geometry} as Lefschetz duals of the homology generators.}
\label{tab:duality_dictionary}
\end{table}
Reading the maps between generators of the homology group through \eqref{eq:dual_sequences}, then gives Table~\ref{tab:maps-twisted-geometry}.

\section{Decoupling of free generators}\label{app: decoupled_gapless} 

In this appendix, we explain why we can consistently drop the free factors in the cohomology groups of $\wZ$, whose generators appear in \eqref{eq:passed-through-free-classes}, from our discussion in Section \ref{sec:frozen-setting}.

Let us first summarize the corresponding generators via their dual homology cycles, since this gives a slightly more intuitive understanding. (Note we only consider the free parts of the (co)homology groups here.)
\begin{equation}\label{eq:free-generators_dual-cycles}
    \begin{array}{lcl}
        H^4(\wZ, \partial \wZ) = H_1(\wZ) \supset \Z \langle  c \rangle & \longleftrightarrow & \Z\langle Y^\text{ell}_{I_2} \rangle \subset H_4(\wZ, \partial \wZ) = H^1(\wZ) \, ,\\
        H^3(\wZ, \partial \wZ) = H_2(\wZ) \supset \Z \langle F \rangle & \longleftrightarrow & \Z\langle \tau(D) \rangle \subset H_3(\wZ, \partial \wZ) = H^2(\wZ) \, , \\
        H^2(\wZ, \partial \wZ) = H_3(\wZ) \supset \Z \langle \tau(F) \rangle & \longleftrightarrow & \Z\langle D \rangle \subset H_2(\wZ, \partial \wZ) = H^3(\wZ) \, ,
    \end{array}
\end{equation}
where $c \cong S^1/\Z_2$ is the twisted circle in $\wZ$ over which $Y^\text{ell}_{I_2}$ is fibered, $F$ is the generic elliptic fiber of $Y^\text{ell}_{I_2}$, $D$ is its zero-section, and $\tau( \cdot )$ denoting the 3-cycle obtained by fibering an invariant 2-cycle around the twist circle.
The generators in each line are UCT-dual to each other.
It is easy to see that the compactly supported classes on the left have no compactly supported classes with which they can pair non-trivially: for $c \in H_1(\wZ)$, any such class must be a free element in $H_4(\wZ) = 0$; for $F \in H_2(\wZ)$, the dual class must be in $H_3^\text{free}(\wZ) = \Z\langle\tau(F)\rangle \oplus \Z\langle\tau(E)\rangle$, but $F \cdot_{\wZ} \tau(\Gamma) = F \cdot_{Y^\text{ell}_{I_2}} \Gamma = 0$ for $\Gamma = E,F$; and likewise for $\tau(F) \in H_3(\wZ)$.
This means that the reduction on each of these free factors gives a (higher-form) $U(1)$ gauge field that has no non-zero kinetic term in the 6d action, and is also not coupled to the gauge sector (whose Cartan fluxes are supported on $E$ and $\tau(E)$).
Since they also do not pair with torsion forms, the $U(1)$s obtained in this way also do not couple to the discrete 6d gauge fields that we focused on in Section~\ref{sec: frozen_physical_boundary}.\footnote{A caveat concerns a potential mixed anomaly between the electric 1-form symmetry of the gauge sector and one of these $U(1)$s; we will comment on this below.}

These $U(1)$s also do not couple directly to the SymTFT sector obtained from reducing only on the torsional classes on $\partial \wZ$.
More precisely, the generators in \eqref{eq:free-generators_dual-cycles} have ``identical'' (in the sense that their are related by isomorphisms \eqref{eq:passed-through-free-classes}) counterparts in the (co)homology of $\partial \wZ$ (cf.~Table~\ref{tab:homology-maps-resolved}):
\begin{align}
\begin{array}{lcl}
    H_1(\partial \wZ) \ni c \stackrel{\sim}{\mapsto} c \in H_1(\wZ)  & \longleftrightarrow & H_4(\wZ, \partial \wZ) \ni Y_{I_2}^\text{ell} \stackrel{\sim}{\mapsto} \partial Y_{I_2}^\text{ell} \in H_3(\partial \wZ) \, , \\
    H_2(\partial \wZ) \ni F \stackrel{\sim}{\mapsto} F \in H_2(\wZ) & \longleftrightarrow & H_3(\wZ, \partial \wZ) \ni \tau(D) \stackrel{\sim}{\mapsto} \tau(\gamma) \in H_2(\partial \wZ) \, , \\
    H_3(\partial \wZ) \ni \tau(F) \stackrel{\sim}{\mapsto} \tau(F) \in H_3(\wZ) & \longleftrightarrow & H_2(\wZ, \partial \wZ) \ni D \stackrel{\sim}{\mapsto} \gamma \in H_1(\partial \wZ) \, .
\end{array}
\end{align}
In each line the free boundary cycles are UCT-dual on $\partial \wZ$, and in particular they do not pair non-trivially with the torsion generators of $H_n(\partial \wZ)$.
Thus, there are no quadratic terms in the action involving one of these $U(1)$ gauge fields and one of the discrete gauge fields that were our main focus in Section \ref{sec:frozen-setting}

Let us also give a physical characterization for these gauge fields and their charged objects.
We begin with the class $F$, which is the generic fiber of a torus-fibration structure on $\wZ$ that it inherits from $Y^\text{ell}_{I_2}$.
Using intuition from M-/F-theory duality, we can identify a tower of BPS particles from M2-branes (multi-)wrapping $F$ that naturally has the interpretation of a KK-tower that emerges in a $S^1$-reduction from a 7d theory.
Consequently, we should identify the 2-form ${\cal F}_2$ in the expansion $G_4 = {\cal F}_2 \wedge \text{PD}(\tau(D)) + ...$ with the KK-$U(1)$ field strength.
Wrapping M5-branes on $\tau(D)$ then gives the magnetic KK-``monopoles''; reducing $G_7$ on $\text{PD}(F)$ gives the magnetic field strength of the KK-$U(1)$ which, as mentioned before, has no kinetic term in the 6d theory.
In the 7d SymTFT, the classes $F, \tau(\gamma) \in H_2(\partial \wZ) \stackrel{\text{PD}}{=} H^2(\partial \wZ)$ play a similar role, in that they give rise to the KK-$U(1)$ and its magnetic partner upon reducing $G_{4,7}$ on them.
The maps \eqref{eq:passed-through-free-classes} then identify 6d fields as the restriction of the SymTFT fields to the boundary.

Next we turn to the class $\text{PD}(Y_{I_2}^\text{ell}) \in H^1(\wZ)$.
Reducing $G_4 = H_3 \wedge \text{PD}(Y_{I_2}^\text{ell}) + ...$ on this class gives the field strength $H_3$ for a background 2-form gauge field $B_2$.
The charged objects for it are M2-branes wrapped on $c \in H_1(\wZ)$ which are strings, which we claim are the gauge instantons of the 6d SYM sector (which is non-dynamical with ${\cal N}=(1,1)$ supersymmetry).
To see this, we can think of $c$ as a section of the fibration $Y^\text{ell}_{I_2} \hookrightarrow \wZ \rightarrow S^1/\Z_2$, i.e., it is homologically a point in $Y^\text{ell}_{I_2}$ dragged around the twist circle.
A point in the elliptic surface, however, is homologically equivalent to $[\text{pt}] \sim -\frac12 E \cdot_{Y} E$, where $E$ is the exceptional curve of the $I_2$-fiber.
In $\wZ$ we therefore have $\text{PD}(\tau(E))^2 \propto \text{PD}(c)$.
Since the Cartan flux $f_2$ of the gauge sector arises from the reduction $G_4 = f_2 \wedge \text{PD}(\tau(E)) + ...$, this intersection property implies that the $G_4^3$-term will reduce to a term $B_2 \wedge f_2^2$ with a non-zero coefficient in the 6d action, where $f_2^2$ is proportional to the instanton current on the Coulomb branch of the 6d gauge sector.
Importantly, this is precisely the coupling one would expect also to appear in an $S^1$-reduction of a 7d SYM, where the instanton current couples to a background 3-form $U(1)$ gauge field.

Interestingly, from a bottom-up perspective this coupling would clearly induce a mixed anomaly of the form $\# H_3 \cup E_2^2$, with $E_2$ the background field for the electric 1-form symmetry of the gauge sector, and $\#$ a fractional number \cite{Kapustin:2014gua, Gaiotto:2014kfa, Gaiotto:2017yup, Cordova:2019uob}.
Since in the SymTFT, $E_2$ arises from reduction on $\text{Tor}(H^2(\partial \wZ))$ whereas $H_3$ is the reduction on $\text{PD}(\partial Y_{I_2}^\text{ell}) \in H^1_\text{free}(\partial \wZ)$, it would require a \emph{cubic} pairing involving both torsion and free classes on $\partial \wZ$ to make this term manifest in the SymTFT derived from geometry.
We leave a detailed investigation of this pairing for future works.

Lastly, reducing $G_4$ on $\text{PD}(D) \in H^3(\wZ)$ appears to give an axionic scalar which is non-normalizable (because the cohomology class is non-compactly supported).
In other words, this gives a \emph{theta parameter} of the theory which is the $C_3$-holonomy on the 3-cycle $\tau(F)$.
Naively, such a paramter can be identified with the holonomy of a 7d $U(1)$ background gauge field reduced an $S^1$.
However, it is currently unclear how this background field originates in the frozen 7d theory (i.e., M-theory on a frozen $D_5$ singularity).

\bibliographystyle{JHEP}
\bibliography{ref}
\end{document}